\documentclass[letterpaper]{article} %
\usepackage[draft]{aaai25}  %
\usepackage{times}  %
\usepackage{helvet}  %
\usepackage{courier}  %
\usepackage[hyphens]{url}  %
\usepackage{graphicx} %
\usepackage{natbib}  %
\usepackage{caption} %
\usepackage{algorithm}
\usepackage{algorithmic}

\usepackage{booktabs}
\usepackage{amsmath,amssymb,amsthm}
\usepackage{dsfont}
\usepackage{cleveref}

\newboolean{releaseMode}

\setboolean{releaseMode}{true}

\newcommand{\bel}{Public Belief State}
\newcommand{\bels}{PBS}

\ifthenelse{\boolean{releaseMode}}{
    \usepackage[colorinlistoftodos,prependcaption,disable]{todonotes}
    \newcommand{\para}[1]{}
}{ %
    \usepackage[colorinlistoftodos,prependcaption]{todonotes}
    \presetkeys{todonotes}{inline}{} %
    \newcommand{\para}[1]{\textbf{\color{blue!70!black} § #1:}}
} %

\DeclareMathOperator*{\Exp}{\mathbb{E}}

\newtheorem{definition}{Definition}
\newtheorem{lemma}{Lemma}

\newtheorem{cor}{Corollary}
\newtheorem{example}{Example}
\newtheorem{theorem}{Theorem}
\newtheorem{proposition}{Proposition}

\newcommand{\R}{\mathbb{R}}
\newcommand{\Pro}{\mathbb{P}}
\DeclareMathOperator*{\argmax}{arg\,max}
\renewcommand{\epsilon}{\varepsilon}
\newcommand{\bs}{\boldsymbol}

\usepackage{newfloat}
\usepackage{listings}
\DeclareCaptionStyle{ruled}{labelfont=normalfont,labelsep=colon,strut=off} %
\floatstyle{ruled}
\newfloat{listing}{tb}{lst}{}
\floatname{listing}{Listing}
\title{Computing Perfect Bayesian Equilibria\\ in Sequential Auctions with Verification}
\author{
    Vinzenz~Thoma\textsuperscript{\rm 1,\rm 3},
    Vitor~Bosshard\textsuperscript{\rm 2},
    Sven~Seuken\textsuperscript{\rm 2,\rm 3}
}
\affiliations{
    \textsuperscript{\rm 1}ETH Zurich,\\
    \textsuperscript{\rm 2}University of Zurich,\\
    \textsuperscript{\rm 3}ETH AI Center\\
   
vinzenz.thoma@ai.ethz.ch,
\{bosshard, seuken\}@ifi.uzh.ch
}

\begin{document}

\maketitle

\begin{abstract}
We present an algorithm for computing pure-strategy $\varepsilon$-perfect Bayesian equilibria in sequential auctions with continuous action and value spaces. Importantly, our algorithm includes a \textit{verification phase} that computes an upper bound on the utility loss of the found strategies.
Prior work on equilibrium computation in auctions with verification has focussed on the single-round case, but the methods do not work for sequential auctions because of two main challenges: (1) there are infinitely many subgames, and (2) the setting has no optimal substructure as bidders' beliefs and best response strategies depend on the strategies of previous rounds. We make two contributions. First, we introduce a tailor-made \textit{game abstraction} that discretizes the auction and augments the state space with the \emph{public beliefs}, such that an approximate equilibrium can be computed via dynamic programming. Second, we prove a \textit{decomposition theorem} to upper bound the utility loss of the computed equilibrium. This is essential because it is neither guaranteed that the auction has an equilibrium nor that any algorithm converges to it.
We validate our algorithm on multiple settings with known equilibria and apply it to a new multi-round combinatorial auction.
\end{abstract}

\section{Introduction}

In sequential auctions, multiple items are sold or bought over multiple rounds. Such formats are widely used, for example for selling art, fish, timber, mineral rights, broadcast licences, and government debt \cite{Gale2001Sequential}. Sequential auctions are easy to run in practice and are particularly useful in settings where not all bidders or items are present in all rounds \cite{Parkes2003MDP}.

\subsection{Equilibrium computation in auctions}
As many auctions used in practice are not strategyproof, it is important to study their equilibrium properties. This is necessary to predict how bidders are likely to bid, for providing strategic recommendations to bidders, and to assess the efficiency and revenue properties of new auction designs \cite{Buenz2022Designing}. Consider the famous example of the 2000 Swiss 3G sequential auction that only produced a fraction of the expected revenue due to a poor design and a failure to anticipate the equilibrium outcome \cite{Klemperer2002How}. 

Few analytical results are known for sequential auctions, which motivates computational methods. However, designing scalable or converging algorithms is challenging, given the existing hardness results for computing equilibria \cite{Daskalakis2009complexity,cai2014simultaneous}. Prior work has used iterative best-response algorithms to compute Bayes-Nash equilibria in single-round auctions \cite{Vorobeychik2008Stochastic,Reeves2004Computing,Rabinovich2013ComputingBNEs,Bosshard2020JAIR}.
This does not extend to sequential auctions, due to two complications: (1) the continuous bidding space leads to infinitely many subgames, in each of which the bidders need to play an equilibrium strategy and form beliefs by Bayesian updating, and (2) the auction does not have an optimal substructure, i.e. the equilibrium of a subgame depends on bidding strategies in previous rounds. Therefore, the auction cannot be solved by backward induction.

\subsection{Our Contribution}
We study the problem of computing perfect Bayesian equilibria (PBEs) with verification in sequential auctions. We make two technical contributions. First, we introduce a bespoke game abstraction that reduces the state space to finite cardinality. It further endows the game with an optimal substructure, such that we can compute PBEs by backward induction~(\Cref{sec:abstraction}). We achieve this by (1) introducing public belief states, on which the bidders condition their strategy, (2) restricting them to bid piecewise constant, and (3) choosing how bidders update their beliefs off-equilibrium. To find a PBE candidate in the abstraction, we recursively compute Bayes-Nash equilibria~(BNEs) in each round of the auction via an iterated best-response algorithm~(\Cref{sec:ibralg}). 

Second, we prove a bound on the utility loss of our computed PBEs in the unabstracted game (Section \ref{sec:verification}). Our proof exploits the optimal substructure and the piecewise convexity of bidders' utilities to bound the utility loss over the infinite set of possible types and histories. To approximate the bound for a given set of strategies, we provide an easy-to-implement verification procedure~(Section~\ref{sec:ver_proc}). 

Our algorithm reproduces existing analytical results in multiple settings, including sequential sales with and without reserve prices, as well as procurement auctions. It further finds interesting strategic behavior in a novel sequential auction with combinatorial values (\Cref{sec:experiments}). %

\section{Related Work}
\label{sec:rel_work}
\subsection{Analytical Equilibria in Sequential Auctions}
 The seminal work of \citet{Weber2000Theory} derived equilibria for sequential auctions with single-minded bidders. This was later extended to multi-unit demand \cite{Katzman1999Two,Rodriguez2009Sequential,Gale2001Sequential} and round-dependent reserve prices \citep{Gong2013Ordering,Landi2018Sequential}. Another line of work has focused on equilibria of sequential procurement auctions~\cite{Gale2000Sequential,McAfee1993Declining,Ashenfelter1989How,kokott2019beauty}.

\subsection{Equilibrium Computation}
\label{sec:rel_eq_comp}

There exists a vast literature on equilibrium computation in extensive-form games. The main approaches are: computing best-responses
\cite{Hendon1996Fictitious,McMahan2007Fast,Ganzfried2009Computing,Heinrich2015Fictitious}, counterfactual regret minimization \cite{Zinkevich2007Regret,Moravcik2017DeepStack,Brown2019Superhuman},
and reinforcement learning (RL) \cite{Silver2016Mastering,Brown2020Combininga}.

These methods generally assume finite actions and often restrict to two-player zero-sum games. In contrast, auctions are general-sum and have a continuous type and action space. This gap has motivated auction-specific equilibrium solvers. Ideas include using deep learning \cite{Martin2023Finding,Bichler2021Learning}, gradient-based optimization \cite{Bichler2023Computing,Kohring2023Enablinga}, and iterated best response computation~\cite{Vorobeychik2008Stochastic,Reeves2004Computing,Rabinovich2013Computing,Li2021Evolution}. Generally, these approaches compute equilibria in an abstracted version of the game, without guarantees on the \textit{utility loss} in the original auction. Bosshard {et al.} \shortcite{bosshard2017fastBNE,bosshard2018nondecreasing,Bosshard2020JAIR} provide a verification procedure to bound this loss. %
However, all the above approaches for equilibrium computation are limited to single-round auctions.

 There are few results on equilibrium computation in {sequential} auctions. The concurrent work of \citet{pieroth2023equilibrium} and an earlier work by \citet{Greenwald2012Approximating}  both iteratively use RL to compute \textit{ex-ante} Nash equilibria, such that bidders's strategies are best responses to each other \textit{in expectation} before the auction starts and bidders observe their own type. In contrast, we focus on \textit{ex-interim} perfect Bayesian equilibria, requiring that bidders best respond in each subgame after having observed their type and updating their beliefs. \citet{dEon2024Understanding} use RL to compute equilibria in auctions with demand instead of value queries, and \citet{Thoma2023Learning} use it to study the exploitability of truthful bidding.

From an algorithmic perspective, these RL approaches iterate over bidders and approximately compute their best responses for the whole auction by solving single-agent Bellman equations. In contrast, we first introduce equilibrium Bellman equations. These decompose the auction into simpler single-round auctions, whose equilibria we subsequently find via iterated best responses.

\section{Preliminaries}
\label{sec:prel}
\subsection{Formal Model}
A set $N$ of $n$ bidders compete over $T$ rounds for a set $K$ of $k$ items. Each bidder $i$ has a type $\theta_i \in \Theta_i\subseteq \R^{d_i}$ with $d_i \in \mathbb{N}_{+}$. $\theta_i$ is a continuous random variable with density $f_i$. We use the subscript $-i$ to denote all bidders except $i$ and no subscript to denote all bidders, e.g. $\theta_{-i}$ denotes all but $i$'s types and $\theta$ all types. 
In every round $t$, bidders can submit an XOR bid ${b}_{i,t} \in \R_{\geq 0}^{q}$, where $q\leq2^{K}$.\footnote{We use the XOR language for ease of exposition, but our results apply without loss of generality to other bidding languages.} Given $b_t=(b_{1,t},\dots,b_{n,t})$, the auctioneer employs an allocation rule $X_t(b_t|x_{<t})$, where $x_{<t}$ denotes past allocations. $X_t$ deterministically assigns a subset $K_t \subseteq K\setminus \bigcup_{\tau<t} K_\tau$ of the remaining items to the bidders, who get charged according to a deterministic payment rule $P_t(b_t|x_{<t})\in \R^n$. The allocation $x_t$ and a deterministic announcement $a_t=A_t(b_t|x_{<t})$ is made public after each round. Examples for $A_t$ are announcing the winning bid, all bids, or the winner's payment.

Bidders can decide to \textit{drop out} of the auction, e.g. if they already won what they are interested in. $N_t(x_{<t})\subseteq N$ denotes the set of bidders remaining at round $t$, given $x_{<t}$.  

We illustrate the model in the following example, which we elaborate on in \Cref{app:example}.
\begin{example}[Two-round FPSB auction]
    Consider a two-round first-price sealed-bid (FPSB) auction with uniform types where the highest bid wins. In this case, $\theta_i\in [0,1]$, $f_i(\theta_i)=1$. The bidding space is $\R_{\geq0}$. The allocation rule assigns the good to the highest bidder, i.e. $X_t = \arg\max_i b_{i,t}$ for $t\in \{1,2\}$. The auctioneer announces the winning bid, such that $A_t(b_t|h_t)=P_t(b_t|h_t)=\max_i b_{i,t}$. As bidders are single-minded in the first round, we have $\forall i: v_i(\theta_i,\{i\}|\emptyset)=\theta_i$, where $v_i$ is the value function of bidder $i$.
\end{example}

A sequential auction $\mathcal{A}$ can be modeled as a Bayesian extensive-form game. The state space is $S=\Theta\times H$, where $\Theta=\{\Theta_1,\dots,\Theta_n\}$ and $H$ is the set of \textit{histories}. A history $h_t=\{x_1,a_1,\dots, x_{t-1},a_{t-1}\}$ consists of past allocations and announcements. Bidders act simultaneously. The resulting allocation and announcement define the transition to the next state. The possible successors of state $s_t$ are denoted by $N^+(s_t)=\{s|\theta^s=\theta^{s_t}\wedge \exists x_{t},a_{t}: h^s=h^{s_t}\cup \{x_{t},a_{t}\}\}$, where $\theta^s,h^s$ denote the types and history of $s$. $N^+$ endows $S$ with a tree structure, which we call \textit{game tree}.
 The game starts with nature choosing $\theta$ and entering the state $s_0=(\theta,\emptyset)$. A \textit{subgame} at $s$ is an auction that starts from $s$ instead of $s_0$.

Each bidder $i$ does not fully observe $s$ but only their respective \textit{information set} $I_i(s)=(\theta_i^s,h^s)$. A (pure) bidding strategy is thus a function $\sigma: \mathcal{I}_i \rightarrow \R^q_{\geq 0}, (\theta_i,h)\mapsto \sigma(\theta_i|h)$, where $\mathcal{I}_i$ is the set of $i$'s information sets. 
In \Cref{sec:search} we work with \textit{piecewise-constant} strategies. Following \citet{Bosshard2020JAIR}, we define a \emph{hyperrectangle} $[y,z)= \prod_{k=1}^d [y_k,z_k)$ in $\R^d$ as a Cartesian product of half open intervals and a \textit{partition} $\mathcal{P}_i$ of $\Theta_i$ as a set of disjoint hyperrectangles $[y,z)_j$ covering $\Theta_i$. A strategy $\sigma_i$ is piecewise constant if there exists a partition $\mathcal{P}_i$ such that $\forall [{y},{z})_j\in \mathcal{P}_i, \forall \theta_i \in [{y},{z})_j: \sigma_{i}(\theta_i|h)=\sigma_{i}({y}|h)$.

As bidders do not know each other's types, they form \textit{beliefs}. Initially, the public belief about $\theta_i$, denoted by $\mu_i(\cdot|\emptyset)$, is given by $f_i$. As the game progresses, $h_t$ reveals information about the other bidders's types---referred to as \textit{signaling}---and is used to update the beliefs using Bayes rule. We denote the common belief about bidder $i$ after history $h_t$ by $\mu_i(\cdot|h_t)$ and definde $\mu=(\mu_1,\dots,\mu_n)$. 
To ensure beliefs are independent and common information, we make two assumptions. First, the announcements cannot correlate bidder types. Moreover, any bidder remaining in the auction has no informational advantage from his privately observed payment $p_{i,t}$ over the public observation $(x_t,a_t)$. Both assumptions are satisfied for most standard auctions.\footnote{We defer the details to \Cref{app:bayes}. However, we briefly note that piecewise-constant strategies---as we use later---in combination with random tiebreaking can cause types to become interdependent, which is why we focus on deterministic mechanisms.}

\subsection{Utility and Equilibrium}

Bidder $i$ has a value function $v_{i,t}(\theta_i,{x}_{t}|{x}_{<t})$, linear in $\theta_i$, denoting the value of assignment ${x}_t$ in round $t$, given ${x}_{<t}$.
The corresponding utility is defined as $ u_{i}(\theta_i,{x}_t,{p}_t|x_{<t})=v_{i,t}(\theta_i,{x}_t|{x}_{<t})-{p}_{i,t}$.
Given $h_t$ and bidding strategies $\sigma$, $i$'s expected utility $\overline{u}_i(\theta_i|{\sigma},{h}_t)$ (over all remaining rounds) is defined as: 
\begin{equation}   
\begin{aligned}
    \label{eq:exputil}
   \Exp_{\theta_{-i}\sim\mu_{-i}}\Biggl[\sum_{\tau=t}^T u_{i}(\theta_i,X_\tau({\sigma}(s_{\tau})|{x}_{<\tau}),P_\tau({\sigma}(s_{\tau})|{x}_{<\tau}) )\Biggr],
    \end{aligned}
\end{equation} 
where $\sigma(s_{\tau})=(\sigma_1(\theta^{s_\tau}_1|h^{s_\tau}_\tau),\dots,\sigma_n(\theta^{s_\tau}_n|h^{s_\tau}_\tau))$ and the expectation is taken with respect to $\mu_{-i}(\cdot|h_t)$. 

Given fixed opponent strategies $\sigma_{-i}$, we define the \textit{best response utility} as
$\overline{u}_i^{BR}(\theta_i|{\sigma}_{-i},{h})=\sup_{{\sigma_i}}\overline{u}_i(\theta_i|{\sigma}_i,\sigma_{-i},{h})$.
If the utility-maximizing strategy exists, we refer to it as the \textit{best response} $\sigma_i^{BR}$.
The related \textit{utility loss} of agent $i$ is denoted as $
l_i^{BR}(\theta_i|{\sigma},h) = \overline{u}_i^{BR}(\theta_i|{\sigma}_{-i},{h}) -\overline{u}_i(\theta_i|{\sigma},{h})$.
The latter is used to characterize equilibrium strategies.
\begin{definition}[$\epsilon$-PBE] 
	\label{def:pbe} A tuple $({\sigma},{\mu})$ of strategies and beliefs constitutes an $\epsilon$-perfect Bayesian equilibrium if it fulfills the following conditions:
\begin{enumerate}
    \item \textbf{Sequential rationality:} $\forall i, \forall \theta_i, \forall {h}:l_i^{BR}(\theta_i|\sigma,{h}) \leq \epsilon$.
    \item \textbf{Correct initial beliefs:} $\forall i: \mu_i(\theta_i|{h})=f_i(\theta_i)$.
    \item \textbf{Bayesian updating:} Whenever possible, beliefs are obtained by Bayesian updating.
\end{enumerate}
\end{definition}

\section{Equilibrium Search}
\label{sec:search}

We focus on finding pure-strategy PBEs. First, pure strategies are more intuitive for bidders to follow and simplify the search space. Second, there always exists an $\epsilon$-PBE in pure strategies for some $\epsilon$, which we can upper bound with our verification procedure introduced in \Cref{sec:ver_proc}.\footnote{While auctions without pure-strategy PBEs exist~\citep{Gong2013Ordering}, they are not guaranteed to have a mixed-strategy equilibrium either (consider a first-price auction with one good and one bidder with bidding space $(0,1]$).} Third, most auction mechanisms do not disclose all bids, such that bidders have fewer incentives to mix strategies in order to conceal their types. Fourth, results in finite games link pure-strategy equilibria to better convergence of learning dynamics \cite{Mertikopoulos2016Learning}.

Our approach involves (1) creating a belief-augmented representation, such that finding a PBE can be decomposed into iteratively finding BNEs of the individual auction rounds, and (2) creating a finite abstraction, such that the resulting dynamic program can be approximately solved.

\subsection{Belief-Augmented Game Representation}

In perfect-information extensive-form games, an equilibrium can be written as a dynamic program and computed by backward induction \cite{Laraki2019Mathematical}. This is not true for sequential auctions, as the best response in a subgame depends on the beliefs induced by previous-round strategies. In comparable cases, these issues were resolved by augmenting the state space with the \textit{public beliefs} (also referred to as \textit{common information}) \cite{Brown2020Combininga,Nayyar2014Common,Zhang2021Subgame}. As the \textit{belief-augmented} representation gives agents all necessary information to make their decision, an equilibrium can be computed via backward induction. In particular, \citet{Ouyang2017Dynamic} propose a dynamic program to compute PBEs, where each timestep consists of finding the BNE of a simpler stage game. We extend the approach to sequential auctions, where---in contrast to their setting---actions are not public, bidders can leave and have continuous type and action spaces.

So far, a strategy depended on $\theta_i$ and $h_t=\{x_{<t},a_{<t}\}$. Next, we augment $h_t$ by $\mu$. In fact, as the main purpose of announcements $a_{<t}$ is signaling, we replace them with $\mu$.
\begin{definition}[\bels]
    A \bel~(\bels) is a tuple $(t,x_{<t},\mu)$, where $t$ is the round, $x_{<t}$ the past allocations and $\mu$ the set of current public beliefs.
\end{definition}
We denote a single PBS by $\beta$. Moreover, $\mathcal{B}_t$ is the set of all possible \bels~in round $t$ and $\mathcal{B}=\bigcup_{t=1}^T \mathcal{B}_t$. To ensure correct initial beliefs, we set $\mathcal{B}_1=\{(1,\emptyset,f)\}$.
We define a \bels~ strategy $\sigma_i(\theta_i|\beta)$ as bidding $\sigma_i(\theta_i|\beta)$ in every history $h$ that has the same allocations $x_{<t}^\beta$, and induces the same public belief $\mu^{\beta}$ as $\beta$. \footnote{Mapping histories to their induced beliefs is not injective. By focussing on \bels~strategies we are restricting the strategy space. This is a mild restriction, as past announcements have no direct influence on the expected utility (the mechanism only depends on $x_{<t}$). Moreover, in \Cref{prop:equtil} we show that the best response to \bels~strategies is a \bels~strategy again.}
As beliefs are part of a \bels, the belief updates and state transitions merge.
\begin{definition}[PBS Transition Function]
    A PBS transition function $\rho$ maps a tuple of the current PBS, allocation, and announcement $(\beta,x_t,a_t)$ to a new PBS $\beta'$. 
\end{definition}

\begin{definition}[Consistent Transition]
    Given PBS strategies $\sigma$, a transition function $\rho$ is \emph{consistent} with $\sigma$ if for all $\beta$ it holds that if $(x_t,a_t)$ has a nonzero probability under $\sigma(\cdot|\beta)$, then $\rho(\beta,x_t,a_t)$ maps $\beta$ to $(x_{<t}\cup \{x_t\},\mu')$, where $\mu'$ is obtained by Bayesian updating.
\end{definition}

To compute PBEs by backward induction, we introduce the concept of a \textit{stage auction}.
\begin{definition}[Stage Auction]
    Given a sequential auction $\mathcal{A}$, round $t$, a function $U_{t+1}:\Theta_i \times \mathcal{B}_{t+1} \rightarrow \R$, and a \bels~transition $\rho_t$, we define the {stage auction} $\mathcal{A}_t(U_{t+1},\rho_t,\beta)$ as the following Bayesian game. There is a set of strategies $\sigma$, assumed to be common knowledge. Bidder $i$ observes $\theta_i$ and $\beta$, bids $b_{i,t}=\sigma_i(\theta_i|\beta)$ and receives the utility 
\begin{align*}
u^{\mathcal{A}_t(U_{t+1},\rho_t,\beta)}_i (\theta_i|b_t)=&v_i(\theta_i,X_t(b_t)|x_{<t}^\beta)-P_{i,t}(b_t)\\&+U_{t+1}(\theta_i,\rho_t(\beta,X_t(b_t),A_t(b_t))),
\end{align*}
where $v_i,X_t,A_t,P_t$ are as described by $\mathcal{A}$.
\end{definition}

Bidder $i$'s expected utility in the stage auction is given by
\[
    \overline{u}^{\mathcal{A}_t(U_{t+1},\rho_t,\beta)}_i(\theta_i|b_{i,t})=\Exp_{\substack{\theta_{-i}\sim \mu_{-i},\\b_{-i}\sim \sigma_{-i}}}\left[u^{\mathcal{A}_t(U_{t+1},\rho_t,\beta)}_i (\theta_i|b_{t})\right].
\] 
Strategies $\sigma(\cdot|\beta)$ form a BNE of the stage auction $\mathcal{A}_t(U_{t+1},\rho_t,\beta)$ if for all $i$ and $\theta_i$ it holds that 
\[
\sup_{b_{i,t}}\ \overline{u}^{\mathcal{A}_t(U_{t+1},\rho_t,\beta)}_i(\theta_i|b_{i,t})\leq \overline{u}^{\mathcal{A}_t(U_{t+1},\rho_t,\beta)}(\theta_i|\sigma_i(\theta_i|\beta)).
\]

To formulate PBEs as a dynamic program, we need a Bellman equation for equilibria.

\begin{definition}[Bellman Utility Update]
The Bellman utility update is given by
\[
\mathcal{T}(t,i,U_{t+1},\sigma_t,\rho_t)(\sigma_i,\beta_t)= \overline{u}^{\mathcal{A}_t(U_{t+1},\rho_t,\beta)}_i(\cdot|\sigma_i(\theta_i|\beta)).
\]
\end{definition}
Next, we extend the PBE dynamic program of \citet{Ouyang2017Dynamic} to sequential auctions.
\begin{proposition} 
    \label{thm:dp}
    A set of strategies $\sigma$ and a transition functions $\rho$ form a PBE of a sequential auction $\mathcal{A}$, if they solve the following dynamic program for $t\in \{1,\dots,T\}$:
    \begin{enumerate}
        \item $\forall i:U^i_{T+1}=0$ and $\forall i \notin N^t(\beta_t):U^i_{t}(\cdot|\beta_t)=0 $.
        \item $\forall i \in N, \forall t:U^i_t = \mathcal{T}(t,i,U_{t+1},\sigma_t,\rho_t) $.
        \item $\forall t, \forall \beta_t \in \mathcal{B}_t:\sigma_t(\cdot|\beta_t)$ is a BNE of $\mathcal{A}_t(U_{t+1},\rho_t,\beta_t)$.
        \item $ \forall t, \forall \beta_t \in \mathcal{B}_t: \rho_t(\beta_t,\cdot,\cdot)$ is consistent with $\sigma_t(\cdot|\beta_t)$.
    \end{enumerate}
\end{proposition}
\noindent The proof works by backward induction and is deferred to \Cref{app:proofs}.

\subsection{Game Abstraction}
\label{sec:abstraction}
The dynamic program defined in \Cref{thm:dp} is not yet computationally solvable. First, $\mathcal{B}_t$ is infinite for all $t>1$, such that there are infinitely many stage auctions to solve. Second, computing the BNE of a single stage auction is hard, especially due to the continuous type space. 
For this reason, we create a finite abstraction of $\mathcal{A}$, compute a PBE of the abstraction and then map it back to $\mathcal{A}$. This is a common approach for solving games with prohibitively large strategy spaces (Sandholm et al. \citeyear{sandholm2012lossy,sandholm2015abstraction}). Below, we seperately describe our approach to create a finite set of \bels~$\mathcal{B}_t$ (to solve only finitely many stage auctions) and to discretize $\Theta$ (to compute approximate BNEs).

\subsubsection{Type Space Abstraction}
Given $\mathcal{A}_t(U_{t+1},\rho_t,\beta)$ and strategies $\sigma_{-i}$, the continuity of $\Theta_i$ renders it infeasible to compute a best response for each $\theta_i$. However, \citet{Bosshard2020JAIR} have presented an approach for computing BNEs using piecewise-constant strategies that we extend to stage auctions.

 By restricting PBS strategies to be piecewise constant with respect to a partition $\mathcal{P}$, we are creating an abstracted game, where each bidder $i$ has a finite typespace $\hat{\Theta}_i=\mathcal{P}_i$, with types $\hat{\theta}_i$ corresponding to hyperrectangles $[y,z)_j$.\footnote{The definition of piecewise-constant strategies canonically extends to PBS strategies.} In this case, the initial distributions become probability mass functions $\hat{F}_i(\hat{\theta}_i) = \int_{\hat{\theta}_i} f_i(\theta_i)d\theta_i$, which simplifies Bayesian updating (cf. \Cref{app:example}). The utility of a bidder in the abstraction with type $[y,z)_j$ is the utility of the corresponding bidder in $\mathcal{A}$ with type $y$ (the vertex nearest to the origin). This causes an approximation error, which we bound in \Cref{sec:verification}.

\subsubsection{Belief Space Abstraction}
Given partition $\mathcal{P}$ of $\Theta$ and initial distribution $f$, we denote by $\mathcal{B}^{f,\mathcal{P}}_t\subset \mathcal{B}_t$ the set of \bels~in round $t$ that can be reached by Bayesian updating from $f$. We further define $\mathcal{B}^{f,\mathcal{P}}=\bigcup_{t=1}^T \mathcal{B}^{f,\mathcal{P}}_t$. $\mathcal{B}^{f,\mathcal{P}}$ is finite, which is crucial to solving only finitely many stage auctions.
\begin{proposition}
    \label{prop:finitebeliefs}
    Given a partition $\mathcal{P}$ of $\Theta$ and initial type distribution $f$, it holds that $|\mathcal{B}^{f,\mathcal{P}}|<\infty$.
\end{proposition}
\noindent This follows from the finiteness of the $\mathcal{P}$ and the auction being deterministic. The full proof is deferred to \Cref{app:proofs}.

If all bidders follow a piecewise-constant strategy $\hat{\sigma}$, the number of bids and thus of payments, announcements and histories with positive probability is finite. In particular, all \bels~with positive probability are in $\mathcal{B}^{f,\mathcal{P}}$. However, for a PBE we need counterfactual reasoning. In order for $\hat{\sigma}_i$ to be a best response to $\hat{\sigma}_{-i}$ in a given stage auction, we need to know what utility $i$ would achieve by deviating. There are infinitely many such deviations that have probability zero under $\hat{\sigma}$ and thus infinitely many possible PBS and stage auctions to solve. However, recalling \Cref{def:pbe}, in a PBE we are free to choose beliefs on histories with zero probability under $\hat{\sigma}$. In particular, we can choose a consistent transition function $\rho$ that projects all those outcomes onto the finite set $\mathcal{B}^{f,\mathcal{P}}$. We call a transition function $\rho$ consistent with a piecewise-constant strategy $\hat{\sigma}$ and \textit{finite} if it performs Bayesian updating on all histories with positive probability under $\hat{\sigma}$ and projects all other histories onto $\mathcal{B}^{f,\mathcal{P}}$. There are several ways to implement a finite and consistent transition function. We discuss this in \Cref{app:example}. Next, we formalize the auction abstraction.
\begin{definition}[Abstracted Belief-Based Sequential Auction]
    \label{def:abstract}
A partition $\mathcal{P}$, strategies $\hat{\sigma}=(\hat{\sigma}_1,\dots,\hat{\sigma}_n)$, and transition $\hat{\rho}$ induce a finite belief-augmented abstraction $\hat{\mathcal{A}}$ of a sequential auction $\mathcal{A}$ if the following hold:
\begin{itemize}
    \item For all $i$, $\hat{\sigma}_i:\mathcal{P}_i\times \mathcal{B}^{f,\mathcal{P}} \rightarrow \R^q$  is a well-defined piecewise-constant (with respect to $\mathcal{P}$) \bels~strategy. 
    \item $\rho$ is finite and consistent with $\hat{\sigma}$.
\end{itemize}
\end{definition}

\subsection{A Recursive Best Repsonse PBE Algorithm}
\label{sec:ibralg}
\subsubsection{Backward induction}
In \Cref{alg:backwards}, we combine the dynamic program from \Cref{thm:dp} with the abstraction from \Cref{def:abstract} to compute a PBE of $\hat{\mathcal{A}}$. This is done by recursively computing the BNEs of its finitely many stage auctions, starting in the final round $T$.
\begin{algorithm}
    \caption{Recursive PBE Computation}
    \label{alg:backwards}
    \begin{algorithmic}[1]
    \STATE \textbf{Input:} Auction $A$, Partition $\mathcal{P}$
    \STATE \textbf{Compute} $\mathcal{B}^{f,\mathcal{P}}_t$ for all $t$
    \STATE Set $U_{T+1} = 0$ and $U_{t}^i(\cdot|\beta) = 0$ for all $i \notin  N^t(\beta)$
    \FOR{$t = T$ to $1$ }
        \FOR{all $\beta_t \in \mathcal{B}^{f,\mathcal{P}}_t$ }
            \STATE $\hat{\sigma}_t(\cdot|\beta_t), \hat{\rho}_t(\cdot,\cdot |\beta_t) \gets \texttt{FindBNE}(\mathcal{A}_t(U_{t+1},\cdot, \beta_{t}))$
            \STATE $U_t(\cdot|\beta_t) \gets \mathcal{T}(t,i,U_{t+1},\hat{\sigma}_t, \hat{\rho}_t)$
        \ENDFOR
    \ENDFOR
    \STATE \textbf{return} $\hat{\sigma}, \hat{\rho}$
    \end{algorithmic}
    \end{algorithm}

\subsubsection{Computing BNEs of Stage Auctions}
To compute the BNEs of the abstracted stage auctions, we build upon the work of \citet{Bosshard2020JAIR} for single-round auctions. We describe the main ideas of how we adapt their algorithm below and defer the details to \Cref{app:algorithms}.

In brief, the approach relies on iteratively computing the best response of each bidder in the stage auction and updating their strategies accordingly. While such \emph{best response dynamics} are not guaranteed to converge, when they do, the obtained strategies form a BNE by definition. 

As the expected utility is not differentiable, we cannot use gradient-based methods to compute best responses.\footnote{Indeed a small change in the bid could mean the difference between winning or loosing.} Instead, we use pattern search---a zeroth order method proposed by \citet{hooke1961direct}. To evaluate the expected utility, we use Monte Carlo (MC) integration, sampling types from the belief distribution $\mu_{-i}^\beta$.
Having computed a best response $\hat{\sigma}^{BR}_i$ to $\hat{\sigma}_i$, we do not directly use $\hat{\sigma}^{BR}_i$ in the next iteration but instead update according to ${\sigma}_i\gets(1-\gamma){\sigma}_i+\gamma {\sigma}^{BR}_i$,
where $\gamma$ decreases over time to avoid oscillations. Additionally, we design a setting-specific rule on how to choose a finite and consistent transition function for the stage auction. The pseudocode is given in \Cref{app:algorithms}~(\Cref{alg:BNE}), along with further details.

\section{Equilibrium Verification}
\label{sec:verification}
There are two issues with using \Cref{alg:backwards} as a standalone method. 
First, even if a pure-strategy PBE exists, the algorithm might not converge. The lack of convergence guarantees for equilibrium solvers in auctions has been well-documented~\cite{Bichler2023Convergence}. Second, even if \Cref{alg:backwards} converges in the abstraction $\hat{\mathcal{A}}$, the utility loss in $\mathcal{A}$ might be much larger, such that the found strategies only form an $\epsilon$-PBE in $\mathcal{A}$ for impractically large $\varepsilon$. We refer to the problem of bounding the utility loss in $\mathcal{A}$ as \textit{verification}.

The difficulty with verification is that we need to argue over the set of all possible strategies and cannot simply restrict the search space as in \Cref{sec:search}.\footnote{We can restrict to all pure strategies. Fixing all other bidders,  $i$ faces an MDP with a pure optimal strategy~\cite{Puterman2005Markov}.} To address this, we show in \Cref{thm:main} that the utility loss in $\mathcal{A}$ can be bound by only computing best responses at finitely many vertices and in finitely many PBS. Based on this result, we introduce a verification procedure to run after \Cref{alg:backwards}.

\subsection{Decomposing the Utility Loss}
\label{sec:decomp_util_loss}
All strategies $\hat{\sigma}$ found by \Cref{alg:backwards} satisfy Bayesian updating and correct initial beliefs. To verify them, we only have to bound the utility loss, i.e. find $\varepsilon$, such that $\forall \theta_i,\forall h: l_i^{BR}(\theta_i|\hat{\sigma},h) \leq \varepsilon$. This revives the issue that $H$ and $\Theta_i$ are continuous. To tackle the continuity of $H$, we show that the best response to a PBS strategy is again a PBS strategy.

\begin{proposition}[Closedness of \bels~strategies]
    \label{prop:equtil}
        Given a set of PBS strategies ${\hat{\sigma}}$, a transition function $\rho$, and a history ${h}_{t}$. Let  $[{h}_{t}]$ denote the equivalence class of all histories that induce the same \bels~according to $\rho$. Then $\forall \theta_i\in \Theta_i, \forall{h'}\in [{h}_{t}], \forall b_{i,t}\in \R^q_{\geq0}$, it holds that
        \[
        \overline{u}_i(\theta_i|{\sigma}_{-i},b_{i,t},{\sigma}_{i,>t},{h}_{t})=\overline{u}_i(\theta_i|{\sigma}_{-i},b_{i,t},{\sigma}_{i,>t},{h'}).
        \]
        In particular, for any strategy, there exists a corresponding \bels~strategy with the same expected utility.
    \end{proposition}

The proof is deferred to \Cref{app:proofs}. Intuitively, the result holds because $\overline{u}_i$ is independent of $a_{<t}$. \Cref{prop:equtil} shows if a best response to PBS strategies $\hat{\sigma}_{-i}$ exists, there is an equivalent PBS best response. In particular, the utility loss is the same for all $h$ inducing the same PBS. Therefore, we can restrict to bounding the utility loss in the finitely many PBS in $\mathcal{B}^{f,\mathcal{P}}$. We denote by $l_i^{BR}(\theta_i|\hat{\sigma},\beta)=\sup_{\tilde{\sigma}_{i}}\overline{u}_i(\theta_i|\hat{\sigma}_{-i},\tilde{\sigma}_{i},\beta)-\overline{u}_i(\theta_i|\hat{\sigma},\beta)$ $i's$ utility loss at $\beta$ from not playing a PBS best response. Similarly, we define the \textit{immediate utility loss} $l_i^{IBR}(\theta_i|\hat{\sigma},\beta)= u^{IBR}_i(\theta_i|\hat{\sigma},\beta)-u_i(\theta_i|\hat{\sigma},\beta)$, where $u^{IBR}_i(\theta_i|\hat{\sigma},\beta)$ is the highest utility $i$ can reach by changing bids in the stage auction $\mathcal{A}_t(U_{t+1},\rho,\beta)$, with $\rho, U_{t+1}$ given by \Cref{alg:backwards}. Below, we show one of our main contributions: that the utility loss decomposes into a finite sum of immediate utility losses.

\begin{theorem}[Decomposition]
    \label{thm:decomp}
    Given piecewise-constant PBS strategies $\hat{\sigma}$, it holds for all $i,\beta_t,\theta_i$  that
    \begin{align*}
        l_i^{BR}(\theta_i|\hat{\sigma},\beta_t) \leq &l_i^{IBR}(\theta_i|\hat{\sigma},\beta_t) \\&+ \max_{\beta_{t+1}\in \mathcal{B}^{f,\mathcal{P}}_{t+1}(\beta_t,\rho)} l_i^{BR}(\theta_i|\hat{\sigma},\beta_{t+1}),
    \end{align*}
    where $\mathcal{B}^{f,\mathcal{P}}_{t+1}(\beta_t,\rho) \subseteq \mathcal{B}^{f,\mathcal{P}}_{t+1}$ is the set of possible successors of $\beta_t$ under $\rho$.
    \end{theorem}
    The proof works by backward induction and is deferred to \Cref{app:proofs}. \Cref{prop:equtil} and \Cref{thm:decomp} let us tackle the continuity of $H$. Fixing $\theta_i$, we can bound the utility loss by calculating the immediate utility loss at the finitely many possible \bels. However, to verify $\hat{\sigma}$, we also need to upper bound the utility loss over the continuous type space. To do so, we extend an idea by \citet{Bosshard2020JAIR} for verifying BNEs in single-round auctions. We can show that $\overline{u}_i^{IBR}$ is convex in each hyperrectangle of $\mathcal{P}_i$ and thus upper bound by a linear function. $\overline{u}_i$ is linear as well, so that the immediate utility loss can be upper bound by the difference of two linear functions evaluated at the hyperrectangle's finitely many vertices.\footnote{The argument in the single-round case of \citet{Bosshard2020JAIR} is simpler, as there are no subsequent rounds, and the best response utility is convex instead of piecewise convex.}

\begin{proposition}
    \label{prop:immediateloss}
        Given a piecewise-constant~(on $\mathcal{P}$) PBS strategy $\hat{\sigma}$, the following holds for all $i,\beta$:
        \begin{align*}
                   & \sup_{\theta_i \in \Theta_i}l_i^{IBR}(\theta_i|\hat{\sigma},\beta)\\
             \leq &\max_{[y,z)_j \in \mathcal{P}_i} \max_{w \in \text{V}([y,z)_j)} \overline{u}_i^{IBR}(w|\hat{\sigma}_i(y),\hat{\sigma}_{-i},\beta)\\ &-   \overline{u}_i(w|\hat{\sigma}_i(y),\hat{\sigma}_{-i},\beta),
        \end{align*}
        where the last term is $i$'s expected utility with type $w$ bidding $\hat{\sigma}_{i}(y)$, and $V([y,z)_j)$ denotes the vertex set of $[y,z)_j$.
   \end{proposition}
The full proof is deferred to \Cref{app:proofs}. Combining \Cref{prop:equtil}, \Cref{prop:immediateloss}, and \Cref{thm:decomp} yields the main result of this section: that we can verify an $\epsilon$-PBE in the full auction by computing the immediate utility loss at finitely many vertices in finitely many abstracted stage auctions.
\begin{theorem}[$\epsilon$-PBE Verification]
    \label{thm:main}
        A set of piecewise-constant~(on partition $\mathcal{P}$) \bels~strategies $\hat{\sigma}$ and a finite and consistent transition $\rho$ form an $\epsilon$-PBE in a sequential auction $\mathcal{A}$, with $\epsilon$ being upper bound by
             \begin{align*}
              \epsilon\leq& \max_{i \in N} \max_{\beta_{t:T}} \sum_{\beta\in \beta_{t:T}} \max_{[y,z) \in \mathcal{P}_i} \max_{w \in {V}([y,z))} \overline{u}_i^{IBR}(w|\hat{\sigma},\beta)\\
              &-\overline{u}_i(w|\hat{\sigma}_{i}(y),\hat{\sigma}_{-i},\beta),
         \end{align*}
        where $\beta_{t:T}$ denotes a vector of successor \bels~under $\rho$ going from $t$ to $T$.\footnote{The set of $\beta_{t:T}$ is finite, since $\mathcal{B}^{f,\mathcal{P}}$ is finite.}
    \end{theorem}
The proof is deferred to \Cref{app:proofs}.

\subsection{Verification Procedure}
\label{sec:ver_proc}

\Cref{thm:main} gives a strict upper bound on the utility loss of a set of piecewise-constant \bels~strategies $\hat{\sigma}$, expressed in terms of the immediate utility losses evaluated at finitely many vertices of $\mathcal{P}$ and finitely many \bels. The result directly motivates an implementable verification procedure for PBEs. We have already developed an algorithm to compute best responses (and hence the corresponding immediate utility loss) in stage auctions in \Cref{sec:search}. Analogously, we can use pattern search to approximate the immediate utility loss at the finitely many vertices of $\mathcal{P}$ and then find the largest possible utility loss for any $\beta\in \mathcal{B}^{f,\mathcal{P}}$ by backward induction over the rounds. The detailed pseudocode is given in \Cref{app:algorithms} (\Cref{alg:verify}).

\section{Experiments}
\label{sec:experiments}
We evaluate our algorithm in a range of settings with and without known analytical results.\footnote{We provide our source code in the supplementary material. We ran all experiments on a Debian compute cluster with 20 nodes, each node having 128 GB RAM and two Intel E5-2680 2.80GHz  processors for a total of 40 cores. To speed up computation, we enforced monotonicity in the strategies after each iteration, as this makes the number of possible beliefs linear instead of exponential in the grid size.} Unless otherwise specified, we perform 10 runs per setting, using a partition with uniform intervals and a grid size of 100 for our piecewise-constant strategies, 100 iterations of iterated best responses to compute BNEs, and up to 100,000 MC samples for search and 200,000 for verification. Our algorithm is embarrassingly parallelizable, as inherently all $\beta_t$ in round $t$ can be solved independently of each other. Therefore, the wall clock time for solving each setting is a small fraction of the reported core hours. Below we describe the different settings and our corresponding results. Our code can be found at https://github.com/VnznzT/Auction-PBE-Solver.

\subsection{Sequential Sales}
An auctioneer is selling $K$ identical goods to $n$ single-minded bidders with values uniformly distributed on $[0,1]$. There are $K$ rounds. Each round, the auctioneer runs a first-price or second-price sealed-bid auction for one item, breaking ties by a predefined order.
For the first-price auction, the payments are made public (i.e. $a_t=p_t$) and for the second-price auction, the winning bid (i.e. $a_t=\max_i b_{i,t}$). The winning bidder leaves the auction. The equilibrium strategy in round $k$ is given by $\sigma_i^*(\theta_i) = ({n - K})\theta_i/({n - k + 1})$ and $\sigma_i^*(\theta_i) = ({n - K})\theta_i/({n - k })$, for first-price and second-price auctions, respectively \cite{Krishna2002AuctionTheory}.

\begin{table}[]

\centering

\setlength{\tabcolsep}{1mm}

\begin{tabular}{llllllllr}
 
\toprule
    \textbf{Env.}  & \textbf{Pay} & $\bs{n}$ & $\mathbf{|K|}$ & $\bs{L_2^{t=1}}$ & $\bs{L_2^{t=2}}$ & $\bs{L_2^{t=3}}$ & $\bs{L_2^{t=4}}$ & \textbf{Hrs.} \\ \midrule
    Seq. & 1st & 3 & 2 & 0.008 & 0.010 &  &  & 22 \\
    Seq. & 1st & 4 & 3 & 0.007 & 0.007 & 0.012 &  & 110 \\
    Seq. & 1st & 5 & 4 & 0.010 & 0.005 & 0.007 & 0.012 & 176 \\
    Seq. & 2nd & 3 & 2 & 0.008 & 0.006 &  &  & 4 \\
    Seq. & 2nd & 4 & 3 & 0.012 & 0.008 & 0.006 &  & 76 \\ 
    Seq. & 2nd & 5 & 4 & 0.014 & 0.009 & 0.008 & 0.006 & 154 \\ \midrule
    Res. & 1st & 3 & 2 & 0.004 & 0.005 &  &  & 28 \\
    Res. & 1st & 4 & 2 & 0.008 & 0.007 &  &  & 47 \\
    Res. & 2nd & 3 & 2 & 0.005 & 0.006 &  &  & 3 \\
    Res. & 2nd & 4 & 2 & 0.006 & 0.006 &  &  & 5 \\ \midrule
    Split & 1st & 3 & 2 & 0.003 & 0.003 &  &  & 85 \\
    Split & 1st & 4 & 2 & 0.006 & 0.005 &  &  & 139 \\ \bottomrule
\end{tabular}
    \caption{Average $L_2$ distances (with standard error below $10^{-5}$) to the known equilibrium and the runtime (in core hours). Seq = Sequential, Res = Reserve price, Split = Split-award. 1st = First-price, 2nd = Second-price. $n$ = number of bidders, $|K|$ = number of goods.}
    \label{tab:results}
    \end{table}
In Table \ref{tab:results}, we report $L_2$ distances to the known analytical equilibrium for $K\in\{2,3,4\}$ with $n=K+1$. We plot the found strategies in Figures~\ref{fig:krishna_3_2_fp_1}--\ref{fig:krishna_5_4_sp_4} in Appendix \ref{app:exp}. Our algorithm converges to the known PBEs, achieving small $L_2$ distances across all settings.

\subsection{Sequential Sales with Ascending Reserve Price}
\citet{Gong2013Ordering} present a complication of the sequential sales setting, where the auctioneer further chooses ascending reserve prices $r_1\leq\dots \leq r_K$ for each round. The equilibrium strategies are explained in \Cref{app:eq_strats}. We consider the cases where $n \in \{3,4\}$, $K=2$ and $(r_1,r_2)=(0,0.5)$. The $L_2$-distances are given in Table \ref{tab:results} and the {first}-round bidding strategy for the three bidder case is plotted in Figure \ref{fig:res_3_2_fp_1} (see Figures \ref{fig:res_3_2_fp_1_re}--\ref{fig:res_4_2_sp_3} in Appendix \ref{app:exp} for the other plots). Across all settings, the $L_2$ distance to the known PBE is very small; in most settings, the non-zero distance is due to the finite gridsize of the piecewise-constant strategies.

\begin{figure}
    \centering
    \includegraphics[width=\columnwidth]{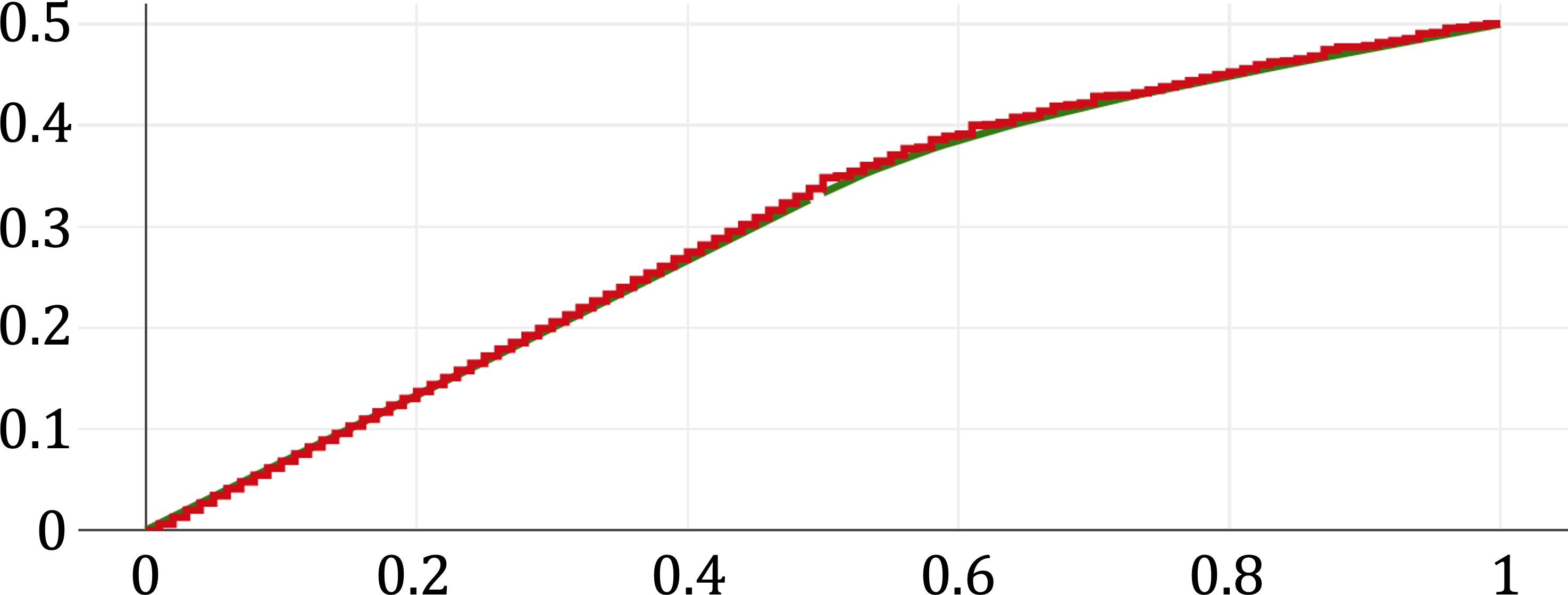}
    \caption{Bidding strategies for the first round of the first-price sequential sales auction with 3 bidders, 2 goods, and reserve prices $r_1=0,r_2=0.5$. Green denotes the analytical solution and red our found strategies. Types are plotted on the x-axis, bids on the y-axis.}
    \label{fig:res_3_2_fp_1}
\end{figure}

\subsection{Two-Round Split-Award Auction}
Our algorithm can also handle reverse auctions. Consider an auctioneer that can either procure 100\% of a business from one bidder or 50\% each from two bidders. Their type $\theta_i $ is the cost of providing 100\%. All bidders share a cost parameter $C\in (0,1)$ determining their cost $C\theta_i$ for providing the first 50\%. For $C>0.5$ ($C<0.5$), the setting has economies (diseconomies) of scale. The auctioneer, unaware of $C$, cannot decide a priori whether to buy 100\% from one or 50\% each from two bidders and instead performs a combinatorial auction. Each bidder $i$ submits both a split bid $b_{i}^{(sp)}$ for 50\% and a sole bid $b_{i}^{(so)}$ for 100\%. The first round has two possible outcomes:
\begin{itemize}
    \item \textbf{Split award:} If $2 \min_i b_{i}^{(sp)} \leq \min_{i} b_{i}^{(so)}$, the auctioneer procures the first half of the business from the bidder with the lowest split bid. Next, the auctioneer runs a first-price sealed-bid auction for the remaining split. In the second round the first-round winner $w$ has cost $(1-C)\theta_w$.
    \item \textbf{Sole award:} If $2 \min_i b_{i}^{(sp)} > \min_{i} b_{i}^{(so)}$, The auctioneer procures the whole business from the most competitive sole bidder and the auction ends.
\end{itemize}

We study the setting for $n\in\{3,4\}$. The bidders' costs are distributed uniformly on $[1,2]$. We choose $C=0.2$; this ensures \textit{strong diseconomies of scale} for which \citet{kokott2019beauty} derive the PBE that can be found in Appendix \ref{app:eq_strats}.  %
In Table \ref{tab:results}, we provide the $L_2$ distances. Figures \ref{fig:kokott3_2_1}--\ref{fig:kokott_4_2_2} in Appendix \ref{app:exp} show the plotted strategies. As before, our algorithm closely recovers the known equilibria.

\subsection{Two-Round Local-Local-Global Auction}

\begin{figure}[tbp]
    \centering
    \includegraphics[width=\columnwidth]{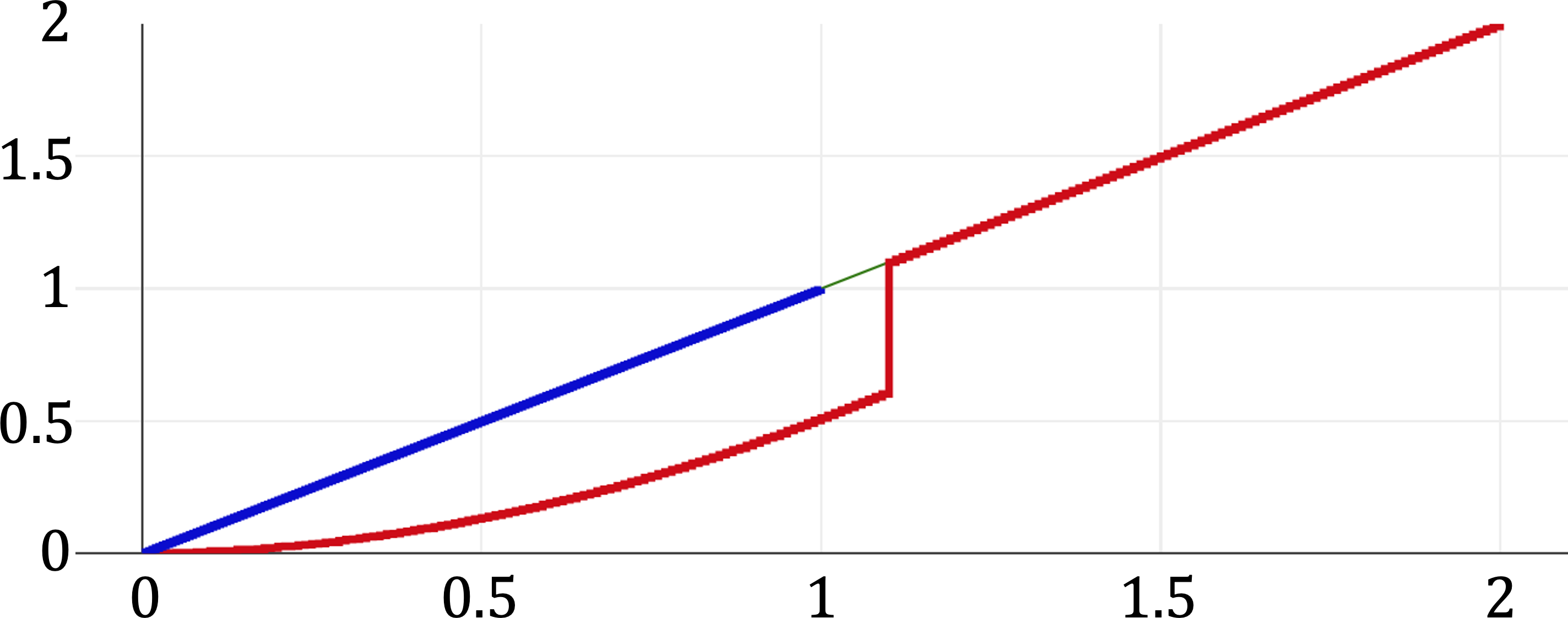}
    \caption{Bidding strategies in the first round of the sequential LLG auction. Green denotes truthful bidding, blue the strategy of the local bidder and red the strategy of the global bidder. Types are plotted on the x-axis, bids on the y-axis.}
    \label{fig:llg_1}
\end{figure}

We study a sequential version of the Local-Local-Global (LLG) auction \cite{ausubel2006lovely}, without previous analytical results. There are two goods $A$ and $B$ sold sequentially via second-price sealed-bid auctions to three bidders. The two local bidders are single-minded. Bidder 1 wants $A$ and bidder 2 wants $B$. Their values are uniformly distributed on $[0,1]$. Bidder 3 is the global bidder whose value for the bundle $\{A,B\}$ is distributed uniformly on $[0,2]$. In the first round, only bidders 1 and 3 bid; in the second, only bidders 2 and 3 bid. 

We run our algorithm with a grid size of 200 and 150 best response iterations in each round.
Figure \ref{fig:llg_1} shows the computed first-round strategies (see Figure \ref{fig:llg_2} in Appendix \ref{app:exp} for the second-round strategies). Our verification procedure computes a small $\epsilon$ of $1.47 \cdot 10^{-4}$, indicating near-perfect convergence. In the second round, both bidders bid truthfully. In the first round, the global bidder shades the bids for types less than $1.1$ due to the uncertainty of winning $B$ in the second round. In contrast, for types larger than $1.1$ the global bidder---likely to win $B$ as well---bids truthfully. This interesting, non-continuous behavior shows the potential of our approach for studying less-understood auction formats.

\section{Conclusion}
In this work, we have introduced a new algorithm for computing pure-strategy $\varepsilon$-perfect Bayesian equilibria in sequential auctions. We have shown how to abstract and decompose a sequential auction such that its PBEs can be computed via backward induction, and how the output can be verified.

Our method complements concurrent works using RL to compute equilibria in auctions. Instead of first fixing all bidders except $i$ and using deep RL to compute $i$'s best response (solving the Bellman equation and abstracting the game using neural networks), we start with an equilibrium Bellman equation and abstraction that simplify our problem to computing a finite number of BNEs. Only then do we iteratively compute best responses---in single-round games rather than multi-round ones. By reversing this order, we obtain a simple verification procedure and a stronger equilibrium concept (ex-interim PBE vs ex-ante Nash).

Our approach opens up exciting future directions. Combining our decomposition and verification with other methods (such as double oracle or deep RL) could lead to scalable, yet verifiable approaches. On the theory side, an interesting open question is how to address interdependent types and mixed strategies and whether there is an auction class for which these algorithms provably converge.

\section*{Acknowledgments}
V. Thoma is supported by an ETH AI Center Doctoral Fellowship and acknowledges funding from the Swiss National Science Foundation (SNSF) Project Funding No. 200021-207343. This paper is part of a project that has received funding from the European Research Council (ERC) under the European Union’s Horizon 2020 research and innovation program (Grant agreement No. 805542).

\bibliography{aaai25}

\appendix

\clearpage

\section{Algorithms}
\label{app:algorithms}

\begin{algorithm}

    \caption{\texttt{FindBNE}, BNE Algorithm for Stage Auctions}
    \begin{algorithmic}[1]
    \label{alg:BNE}
    \STATE \textbf{Input:} Stage Auction $\mathcal{A}_t(U_{t+1},\cdot, \beta_{t})$, transition selection rule $d$, partition $\mathcal{P}$, \texttt{num\_iter}, stepsize $\gamma$.
    \STATE \textbf{Intialize:} $\hat{\sigma}=\texttt{TruthfulPC}(\mathcal{A}_t(U_{t+1},\cdot, \beta_{t}),\mathcal{P})$
    \FOR{$\texttt{iter}=1$ \TO $\texttt{num\_iter}$}
            \FOR{$i=1$ \TO $n$}
                \STATE $\rho^{\hat{\sigma}}=d(\hat{\sigma})$
                \FOR{$\hat{\theta}_i \in \mathcal{P}_i$}
                        \STATE ${\hat{\sigma}}^{BR}_i(\hat{\theta}_i|\beta_t) \gets \texttt{FindBR}(i,\hat{\theta}_i,\hat{\sigma},\mathcal{A}_t(U_{t+1},\rho^{\hat{\sigma}}, \beta_{t}))$
                \ENDFOR
                 \STATE ${\hat{\sigma}}_i=(1-\gamma){\hat{\sigma}}_i+\gamma {\hat{\sigma}}^{BR}_i$
            \ENDFOR
        \ENDFOR
    \RETURN $\hat{\sigma}, \epsilon$
    \end{algorithmic}
    \end{algorithm}

    \begin{algorithm}[H]

        \caption{\texttt{FindBR}}
        \begin{algorithmic}[1]
        \label{alg:IBR}
        \STATE \textbf{Input:} Bidder $i$, Type $\hat{\theta}_i$, strategy $\hat{\sigma}$, stage auction $\mathcal{A}_t(U_{t+1},\rho^{\hat{\sigma}}, \beta_{t})$
        \STATE \textbf{Parameters:}$\texttt{n\_steps, pattern,n\_mc\_samples}$
        \STATE $b^*={\hat{\sigma}}(\hat{\theta}_i|\beta_t)$
        \STATE $\hat{u}_i^*\gets 0$
                \FOR{$1$ \TO \texttt{n\_mc\_samples}}
                    \STATE $b_{-i}\gets \texttt{BidSampler}(\hat{\sigma}_{-i},\mu^{\beta_t}_{-i})$
                    \STATE $b=(b^*,b_{-i})$
                    \STATE $\hat{u}_i^* \gets \hat{u}_i^*+  v_i(\hat{\theta_i},{X}_t(b)|x_{<t}^{\beta_t})-{P}_{i,t}(b|x_{<t}^{\beta_t}) + U_{t+1}(\hat{\theta_i},\rho^{\hat{\sigma}}(\beta_t,X_t(b),A_t(b)))$
                \ENDFOR
                \STATE $\hat{u}_i^* \gets \hat{u}_i^*/\texttt{mc\_samples}$
        \FOR{$1$ \TO $\texttt{n\_steps}$}
            \FOR{$b_i \in \texttt{pattern}(b^*)$}
                \STATE $\hat{u}_i \gets 0 $
                \FOR{$1$ \TO \texttt{n\_mc\_samples}}
                    \STATE $b_{-i}\gets \texttt{BidSampler}(\hat{\sigma}_{-i},\mu^{\beta_t}_{-i})$
                    \STATE $b=(b_i,b_{-i})$
                    \STATE $\hat{u}_i \gets \hat{u}_i+  v_i(\hat{\theta}_i,{X}_t(b)|x_{<t}^{\beta_t})-{P}_{i,t}(b|x_{<t}^{\beta_t})  + U_{t+1}(\hat{\theta}_i,\rho^{\hat{\sigma}}(\beta_t,X_t(b),A_t(b)))$
                \ENDFOR
                \STATE $\hat{u}_i \gets \hat{u}_i/\texttt{mc\_samples}$
                         \IF{$\hat{u}_i>u_i^*$}
                 \STATE $b^*\gets b_i$
                 \STATE $\hat{u}_i^*\gets \hat{u}_i$
                 \ENDIF
            \ENDFOR
        
        \ENDFOR
        \STATE $\texttt{Cache.put}(U_{t}(\hat{\theta}_i,\beta_t)\gets \hat{u}_i^*)$
                    \RETURN $b^*$
        \end{algorithmic}
        \end{algorithm}

        \begin{algorithm}[H]
            \caption{\texttt{Verify}}
            \algsetup{indent=0.25em}
            \begin{algorithmic}[1]
            \label{alg:verify}
            \STATE \textbf{Input:} Strategies $\hat{\sigma}$, partitions $\{\mathcal{P}_i\}_{i=1}^n$
                \FOR{$i=1$ \TO $n$}
                \STATE $\rho^{\hat{\sigma}}\gets d(\hat{\sigma})$
                    \FOR{$t=T$ \TO 1}
                    \STATE $\mathcal{B}^{f,\mathcal{P}}_t\gets  \texttt{AllPBSSates}(f,\mathcal{P},t)$   
                    \IF{$t<T$}
                    \STATE $U_{t+1}\gets \texttt{Cache.get\_u}(t+1)$
                \ENDIF
                    \FOR{$ \beta_t \in \mathcal{B}^{f,\mathcal{P}}_t$}
                            \STATE $\epsilon^i_{\beta_t}\gets 0$
                                \FOR{$[y,z)_j \in \mathcal{P}_i$}
                                    \FOR{$w_i \in \text{Vert}([y,z)_j)$}
                                        \STATE $b^{BR}\gets  \texttt{FindBR}(y,\hat{\sigma},\mathcal{A}_t(U_{t+1},\rho^{\hat{\sigma}},\beta_t))$
                                        \STATE $\epsilon' \gets \overline{u}_i(w_i|b^{BR},\hat{\sigma}_{i,>t}(y),\hat{\sigma}_{-i},\beta_t) - \overline{u}_i(w_i|\hat{\sigma}_{i}(y),\hat{\sigma}_{-i},\beta_t)$
                                        \STATE $\epsilon^i_{\beta_t}\gets \max(\epsilon^i_{\beta_t},\epsilon')$
                                        \STATE $\texttt{Cache.put\_u}(t,\beta_t: \overline{u}_i(w_i|\hat{\sigma}_{i}(y),\hat{\sigma}_{-i},\beta_t))$
                                    \ENDFOR    
                                \ENDFOR
                                 \STATE $\epsilon^i_{\beta_t}\gets \epsilon^i_{\beta_t}+\max_{\beta_{t+1} \in \mathcal{B}^{f,\mathcal{P}}_{t+1}(\beta_t,\rho^{\hat{\sigma}})} \epsilon^i_{\beta_{t+1}}$
                        \ENDFOR
                      
                    \ENDFOR
                     
                \ENDFOR
            \STATE $\epsilon \gets \max_{1\leq i\leq n} \epsilon^i_{\beta_0}$ 
            \RETURN $\epsilon$
            \end{algorithmic}
            \end{algorithm}

In \Cref{alg:verify} the expected utility is once again computed by MC integration (cf. \Cref{alg:IBR}).
\section{Implementation details}
\label{app:impl_det}

In addition to the pseudocode above, we want to give some more details on our algorithm. 

\paragraph{Transition Selection Rule} In almost all our settings (except the Split-award auction) both the type and action space is one-dimensional. The partition consists of a grid of $J$ intervals spanning the type space. In our search phase, we restrict to strictly monotone strategies. Consider the example of a symmetric first price auction. Here, we run a routine after each iteration in \texttt{FindBR} to ensure that all bidders bid $b_1<\dots<b_J$ on intervals $[0,\theta_1), \dots, [\theta_{J-1},1]$ respectively. If bidders bid exactly those bids, then we can perform Bayesian updating. Consider for instance that bidder 1 wins in a first price auction, bidding $b_j$. Then (assuming tiebreaking by indices) all bidders have to have their types in the interval $[0,\theta_j)$. However, how do we update beliefs, when bidder 1 wins with a bid $b'$ where $b_{j}<b'<b_{j+1}$ ? In this case, we know that the other bidders (having bid according to the strictly monotone strategy) cannot have bid more than $b_{j}$. Therefore, their types have to be in the interval $[0,j)$. We use this kind of reasoning for all our experimental settings to update beliefs and thus to define a consistent and finite transition rule, no matter the current monotone piecewise-constant strategy.

\paragraph{Updating strategies} Concerning our choice for updating the current set of strategies $\hat{\sigma}$, we note that directly following the best response dynamics can be prone to overshooting. On the other hand, averaging over all past strategies, as done in fictitious play, can lead to very slow convergence. 
Instead, we update the new strategy to be a linear combination of the old strategy $\hat{\sigma}_i$ and the best response $\sigma^{BR}_i$, where the magnitude of the update depends on the iteration number. More specifically we choose a maximum and minimum update rate $\gamma\_\max,\gamma\_\min$, such that the actual update rate $\gamma$ is given as
\[
\gamma=\gamma_{\max }-(\gamma_{\max}-\gamma_{\min })*\frac{\texttt{iter}}{\texttt{max\_iters}},
\]
and the strategy update by
\[
\hat{\sigma}_i(\theta_{i}|\beta) = (1-\gamma)\hat{\sigma}_i(\theta_{i}|\beta) + \gamma \sigma^{BR}_i(\theta_{i}|\beta).
\]

\paragraph{Inner and Outer loop} Following \citet{Bosshard2020JAIR} when implementing the algorithm \texttt{FindBNE} we keep use an inner and outer loop. In the inner loop we use fewer MC samples and patter search steps to quickly find an approximate best response. Once the computed approximate utility losses reach a certain threshold or we reach the number of maximum iterations, we switch to the outer loop with more MC samples and pattern search steps to get a more accurate best response.

\paragraph{Dealing with slight asymmetries due to tiebreaking} In the sequential sales (with and without reserve prices) and the reverse auction, we break ties by a predefined ordering on the bidders. One of the main reasons for this is that breaking ties at random in combinatin with piecewise-constant strategies can cause bidder types to become interdependent.  When using piecewise-constant strategies this creates slight asymmetries between bidders in later rounds. Indeed, consider the two round FPSB auction with symmetric strictly monotone bidding strategies. If bidder 1 wins in the first round, knowing ties are broken in his favour all other bidders in the second round have types as high or lower than bidder 1 had in the first. However, if bidder 2 wins all bidders $j>2$ have types as high or lower, whereas bidder 1 needs to have type strictly lower than 2. On a discretization with 100 intervals or more, such small asymmetries in the beliefs have little influence on the strategic decisions. Hence when running the experiments we simply solve the subgames once from the point of view of bidder 1 having won and use them no matter which bidder won in the first round. This saves redundant computation time and as the low $L_2$ distances show, has little influence on the results. We emphasize that in the LLG auction, where we have to rely on our own verification, we do not make use of such simplifications as they could lead to an incorrect verification bound.

We are releasing our Java code with this work with additional instructions in the code on how to run it.

\section{Proofs}
\label{app:proofs}

\begin{proof}[\textbf{Proof of \Cref{thm:dp}}]
A version of this Theorem has been stated and proven for a slightly different class of imperfect information games---with public actions, discrete state and actions spaces and a fixed number of agent---by \citet{Ouyang2017Dynamic}. 

    First let us note that the transition $\rho$ gives us a belief set $\mu$ for each history of the game by repeatedly applying $\rho$, i.e. $\mu(\cdot|h_t)= \rho(\rho(\dots,\rho(\mu_0,x_0,a_0),x_1,a_1),\dots,x_{t-1},a_{t-1})$. Saying $\sigma,\rho$ form a PBE is thus a well-defined statement and equivalent to saying $\sigma$ and the induced beliefs $\mu$ are a PBE.

    In order to show that $\sigma,\rho$ define a PBE, we need to verify three things: (a) sequential rationality, (b) correct initial beliefs, and (c) Bayesian updating.

    Note that (b) and (c) hold since the initial \bels~$\beta_0$ has correct initial beliefs and the transition $\rho$ is consistent according to condition (4) of the dynamic program. We now show that the strategies are sequentially rational.

    As shown in \Cref{prop:equtil}, if all other strategies are \bels, then the best response is also a \bels~strategy. Thus, it suffices to show that $\sigma_i$ is the best response of $i$ in the set of \bels~strategies. We show this by induction. For $t=T$, for all $i$, $\sigma_i$ is a best response by definition since $\sigma$ forms a BNE of the last round for all $\beta_t$. Now assume that for round $t+1$ $\sigma$ is a PBE strategy, i.e. $\sigma_i$ is a best response to $\sigma_{-i}$ for each $i$ and each $\beta_\tau$ for $\tau\geq t+1$. We show that $\sigma_i$ is a best response for each $i$ and each $\beta_t$. Take an arbitrary bidder $i$ and PBS $\beta_t$. By the induction hypothesis, $\sigma_i$ is already a best response to $\sigma_{-i}$ in all following rounds. Thus, we only need to show that $\sigma_i$ is a best response to $\sigma_{-i}$ in the stage auction at $\beta_t$. However, this is true by condition (3) of the dynamic program, since $\sigma_t$ forms a BNE of the corresponding stage auction. The claim follows.
\end{proof}

\begin{proof}[\textbf{Proof of \Cref{prop:finitebeliefs}}]
    $\sigma, X_t, A_t$ are all deterministic. Thefore, there exists a deterministic \textit{outcome function} that maps hyperrectangles $\hat{\theta}$ in the the type space to outcomes $x_t,a_t$. In particular, a given hyperrectangle $\hat{\theta}_i$ is either in the preimage of the outcome function or not, which limits the possible Bayesian updates to either include or exclude any given hyperrectangle. For all $i$, the number of possible beliefs obtained by Bayes rule is thus bounded by the number of possible subsets of $\mathcal{P}_i$. Since $\mathcal{P}_i$ is finite, so is its power set and thus $|\mathcal{B}^{f,\mathcal{P}}|<\infty$.
    \end{proof}

\begin{proof}[\textbf{Proof of Proposition \ref{prop:equtil}}]
	The statement follows directly because the expected utility depends on ${\hat{\sigma}}_{> t},\{X_\tau\}_{\tau=t+1}^T,\{P_\tau\}_{\tau=t+1}^T$ all of which do not depend on ${a}_{1:t}$ and on the belief ${\mu}$ at $t$, which is the same for all $h'\in[h_{t-1}]$ since they all induce the same \bels.
\end{proof}

\paragraph{Proof of Proposition \ref{prop:immediateloss}} In order to prove this we first need to show two additional Lemmas.
Note that Lemmas \ref{lem:convex}, \ref{lem:pwconvex} and Proposition \ref{prop:immediateloss} are extensions of Lemmas 1 and 2 and Theorem 2 in \cite{Bosshard2020JAIR} to the sequential setting, taking into the utilities of subseqeuent rounds and extending the argument from convex utility to piecewise-convex utility functions. 
\begin{lemma}
    \label{lem:convex}
        Given a piecewise-constant (on partition $\mathcal{P}$) \bels~strategy $\hat{\sigma}$ and a finite consistent transition $\rho$, let $\vec{b}_i$ denote a vector whose entries are fixed bids for every possible $\beta$, then the expected utility of bidder $i$ bidding according to $\vec{b}_i$ is linear in $\theta_i$.
    \end{lemma}

\begin{proof}
	
	We show this by induction.
	In the base case $t=T$, the expected utility is
\begin{align*}
    &\overline{u}_i(\theta_i|\vec{b}_i,\hat{\sigma}_{-i},\beta_T)\\
    =&\Exp_{\theta_{-i}}[v_i(\theta_i,X_T(\vec{b}_{i},\hat{\sigma}_{-i}(\beta_T)|x_{<T}^{\beta_T})|x_{<T}^{\beta_T})-P_T(\vec{b}_{i},\hat{\sigma}_{-i}(\beta_T)|x_{<T}^{\beta_T})]\\
    =&\sum_{x_T} c_1 v_i(\theta_i,{x}_T|x_{<T}^{\beta_T}) - c_2,
\end{align*}

    where $c_1,c_2$ are constants.
	By assumption $\theta_i$ and $\theta_{-i}$ are indepdendent, such that $c_1,c_2$ are also independent of $\theta_i$. Moreover, by assumption $v_i$ is linear in $\theta_i$ and as we are summing over a finite set of possible allocations $x_T$ so is the expected utility.
	
	For the induction step, let $t\in \{1,\dots,T-1\}$.
\begin{align*}
    &\overline{u}_i(\theta_i|\vec{b}_i,\hat{\sigma}_{-i},\beta_t)\\
=&\Exp_{\theta_{-i}}\left[ v_i(\theta_i,X_t(\vec{b}_{i},\hat{\sigma}_{-i}(\beta_t)|x_{<t}^{\beta_t})|x_{<t}^{\beta_t})-P_t(\vec{b}_{i},\hat{\sigma}_{-i}(\beta_t)|x_{<t}^{\beta_t})\right]\\
&+\sum_{\beta_{t+1}\in \mathcal{B}^{f,p}_{t+1}(\beta_t,\rho)} \Pro(\beta_{t+1}|\beta_t,\rho)\overline{u}_i(\theta_i|\vec{b}_i,\hat{\sigma}_{-i},\beta_{t+1}),
\end{align*}

	which by the induction hypothesis is a sum of linear functions and thus linear.
\end{proof}

\begin{cor}
\label{cor:pcstratpcconv}
Given a piecewise-constant (on partition $\mathcal{P}$) \bels~strategy $\hat{\sigma}$ and a finite consistent transition $\rho$, the expected utility $\overline{u}_i(\theta_i|\mu_{-i},{\sigma},{h})$ is piecewise convex with respect to $\theta_i$, where the regions of convexity are exactly the hyperrectangles of $\mathcal{P}_i$.
\end{cor}

 We can use Corollary \ref{cor:pcstratpcconv} to claim piecewise convexity for $u_i^{BR}$ as well.%
\begin{lemma}
\label{lem:pwconvex}
Given a piecewise-constant (on partition $\mathcal{P}$) \bels~strategy $\hat{\sigma}$ and a finite consistent transition $\rho$, the expected utility $\overline{u}_i^{IBR}(\theta_i|\hat{\sigma},\beta)$ is piecewise convex.
\end{lemma}

\begin{proof}
By definition, 
    \[ \overline{u}_i^{IBR}(\theta_i|\hat{\sigma},\beta)=\sup_{b_{\beta}} \overline{u}_i(\theta_i|b_{\beta},\hat{\sigma},\beta)\]
    is the supremum over a set of piecewise convex functions that are convex on identical hyperrectangles and thus itself also piecewise convex.
\end{proof}

Using the auxiliary Lemmas from above, we can now prove Proposition \ref{prop:immediateloss}. 
\begin{proof}[Proof of Proposition \ref{prop:immediateloss}]
WLOG consider a bidder $i$ with ype $\theta_i$, such that $\theta_i\in [y,z)_j$ for some $[y,z)_j \in \mathcal{P}_i$. Let $\{z_1,\dots,z_k\}=\text{Vert}([y,z)_j)$, then by definition of a hyperrectangle, it holds that 

\[
\exists! \lambda \in \Delta^d : \sum_{k=1}^d \lambda^k z_k =\theta_i
\]
By Lemma \ref{lem:pwconvex}, $\overline{u}^{IBR}_i$ restricted to a hyperrectangle $[y,z)_j$ is convex in $\theta_i$. It therefore holds that
\[\overline{u}_i^{IBR}(\theta_i|\vec{b}_i,\hat{\sigma}_{-i},\beta)\leq \sum_{k=1}^d \lambda^k\overline{u}_i^{IBR}(z_k|\vec{b}_i,\hat{\sigma}_{-i},\beta)  ,\]
where $\vec{b}_i$ is the vector of $i$'s bids (recall that the strategy is piecewise constant on $\mathcal{P}_i$) for different $\beta$ on $[y,z)_j$, i.e. $\vec{b}_{i,\beta}=\hat{\sigma}(y|\beta)$.
Let us identify $\theta_i$ with the unique point $\lambda \in \Delta^d$ described above and define \[f(\theta_i):=\sum_{k=1}^d \lambda^k\overline{u}_i^{IBR}(z_k|\vec{b}_i,\hat{\sigma}_{-i},\beta).\] Note that $f$ is linear, when restricted to the hyperrectangle $[y,z)_j$.

By Lemma \ref{lem:convex} it holds that $\overline{u}_i(\theta_i|\vec{b}_i, \hat{\sigma}_{-i},\beta)$ is linear on $[y,z)_j$. By definition it further holds that
\[
l_i^{IBR}(\theta_i|\hat{\sigma},\beta) \leq f(\theta_i) - \overline{u}_i(\theta_i|\vec{b}_i,\hat{\sigma}_{-i},\beta)
.\]
From our discussion above the latter term is a difference of two functions linear on $[y,z)_j$, which therefore attains its maximum at some $w \in \text{Vert}([y,z)_j)$. It follows that 
     \begin{align*}
l_i^{IBR}(\theta_i|\hat{\sigma},\beta) \leq& \max_{w \in \text{Vert}([y,z)_j)} \overline{u}_i^{IBR}(w|\hat{\sigma}_i(y),\hat{\sigma}_{-i},\beta)\\
&-   \overline{u}_i(w|\hat{\sigma}_i(y),\hat{\sigma}_{-i},\beta).
     \end{align*}
        Here, $\hat{\sigma}_i(y)$ is used to denote the bid of $i$ on $[y,z)_j$ according to the piecewise-constant strategy $\hat{\sigma}$. As this holds for all $\theta$ in the hyperrectangle $[y,z)_j$, taking the maximum over all hyperrectangles in $\mathcal{P}_i$ concludes the proof.
\end{proof}

\begin{proof}[\textbf{Proof of \Cref{thm:decomp}}]
    We denote by $\hat{\sigma}_{i,t}^{BR}$ the component of $\hat{\sigma}^{BR}_i$ at time $t$.%
    \begin{align*}
        &l_i^{BR}(\theta_i|\hat{\sigma},\beta_t) \\=& \overline{u}_i^{BR}(\theta_i|\hat{\sigma}_{-i},\beta_t)-\overline{u}_i(\theta_i|\hat{\sigma},\beta_t)\\
        =& \overline{u}_i^{BR}(\theta_i|\hat{\sigma}_{-i},\beta_t)-\overline{u}_i(\theta_i|\hat{\sigma}_{i,t}^{BR},\hat{\sigma}_{i,>t},\hat{\sigma}_{-i},\beta_t)\\
        &+ \overline{u}_i(\theta_i|\hat{\sigma}_{i,t}^{BR},\hat{\sigma}_{i,>t},\hat{\sigma}_{-i},\beta_t)-\overline{u}_i(\theta_i|\hat{\sigma},\beta_t)\\
        \leq& \mathop{\mathbb{E}}_{\beta_{t+1}\in \mathcal{B}^{f,\mathcal{P}}_{t+1}(\beta_t,\rho)}[\overline{u}_i(\theta_i|\hat{\sigma}_{i,t}^{BR},\hat{\sigma}_{i,>t},\beta_{t+1})-\overline{u}_i(\theta_i|\hat{\sigma},\beta_{t+1})]\\
        &+ l_i^{IBR}(\theta_i|\hat{\sigma},\beta_t)\\
        \leq& l_i^{IBR}(\theta_i|\hat{\sigma},\beta_t) + \max_{\beta_{t+1}\in \mathcal{B}^{f,\mathcal{P}}_{t+1}(\beta_t,\rho)} l_i^{BR}(\theta_i|\hat{\sigma},\beta_{t+1}). \qedhere
    \end{align*} 
 \end{proof}

\begin{proof}[\textbf{Proof of \Cref{thm:main}}]
    By \Cref{def:pbe}, the strategies $\hat{\sigma}$ form an $\varepsilon$-PBE if
    \begin{align*}
        \varepsilon &= \max_i \sup_h \sup_{\theta_i} l_i^{BR} (\theta_i|\hat{\sigma},h)\\
        \underset{\text{(Prop. \ref{prop:equtil})}}{=}& \max_i \sup_\beta \sup_{\theta_i} l_i^{BR} (\theta_i|\hat{\sigma},\beta)\\
        \underset{\text{(Thm. \ref{thm:decomp})}}{\leq}& \max_i \max_{\beta_{t:T}} \sum_{\beta \in \beta_{t:T}} \sup_{\theta_i} l_i^{IBR} (\theta_i|\hat{\sigma},\beta)\\
        \underset{\text{(Prop \ref{prop:immediateloss})}}{\leq}& \max_i \max_{\beta_{t:T}} \sum_{\beta \in \beta_{t:T}} \max_{[y,z) \in \mathcal{P}_i} \max_{w \in {V}([y,z))} \overline{u}_i^{IBR}(w|\hat{\sigma},\beta)\\
              &-\overline{u}_i(w|\hat{\sigma}_{i}(y),\hat{\sigma}_{-i},\beta).
    \end{align*}
    
\end{proof}

\section{Bayesian Updating in Sequential Auctions}
\label{app:bayes}
In this section we clarify how to apply Bayes rule in the context of sequential auctions, where action and type spaces are continuous. We refer the reader to \Cref{app:example} for a detailed example of how to perform Bayesian updating in the abstracted auction.

Denote by $\mu_i(\cdot|h)$ the probability density.\footnote{We assume the probability measure is absolutely continuous wrt to the Lebesgue measure on $\Theta\subseteq \R^{\prod d_i}$.} As both the mechanism $(X_t,P_t,A_t)$ and the bidding strategies $\sigma$ are deterministic. Given $\theta$ and $h$ as input there is a deterministic outcome $x_t,p_t,a_t$. We can thus define a characteristic function as follows:
\begin{equation*}
    \mathds{1}(a,x|\theta,h_t)=\begin{cases}
        1 &\text{ if} (x,a)=(X_t(\sigma(\theta|h_t)),A_t(\sigma()))\\
        0 &\text{else}
    \end{cases}
\end{equation*} 
Assuming $supp(  \mathds{1}(a,x|\theta,h_t))$ is not a null set with respect to the Lebesgue measure the Bayesian update can be written as 
\begin{equation*}
    \mu(\theta|h_t\cup \{x_t,a_t\}) =\frac{    \mathds{1}(a,x|\theta,h_t)\prod_{i=1}^n \mu_i(\theta_i|h_t)}{\int_\theta    \mathds{1}(a,x|\theta,h_t)\prod_{i=1}^n \mu_i(\theta_i|h_t) d\theta}
\end{equation*}
Note that although the prior beliefs about the types might be independent, the mechanism can cause the types to become interdependent over time. Consider an adaptation of the FPSB auction from \Cref{ex:FPSB_bayes} with four bidders. The winner pays the sum of the second and third-highest bid and the payment is announced publicly as $a_1$. In that case, the fourth-highest bidder cannot update beliefs independently, as they know that $\theta_2+\theta_3=a_1$. We leave the analysis of such mechanisms to future work. For now, we assume the mechanism keeps bidders independent, i.e. that there exist $  \mathds{1}_{i}(a,x|\theta_i,h_t)$ ,such that $ \mathds{1}(a,x|\theta,h_t) = \prod_{i=1}^n  \mathds{1}_{i}(a,x|\theta_i,h_t)$ such that we can write the update as
\begin{align*}
    \mu(\theta|h_t\cup \{x_t,a_t\}) &=\prod_{i=1}^n \frac{    \mathds{1}_{i}(a,x|\theta_i,h_t)\mu_i(\theta_i|h_t)}{\int_{\theta_i}     \mathds{1}_{i}(a,x|\theta_i,h_t)\mu_i(\theta_i|h_t) d\theta_i}\\
    &= \prod_{i=1}^n \mu_i(\theta_i|h_t\cup \{x_t,a_t\})
\end{align*}
As we discuss in \Cref{ex:FPSB_bayes} this holds for common auction formats such as first and second-price sealed-bid auctions. 

We emphasize that the bidders only use these common knowledge or public beliefs decscribed above if they do not have an informational advantage from their observed payments. For example if a bidder is charged another bidder's bid and stays in the auction, they could use that information to update their beliefs about the other bidder. This would complicate things, as bidders would start having asymmetric beliefs about each other. We leave the analysis of such mechanisms to future work.

For our definition of Perfect Bayesian Equilibria later on we assume that the bidders perform Bayesian updates, as described above,\textit{ whenever possible}. Note that there are two situations when the above is not “possible”.
\begin{itemize}
    \item If $ \mathds{1}_{i}(a,x|\theta_i,h_t)$ is a null set with respect to the Lebesgue measure, then Bayesian updating is still possible but as the new measure is not absolutely continuous with respect to the Lebesgue measure, there is now well-defined density function, for example if a bidder's winning bid is announced and all bidders know the exact type of the winner. For now we assume, that in this case bidders still perform Bayesian updating. As we later see, this is not an issue as we will operate on an abstracted game with finitely many possible types, where the beliefs become probability mass functions and Bayesian updating is always possible.
    \item The second case is when the outcome is not possible given the current set of strategies and possible types such that $\text{supp}( \mathds{1}(a,x|\theta,h_t))\cap \text{supp}\mu_i=\emptyset$. In this case Bayesian updating is clearly infeasible and bidders are free to choose a new belief. We will later make extensive use of this property to construct a finite abstraction of the sequential auction.
\end{itemize}

\section{Running Example to Illustrate Concepts}
\label{app:example}
We will illustrate the concepts defined in this work on the simple example of a two-round first-price sealed-bid (FPSB) auction with three single-minded bidders and two identical goods sold in two rounds.

\begin{example}[Auction Model]
    \label{ex:FPSB}
    Consider a two-round first-price sealed-bid (FPSB) auction with uniform types where the highest bid wins. In this case, $\theta_i\in [0,1]$, $f_i(\theta_i)=1$. The bidding space is $\R_{\geq0}$. The allocation rule assigns the good to the highest bidder, i.e. $X_t = \arg\max_i b_{i,t}$ for $t\in \{1,2\}$. The auctioneer announces the winning bid, such that $A_t(b_t|h_t)=P_t(b_t|h_t)=\max_i b_{i,t}$. We assume ties are broken by index, i.e. if bidders $i$ and $j$ bid the same but $i<j$ then $i$ wins. As bidders are single-minded in the first round, we have $\forall i: v_i(\theta_i,\{i\}|\emptyset)=\theta_i$. We assume that WLOG $\theta_1>\theta_2>\theta_3$. If bidder 1 wins in the first round (i.e. $x_1=1$) and would stay in the second round then $v_1(\theta_1,\{1\}|\{1\})=0$, whereas e.g. $v_2(\theta_2,\{2\}|\{1\})=\theta_2$. Therefore, 1 leaves the auction and is not part of $N_{2}(\{1\})$. 
\end{example}

\begin{example}[Belief Updating]
    \label{ex:FPSB_bayes}
The prior distributions are uniform on $[0,1]$, i.e. $f_i(\theta)=1$ for $\theta \in [0,1]$. The bidding space is $\R_+$. The allocation rule assigns the goods to the highest bidder and $a_t(b_t|h_t)=P_t(b_t|h_t)=\argmax_i b_{i,t}$. Assuming all bidders bid truthfully their value and that bidder $1$ wins with a bid of $0.5$. %
    What can bidder 3 infer about bidder 2 after the first round? As the auction mechanism and the bidding strategy are deterministic, the observed outcome $x_1=\{1\},a_1=0.5$ was either possible or not given bidder 2's type. This allows bidder 3 to define a characteristic function $\mathds{1}_2(a,x|\theta_2,h)$ of whether the outcome was possible or not given a type $\theta_2$. For a previously empty history $\emptyset$ and our outcome we $a_1,x_1$ have:
    \begin{equation*}
        \mathds{1}_2(0.5,\{1\}|\theta_2,\emptyset) =\begin{cases}
            1 & \text{if } \theta_2\leq 0.5\\
            0 & \text{else}
        \end{cases}
    \end{equation*}
    The Bayesian update can thus be written as:
    \begin{align*}
        &\mu_2(\theta_2|\emptyset \cup \{x_1=\{1\},a_1=0.5\})\\
        =&\frac{\mu_2(\theta_2|\emptyset)\mathds{1}_2(0.5,\{1\}|\theta_2,\emptyset)}{\int_{\Theta_2}\mu_2(\theta_2\emptyset)\mathds{1}_2(0.5,\{1\}|\theta_2,\emptyset)d\theta_2}\\
        =& \frac{f_2(\theta) \mathds{1}_2(0.5,\{1\}|\theta_2,\emptyset)}{\int_0^{0.5} f_2(\theta_2) d\theta_2}\\
        =& 2\quad  \text{ for } \theta_2\in [0,0.5]
        \end{align*}
    and by symmetry for bidder 3 we also get as updated belief a uniform distribution on $[0,0.5]$.
    Note in this example the updated beliefs about the individual types remained independent of each other. This independence property holds for many auction formats used in practice and will be assumed throughout this work.
\end{example}

\begin{example}[Abstraction and PBS Representation]
    \label{ex:FPSB_abstraction}
    Now we turn to illustrate the \textbf{abstraction} performed in \Cref{sec:abstraction}. For this assume we choose a partition for all $i$ of $[0,1]$ into $[0,0.5]$ and $(0.5,1]$. In the first round the belief state is $(0,\emptyset,\hat{F})$. Assume for now that in the first round all bidders follow a truthful piecewise constant PBS strategy, i.e. \[
    \hat{\sigma}_i(\theta_i|(0,\emptyset,\hat{F}))=\begin{cases}
        0 & \text{if } \theta_i\in[0,0.5]\\
        0.5 & \text{if } \theta_i\in(0.5,1]
    \end{cases}
    \]

    Here $\hat{F}$ is the initial uniform distribution on the hyperrectangles, i.e.
\[
\hat{F}_i([0,0.5])=\hat{F}_i((0.5,1])=0.5
\]
Now let us discuss how to update the beliefs and transition to the next PBS. We assume that bidder 1 wins the first round. There are two possible outcomes, if all bidders follow $\hat{\sigma}$, either $1$ wins with a bid of $0$ or with a bid of $0.5$. This leads to two PBS states; $\beta=(1,1,\mu)$ and $\beta'=(1,1,\mu')$ respectively.

Let us start determining the belief $\mu_2$ (By symmetry $\mu_2=\mu_3$) if bidder 1 wins with a bid of $0$. In this case the observed outcome is $(x_1,a_1)=(1,0)$. This is only possible if bidders 2 and 3 bid $0$ as well. The characteristic functions are thus as follows:
\begin{align*}
    &\mathds{1}_2 (1,0|\hat{\theta}_2=[0,0.5]) =1\\
    &\mathds{1}_2 (1,0|\hat{\theta}_2=(0.5,1]) =0
\end{align*}
and we get the following Bayesian update:
\begin{align*}
    &\mu_2([0,0.5]|1,0)\\=&\frac{\mathds{1}_2 (1,0|[0,0.5])\mu_2([0,0.5])}{\mathds{1}_2 (1,0|[0,0.5])\mu_2([0,0.5])+ \mathds{1}_2 (1,0|(0.5,1])\mu_2((0.5,1])}\\=&\frac{1\cdot 0.5}{1\cdot 0.5+0\cdot 0.5}\\
    =&1.
\end{align*}
And similarly 
\begin{align*}
    &\mu_2((0.5,1]|1,0)\\=&\frac{\mathds{1}_2 (1,0|(0.5,1])\mu_2((0.5,1])}{\mathds{1}_2 (1,0|[0,0.5])\mu_2([0,0.5])+ \mathds{1}_2 (1,0|(0.5,1])\mu_2((0.5,1])}\\=&\frac{0\cdot 0.5}{1\cdot 0.5+0\cdot 0.5}\\
    =&0.
\end{align*}

We can repeat the same analysis for the case where bidder 1 wins with a bid of $0.5$ and we want to find the belief $\mu'_2$. In this case the observed outcome is $(x_1,a_1)=(1,0.5)$. This is possible both if the other bidders bid $0$ or $0.5$. The characteristic functions for bidder 2 are thus as follows:
\begin{align*}
    &\mathds{1}_2 (1,0.5|\hat{\theta}_2=[0,0.5]) =1\\
    &\mathds{1}_2 (1,0.5|\hat{\theta}_2=(0.5,1]) =1
\end{align*}
and we get the following Bayesian update:
\begin{align*}
    &\mu'_2([0,0.5]|1,0.5)\\=&\frac{\mathds{1}_2 (1,0.5|[0,0.5])\mu_2([0,0.5])}{\mathds{1}_2 (1,0.5|[0,0.5])\mu_2([0,0.5])+ \mathds{1}_2 (1,0.5|(0.5,1])\mu_2((0.5,1])}\\=&\frac{1\cdot 0.5}{1\cdot 0.5+1\cdot 0.5}\\
    =&0.5.
\end{align*}
And similarly
\begin{align*}
    &\mu'_2((0.5,1]|1,0.5)\\=&\frac{\mathds{1}_2 (1,0.5|(0.5,1])\mu_2((0.5,1])}{\mathds{1}_2 (1,0.5|[0,0.5])\mu_2([0,0.5])+ \mathds{1}_2 (1,0.5|(0.5,1])\mu_2((0.5,1])}\\=&\frac{1\cdot 0.5}{1\cdot 0.5+1\cdot 0.5}\\
    =&0.5.
\end{align*}
\end{example}

\begin{example}[Transition Rule]
    \label{ex:transition}
    In \Cref{ex:FPSB_abstraction} we focused on the case where bidders follow the PBS strategy $\hat{\sigma}$, the observed outcomes have positivie probability and we can perform Bayesian updating. Now let us assume bidder 1 does not win with a bid of 0 or 0.5, but instead with a bid of 0.25. In this case the observed outcome is $(x_1,a_1)=(1,0.25)$. While this has zero probability and in accordance with \Cref{def:pbe} we are free to choose beliefs, it makes sense to choose a finite transition rule that maps the observed outcome to a belief state that has positive probability. Given a PBS $\beta=(t,x_{<t},\mu)$ Simple choices would be to map $\rho(\beta,x_t,a_t)=(t+1,x_{<{t+1}},\mu)$, i.e. to keep the beliefs the same, or to map to the initial belief state, i.e. $\rho(\beta,x_t,a_t)=(t+1,x_{<{t+1}},\hat{F})$. However, in the case of this auction and sequential auctions in general there is a more intuitive approach. Given that bidder 1 won with a bid of 0.25, we know that the other two bidders (bidding according to $\hat{\sigma}$) must have types in $[0,0.5)$---otherwise they would have won. We can thus choose as new belief that the types of bidders 2 and 3 are uniformly distributed on $[0,0.5)$, as if bidder 1 had won with a bid of $0$. 
\end{example}

\section{Equilibrium strategies used in \Cref{sec:experiments}}
\label{app:eq_strats}
\subsection{Sequential sales with Ascending Reserve Prices}
\citet{Gong2013Ordering} give the equilibria for both the first and second price sequential sales auction with ascending reserve prices. We reproduce their statements below for completeness.

\paragraph{First-Price Auction}
In case a first-price payment rule is used the general form of the equilibrium strategy for a two-round auction with $n$ bidders and reserve prices $(0,r)$ the equilibrium strategy for bidder $i$ in the first round is given by
\begin{equation}
    \begin{array}{ll}
    b_{F P A}^{(1)}(\theta_i)= \begin{cases}\frac{1}{F(\theta_i)^{n-1}} \int_0^{\theta_i} s d F(s)^{n-1}, & {\theta_i} \leq r \\
    \frac{1}{F({\theta_i})^{n-1}}\Bigg[\int_r^{\theta_i} b_{F P A}^{(2)}(s) d F(s)^{n-1}\\+b_{F P A}^{(1)}(r) F(r)^{n-1}\Bigg], & {\theta_i}>r\end{cases} 
    \end{array}
\end{equation}
and in the second round for ${\theta_i} \geq r$ by
\begin{equation}
    b_{F P A}^{(2)}({\theta_i})=\frac{1}{F({\theta_i})^{n-2}}\Bigg[\int_r^{\theta_i} s d F(s)^{n-2}+
     r F(r)^{n-2}\Bigg].
    \end{equation}
For our case with uniform values this translates to  
\begin{equation}
    b_{F P A}^{(1)}(\theta_i)= \begin{cases}\frac{n-1}{n} \theta_i, & \theta_i \leq r \\ \frac{n-2}{n} \theta_i+\frac{r^{n-1}}{\theta_i^{n-2}}-\frac{n-1}{n} \frac{r^n}{\theta_i^{n-1}}, & \theta_i>r\end{cases}
    \end{equation}
\paragraph{Second-Price Auction}
In the second-price auction the equilibrium strategy for bidder $i$ in the first round is given by
\begin{equation}
    \begin{aligned}
     b_{S P A}^{(1)}(\theta_i)= \begin{cases}\theta_i, & \theta_i \leq r \\
    \frac{1}{F(\theta_i)^{n-2}}\Bigg[\int_r^{\theta_i} b_{S P A}^{(2)}(s) d F(s)^{n-2}\\+r F(r)^{n-2}\Bigg], & \theta_i>r\end{cases} 
    \end{aligned}
    \end{equation}
and in the second round for $\theta_i \geq r$ by
\begin{equation}
     b_{S P A}^{(2)}(\theta_i)=\theta_i, \theta_i \geq r.
\end{equation}

For our case with uniform values this translates to a first-round strategy
\begin{equation}
    b_{S P A}^{(1)}(\theta_i)= \left[(n-2) \theta_i^{n-1}+r^{n-1}\right] /\left[(n-1) \theta_i^{n-2}\right].
\end{equation}

\subsection{Split-award Auction}

The bidders have a type $\theta_i \sim F$ on $[\underline{\Theta},\overline{\Theta}]$, which is their cost for providing 100\% of the good or service.
\begin{align}
{\sigma}_{i}^{(sp)}\left(\theta_i| {h}_0\right) & =\frac{\int_{\theta_i}^{\bar{\Theta}}{\sigma}_{l}^{(sp)}(t)(n-1)(1-F(t))^{n-2} f(t) d t}{\left(1-F\left(\theta_i\right)\right)^{n-1}} \notag \\
{\sigma}_{w}^{(sp)}\left(\theta_w|{h}_1\right) & =\theta_w(1-C) \label{eq:kokotteq} \\
{\sigma}_{l}^{(sp)}\left(\theta_l|{h}_1\right) & =\theta_l C+C \frac{\int_{\theta_l}^{\bar{\Theta}}(1-F(t))^{n-2} d t}{\left(1-F\left(\theta_l\right)\right)^{n-2}} \notag
\end{align}
Note that the sole bid in the first round has no explicit form. We just require it to be high enough to be uncompetitive.

\section{Plots from Experiments}
\label{app:exp}
\begin{figure}[H]
    \centering
    \includegraphics[trim=1mm 1mm 1mm 1mm, clip,width=\columnwidth]{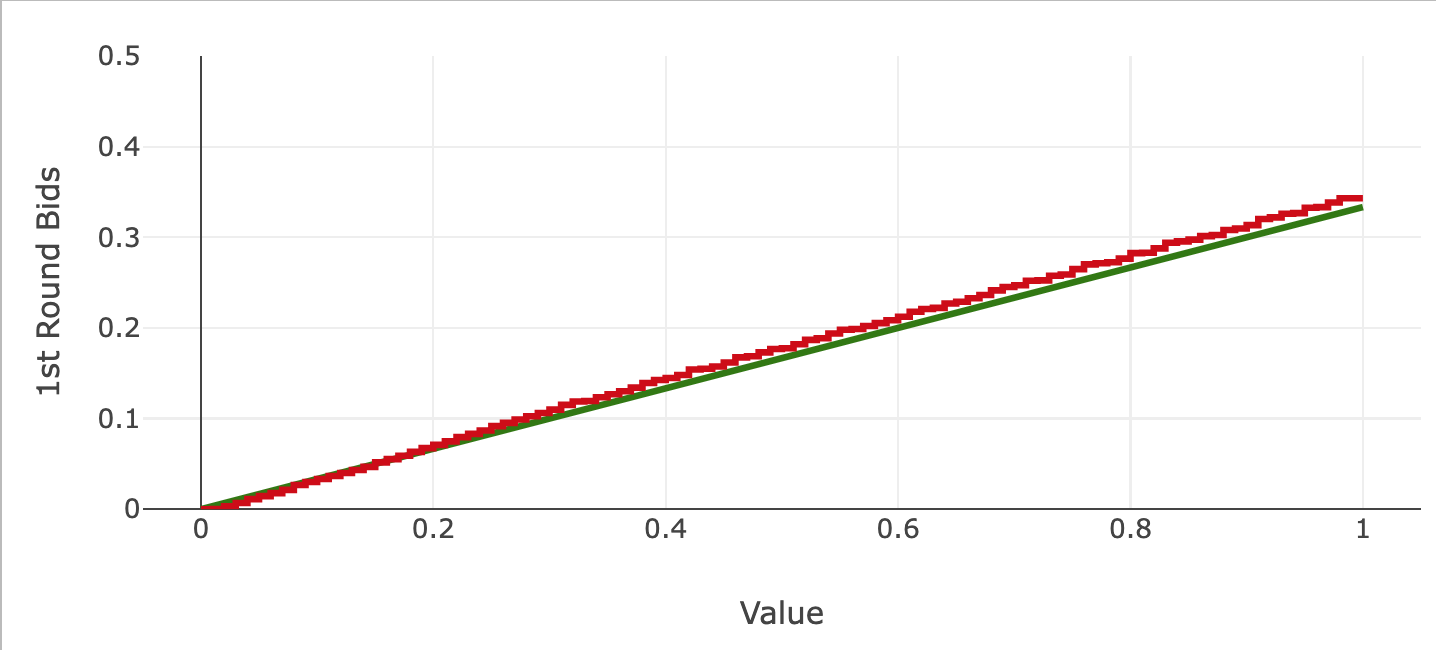}
    \caption{Bidding strategies for first round of the second price sequential sales auction with 3 bidders and 2 goods. Green denotes the theoretical prediction and red our found strategies.}
    \label{fig:krishna_3_2_fp_1}
\end{figure}

\begin{figure}[H]
    \centering
    \includegraphics[trim=1mm 1mm 1mm 1mm, clip,width=\columnwidth]{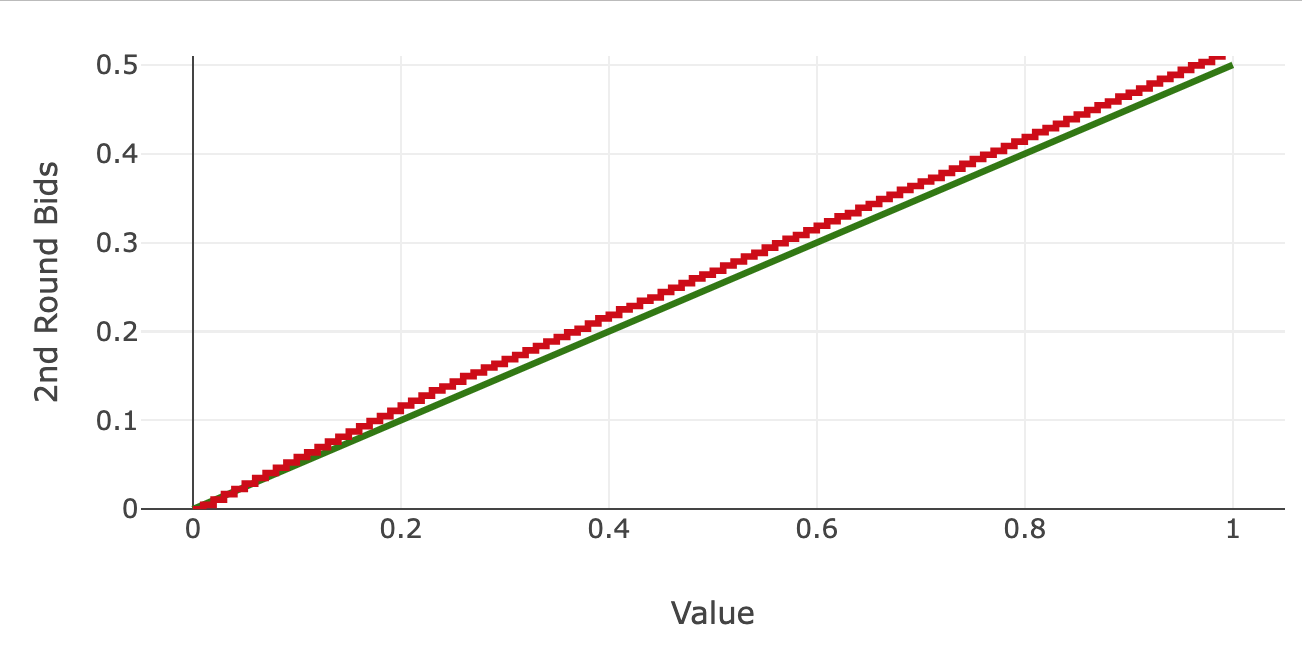}
    \caption{Bidding strategies for second round of the second price sequential sales auction with 3 bidders and 2 goods. Green denotes the theoretical prediction and red our found strategies.}
    \label{fig:krishna_3_2_fp_2}
\end{figure}

\begin{figure}[H]
    \centering
    \includegraphics[trim=1mm 1mm 1mm 1mm, clip,width=\columnwidth]{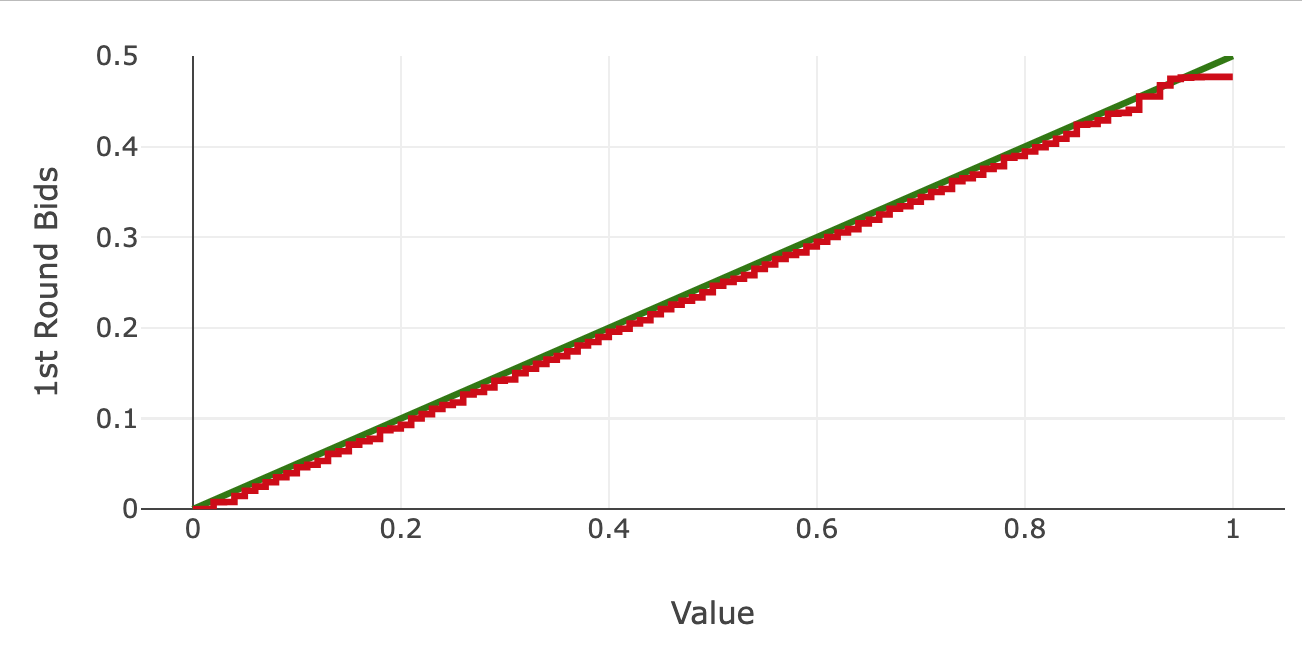}
    \caption{Bidding strategies for first round of the second price sequential sales auction with 3 bidders and 2 goods. Green denotes the theoretical prediction and red our found strategies.}
    \label{fig:krishna_3_2_sp_1}
\end{figure}

\begin{figure}[H]
    \centering
    \includegraphics[trim=1mm 1mm 1mm 1mm, clip,width=\columnwidth]{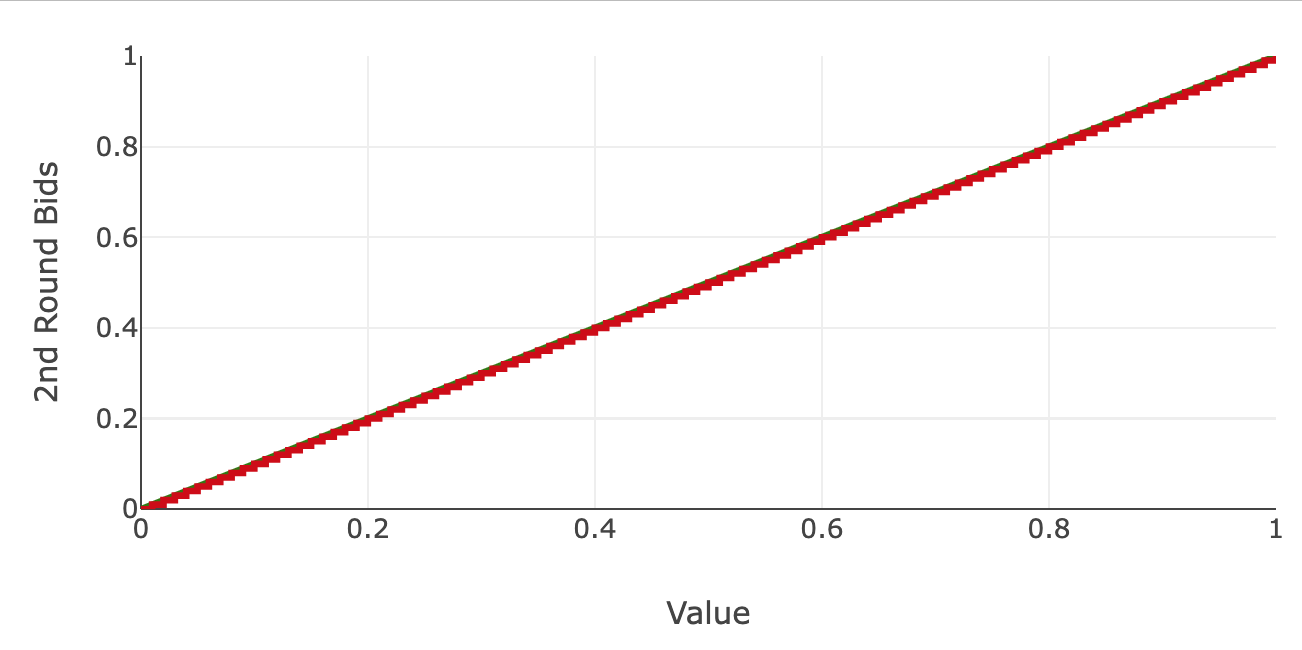}
    \caption{Bidding strategies for second round of the second price sequential sales auction with 3 bidders and 2 goods. Green denotes the theoretical prediction and red our found strategies.}
    \label{fig:krishna_3_2_sp_2}
\end{figure}

\begin{figure}[H]
    \centering
    \includegraphics[trim=1mm 1mm 1mm 1mm, clip,width=\columnwidth]{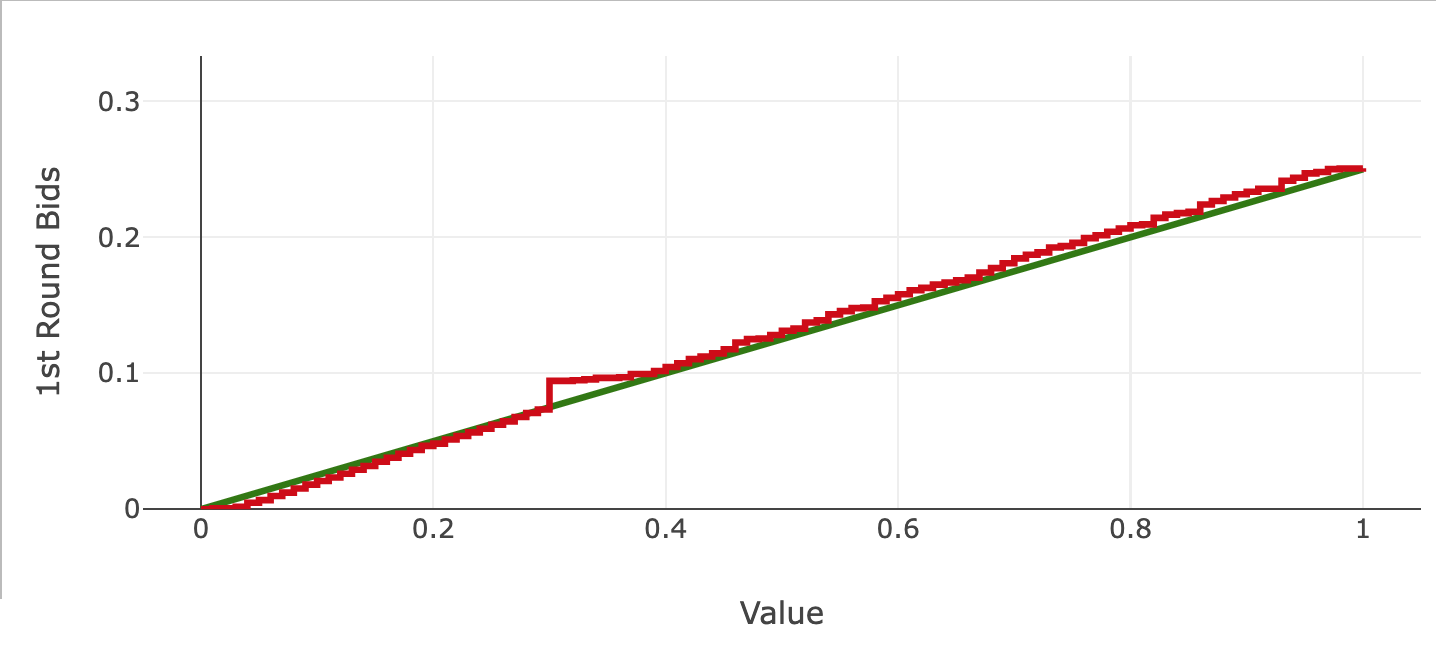}
    \caption{Bidding strategies for first round of the first price sequential sales auction with 4 bidders and 3 goods. Green denotes the theoretical prediction and red our found strategies.}
    \label{fig:krishna_4_3_fp_1}
\end{figure}

\begin{figure}[H]
    \centering
    \includegraphics[trim=1mm 1mm 1mm 1mm, clip,width=\columnwidth]{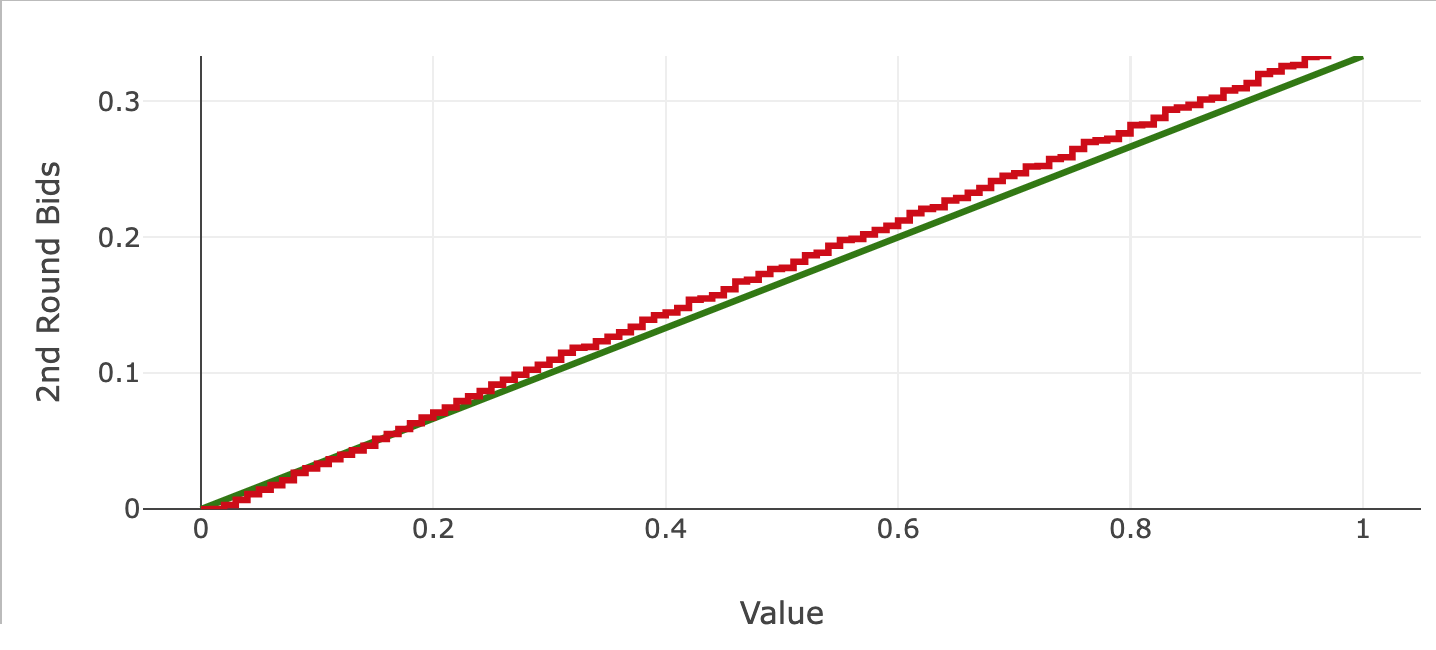}
    \caption{Bidding strategies for second round of the first price sequential sales auction with 4 bidders and 3 goods. Green denotes the theoretical prediction and red our found strategies.}
    \label{fig:krishna_4_3_fp_2}
\end{figure}

\begin{figure}[H]
    \centering
    \includegraphics[trim=1mm 1mm 1mm 1mm, clip,width=\columnwidth]{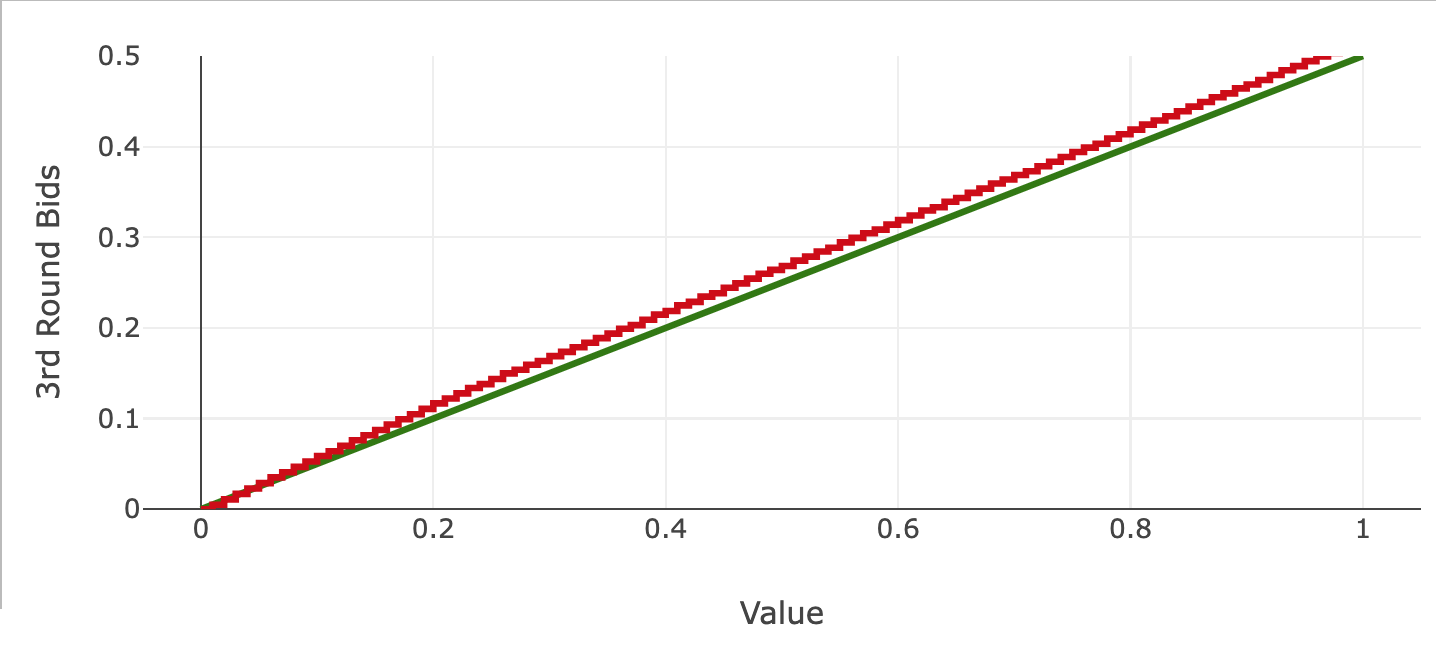}
    \caption{Bidding strategies for third round of the first price sequential sales auction with 4 bidders and 3 goods. Green denotes the theoretical prediction and red our found strategies.}
    \label{fig:krishna_4_3_fp_3}
\end{figure}

\begin{figure}[H]
    \centering
    \includegraphics[trim=1mm 1mm 1mm 1mm, clip,width=\columnwidth]{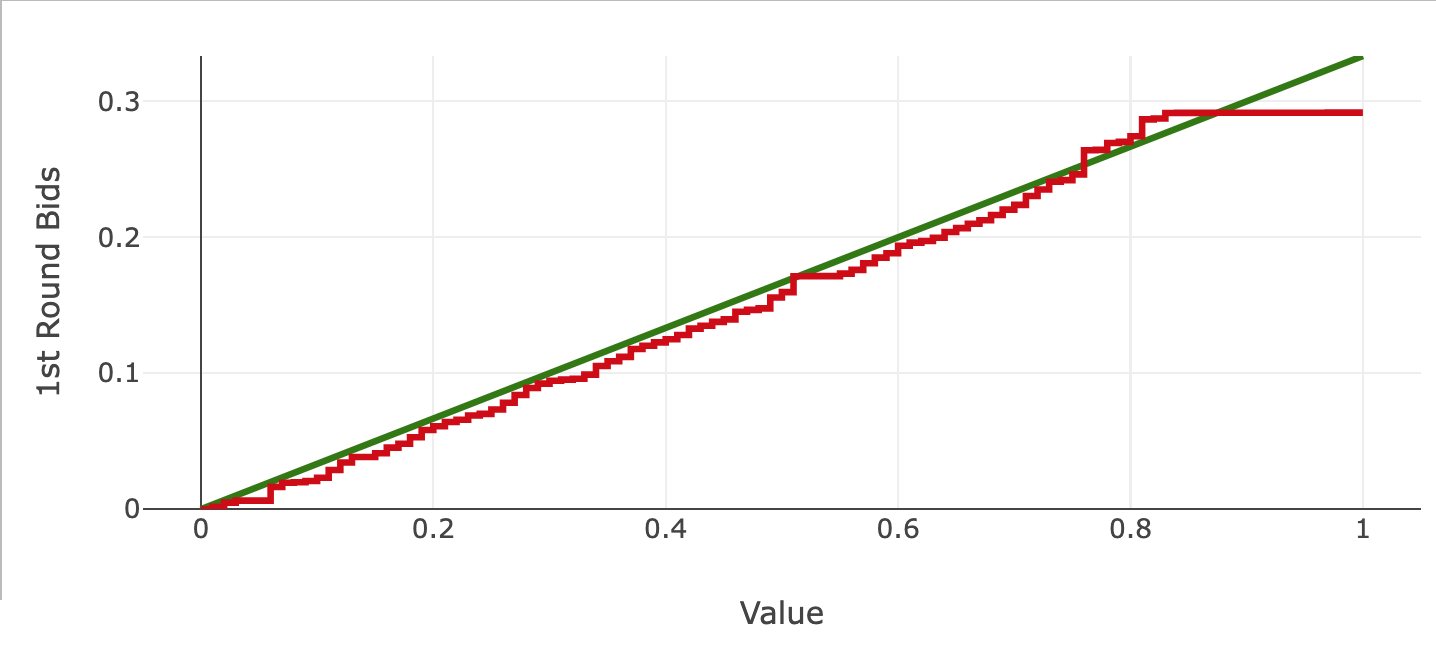}
    \caption{Bidding strategies for first round of the first price sequential sales auction with 4 bidders and 3 goods. Green denotes the theoretical prediction and red our found strategies.}
    \label{fig:krishna_4_3_sp_1}
\end{figure}

\begin{figure}[H]
    \centering
    \includegraphics[trim=1mm 1mm 1mm 1mm, clip,width=\columnwidth]{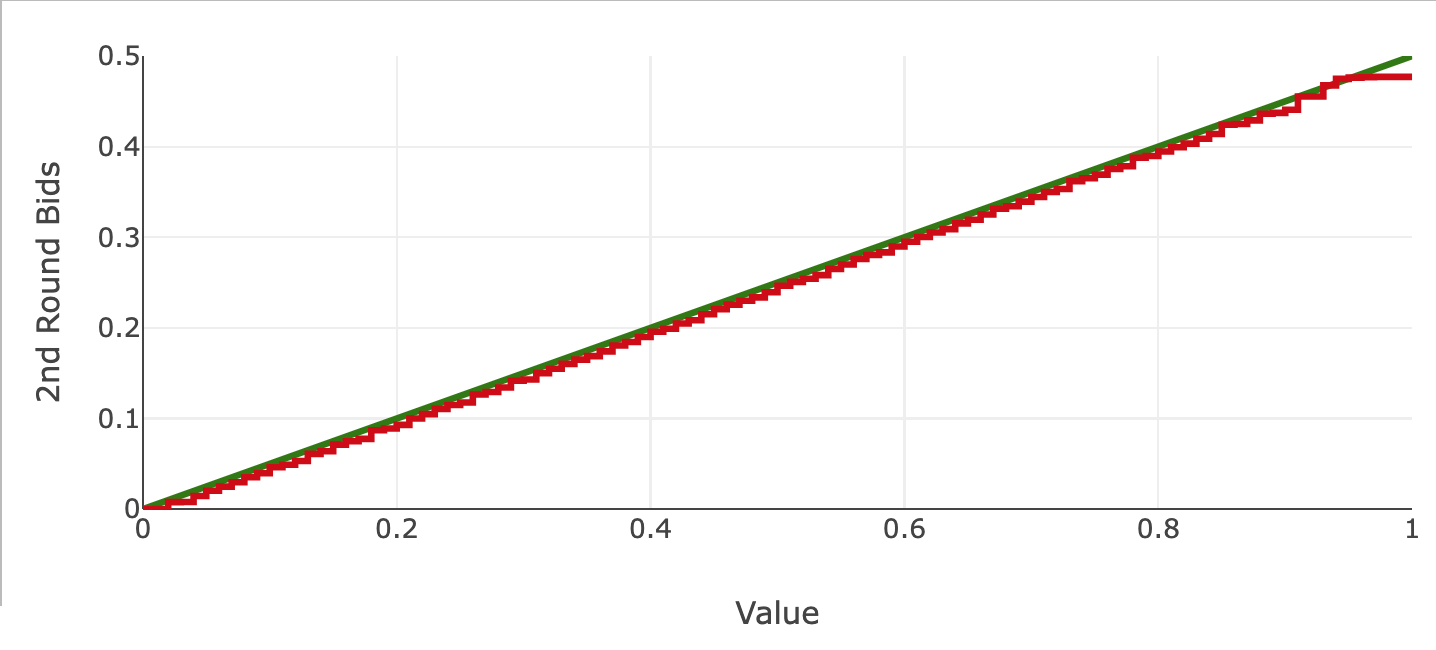}
    \caption{Bidding strategies for second round of the second price sequential sales auction with 4 bidders and 3 goods. Green denotes the theoretical prediction and red our found strategies.}
    \label{fig:krishna_4_3_sp_2}
\end{figure}

\begin{figure}[H]
    \centering
    \includegraphics[trim=1mm 1mm 1mm 1mm, clip,width=\columnwidth]{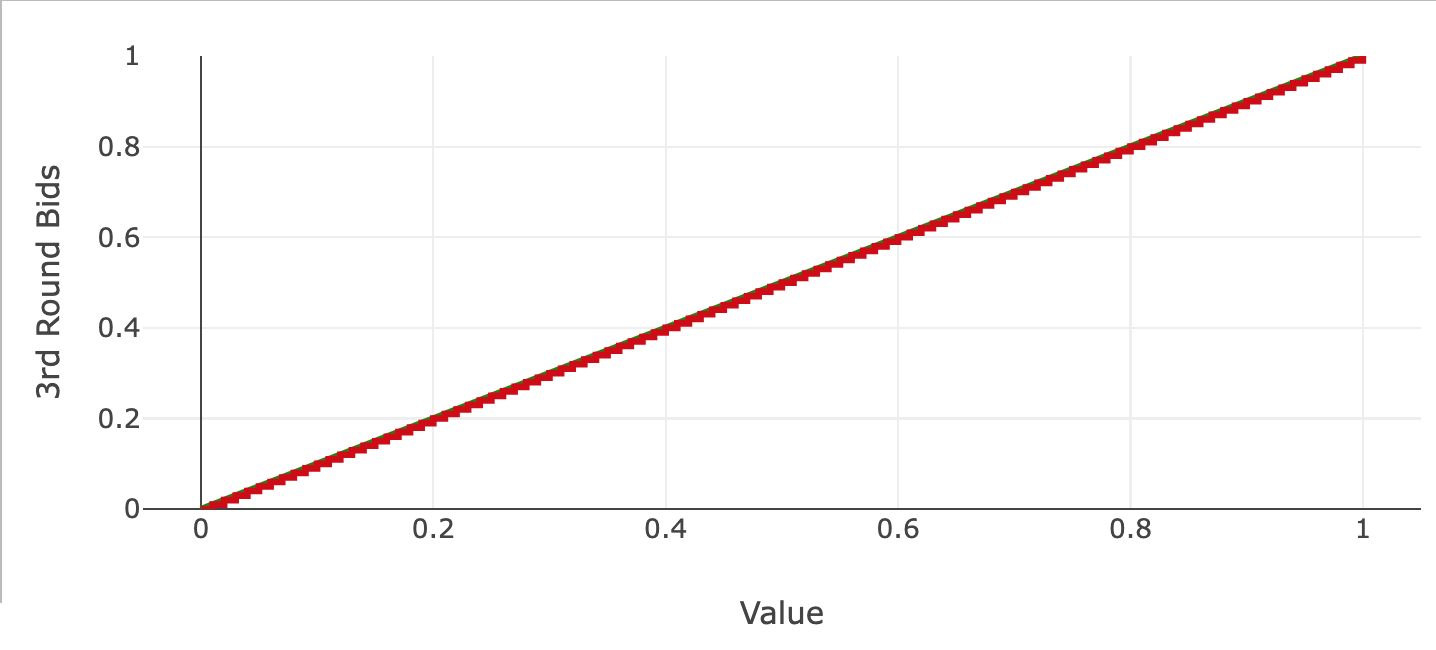}
    \caption{Bidding strategies for third round of the second price sequential sales auction with 4 bidders and 3 goods. Green denotes the theoretical prediction and red our found strategies.}
    \label{fig:krishna_4_3_sp_3}
\end{figure}

\begin{figure}[H]
    \centering
    \includegraphics[trim=1mm 1mm 1mm 1mm, clip,width=\columnwidth]{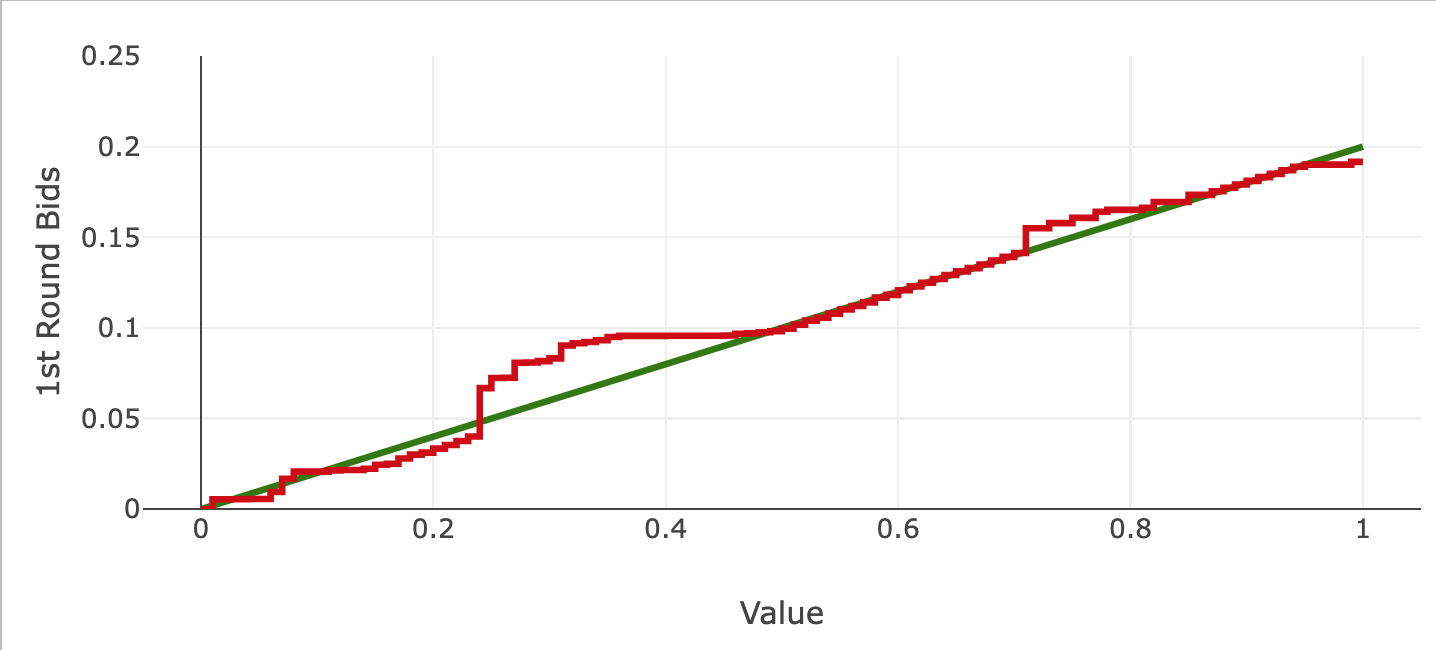}
    \caption{Bidding strategies for first round of the first price sequential sales auction with 5 bidders and 4 goods. Green denotes the theoretical prediction and red our found strategies.}
    \label{fig:krishna_5_4_fp_1}
\end{figure}

\begin{figure}[H]
    \centering
    \includegraphics[trim=1mm 1mm 1mm 1mm, clip,width=\columnwidth]{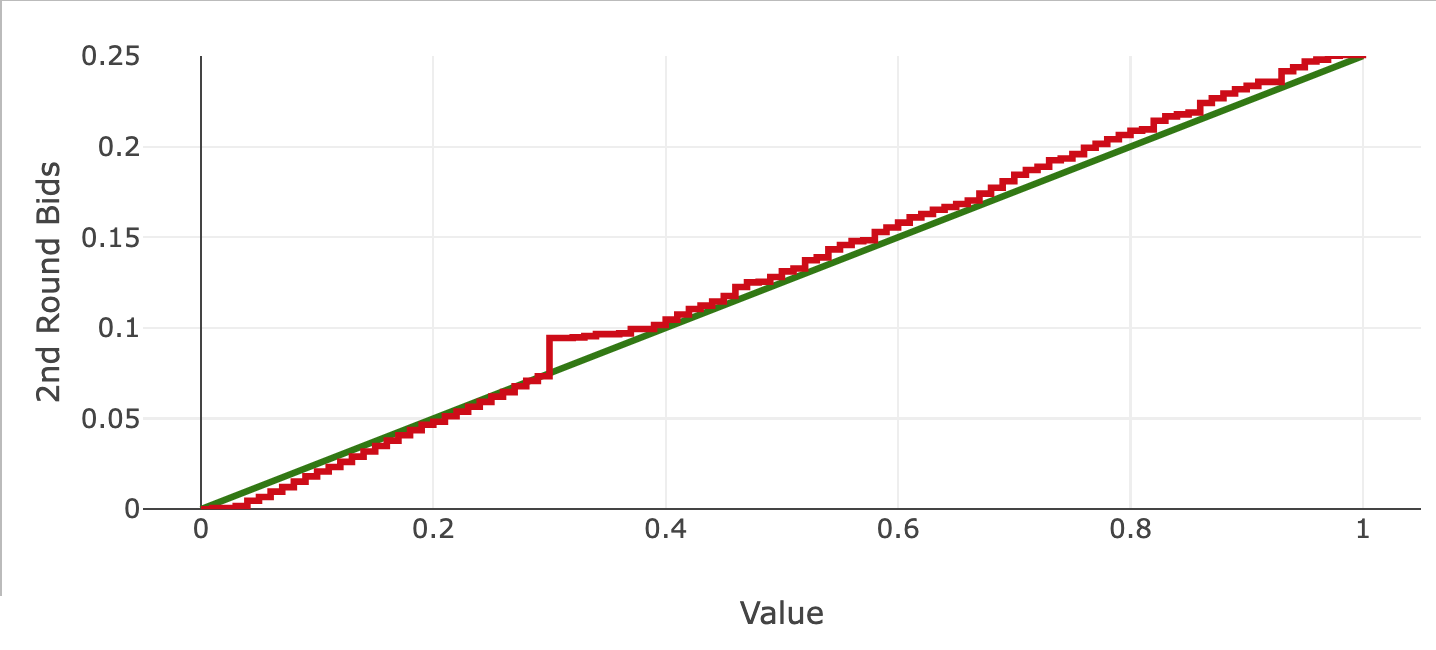}
    \caption{Bidding strategies for second round of the first price sequential sales auction with 5 bidders and 4 goods. Green denotes the theoretical prediction and red our found strategies.}
    \label{fig:krishna_5_4_fp_2}
\end{figure}

\begin{figure}[H]
    \centering
    \includegraphics[trim=1mm 1mm 1mm 1mm, clip,width=\columnwidth]{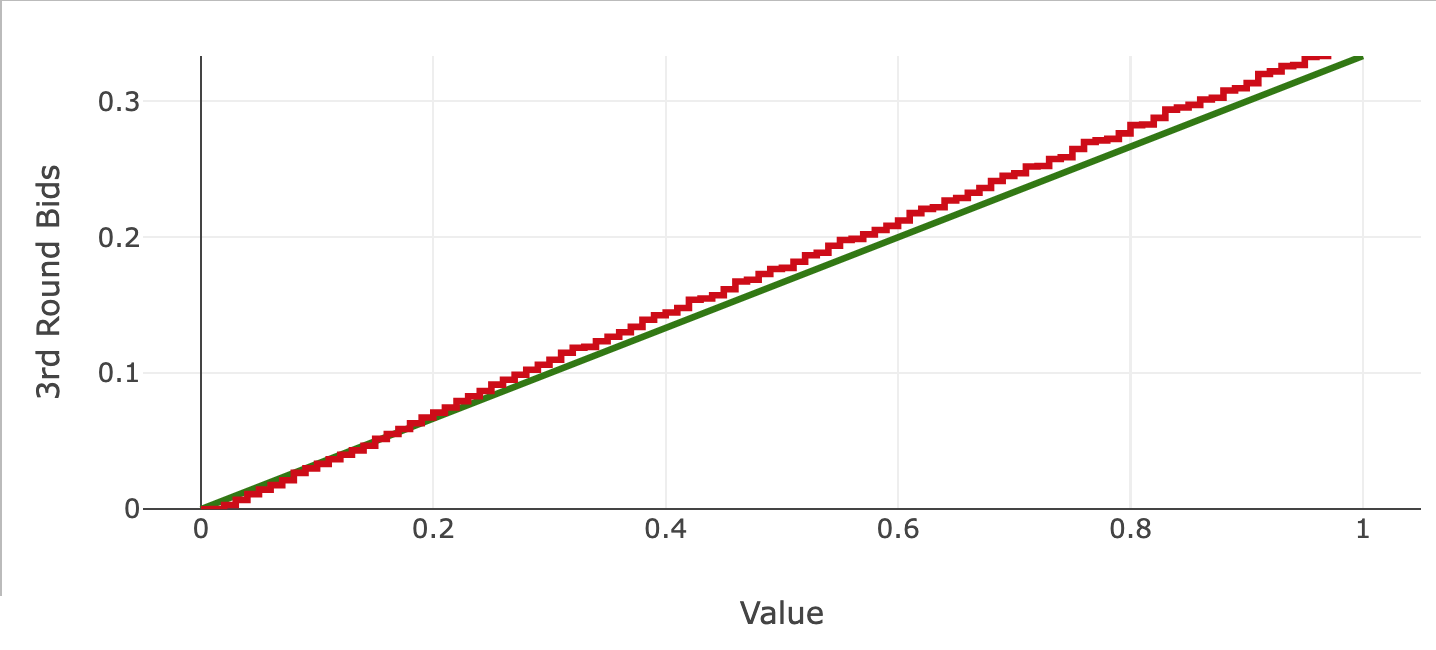}
    \caption{Bidding strategies for third round of the first price sequential sales auction with 5 bidders and 4 goods. Green denotes the theoretical prediction and red our found strategies.}
    \label{fig:krishna_5_4_fp_3}
\end{figure}

\begin{figure}[H]
    \centering
    \includegraphics[trim=1mm 1mm 1mm 1mm, clip,width=\columnwidth]{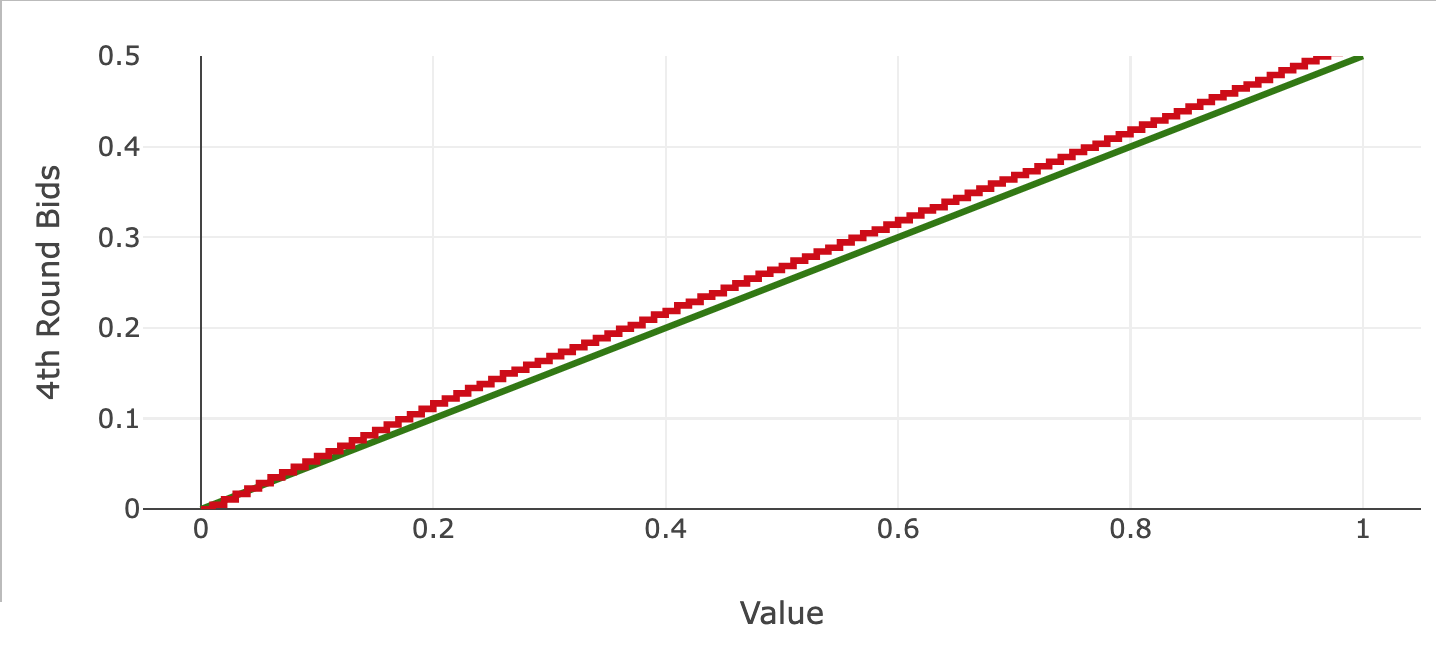}
    \caption{Bidding strategies for fourth round of the first price sequential sales auction with 5 bidders and 4 goods. Green denotes the theoretical prediction and red our found strategies.}
    \label{fig:krishna_5_4_fp_4}
\end{figure}

\begin{figure}[H]
    \centering
    \includegraphics[trim=1mm 1mm 1mm 1mm, clip,width=\columnwidth]{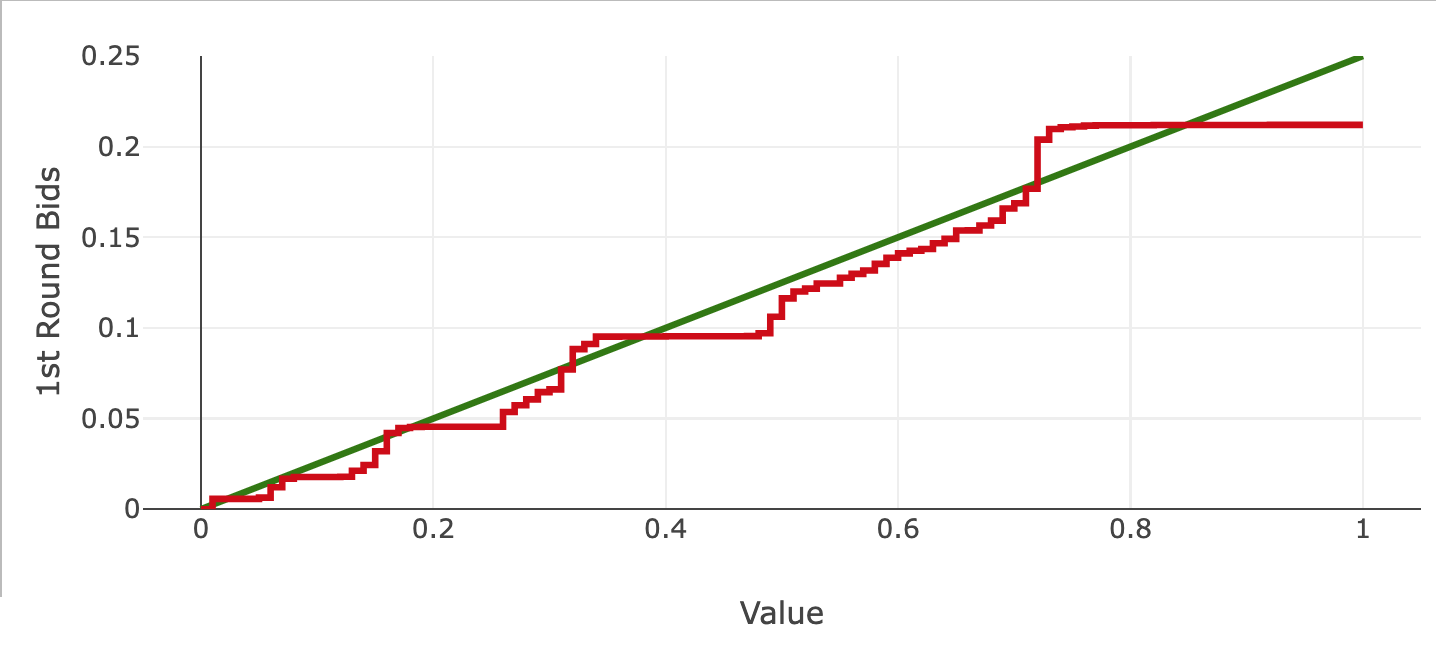}
    \caption{Bidding strategies for first round of the second price sequential sales auction with 5 bidders and 4 goods. Green denotes the theoretical prediction and red our found strategies.}
    \label{fig:krishna_5_4_sp_1}
\end{figure}

\begin{figure}[H]
    \centering
    \includegraphics[trim=1mm 1mm 1mm 1mm, clip,width=\columnwidth]{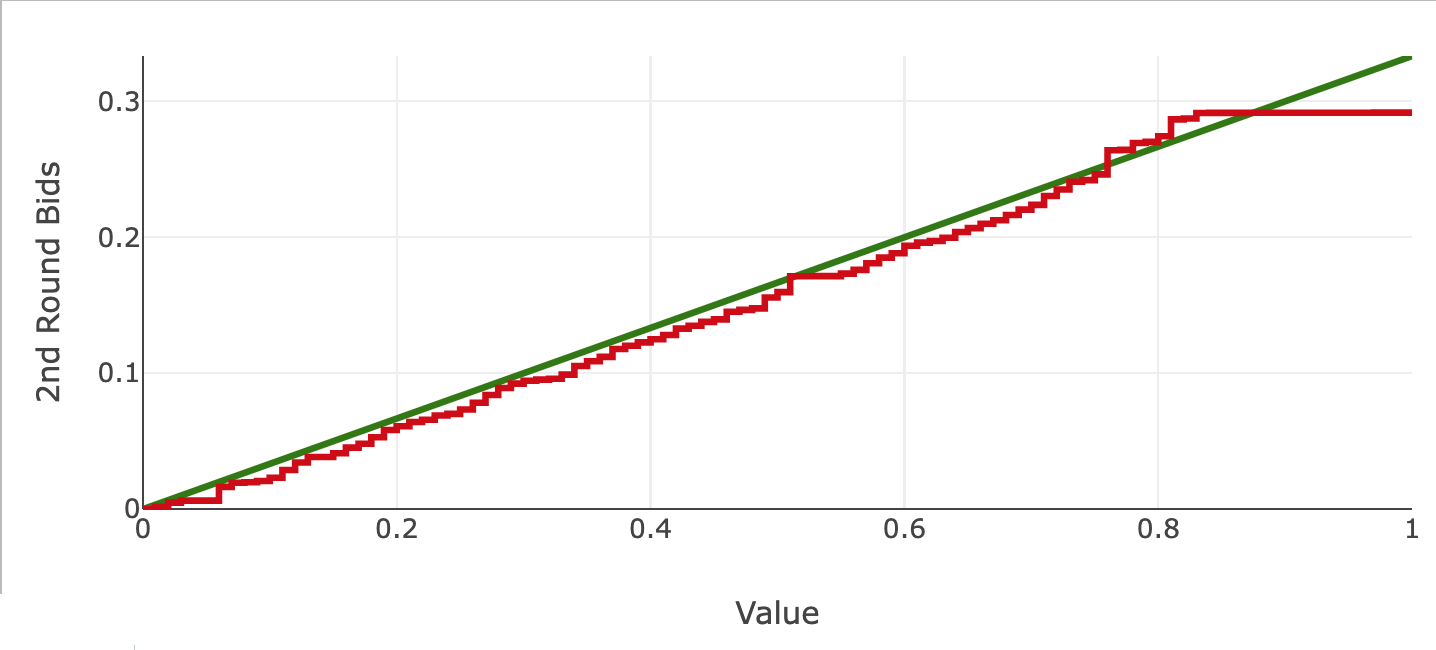}
    \caption{Bidding strategies for second round of the second price sequential sales auction with 5 bidders and 4 goods. Green denotes the theoretical prediction and red our found strategies.}
    \label{fig:krishna_5_4_sp_2}
\end{figure}

\begin{figure}[H]
    \centering
    \includegraphics[trim=1mm 1mm 1mm 1mm, clip,width=\columnwidth]{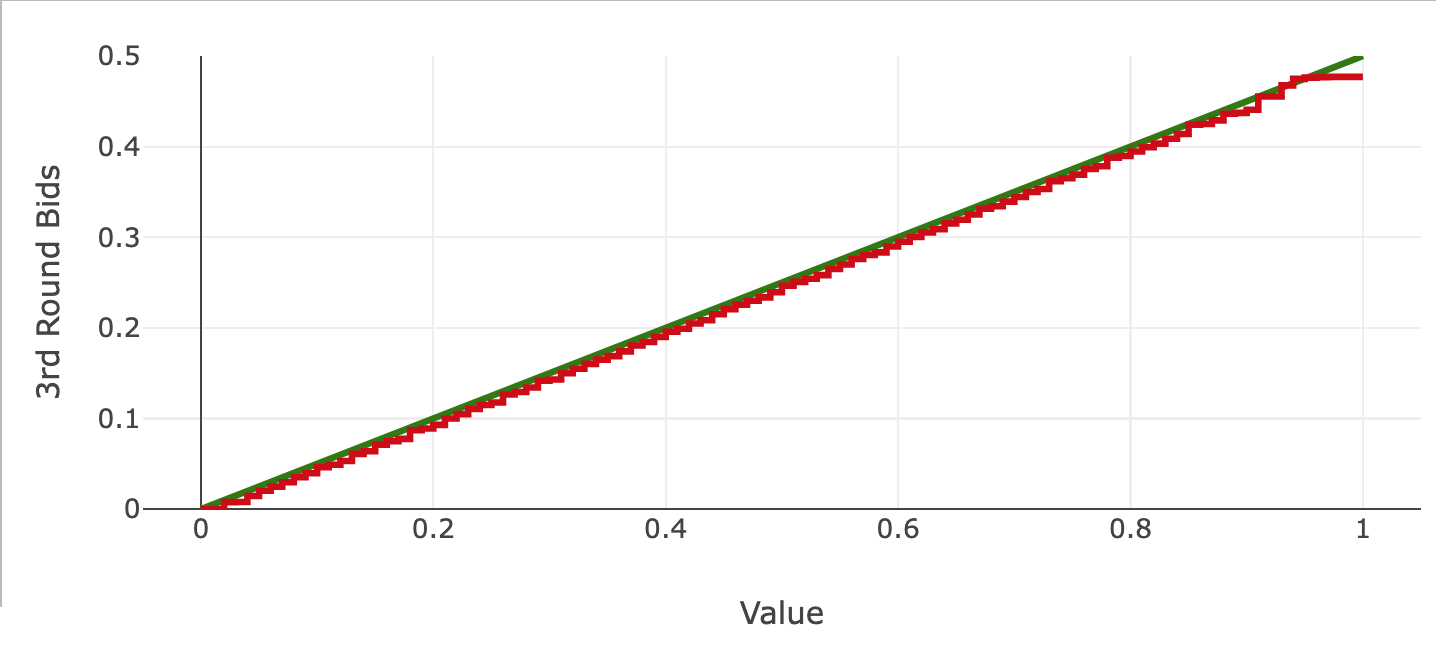}
    \caption{Bidding strategies for third round of the second price sequential sales auction with 5 bidders and 4 goods. Green denotes the theoretical prediction and red our found strategies.}
    \label{fig:krishna_5_4_sp_3}
\end{figure}

\begin{figure}[H]
    \centering
    \includegraphics[trim=1mm 1mm 1mm 1mm, clip,width=\columnwidth]{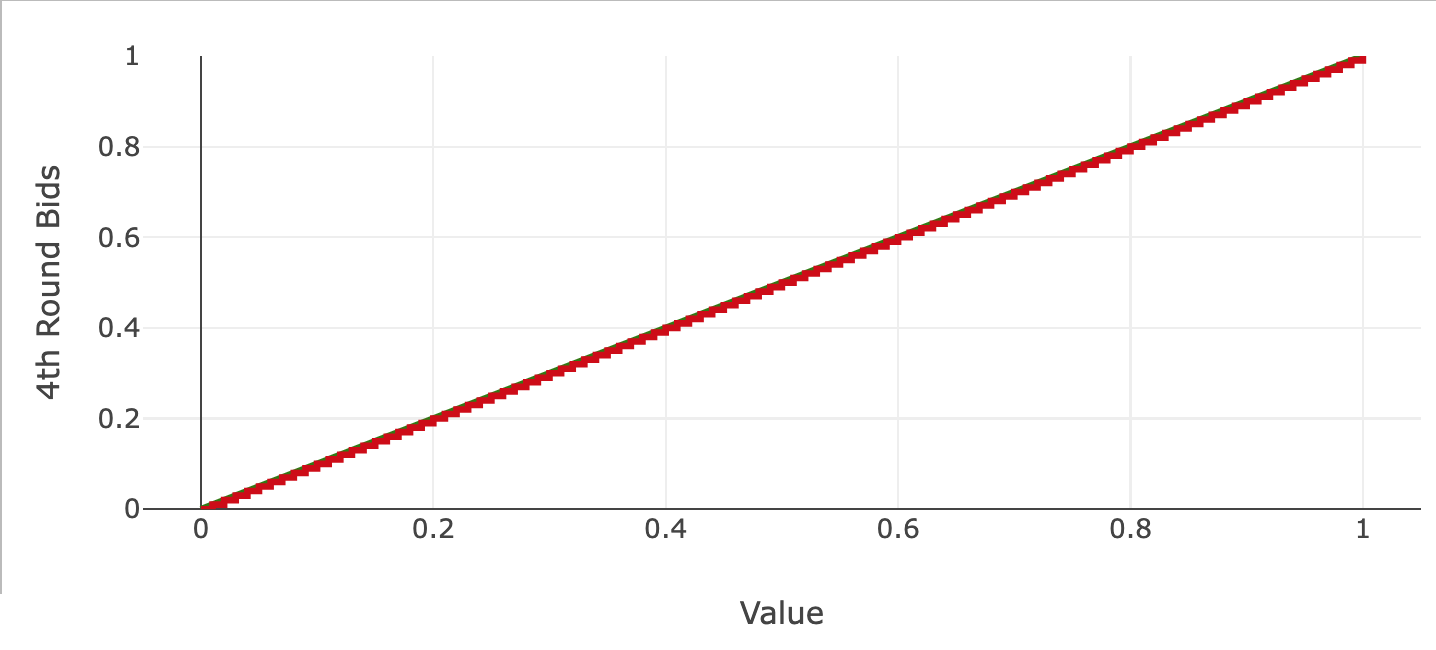}
    \caption{Bidding strategies for fourth round of the second price sequential sales auction with 5 bidders and 4 goods. Green denotes the theoretical prediction and red our found strategies.}
    \label{fig:krishna_5_4_sp_4}
\end{figure}

\begin{figure}[H]
    \centering
    \includegraphics[trim=1mm 1mm 1mm 1mm, clip,width=\columnwidth]{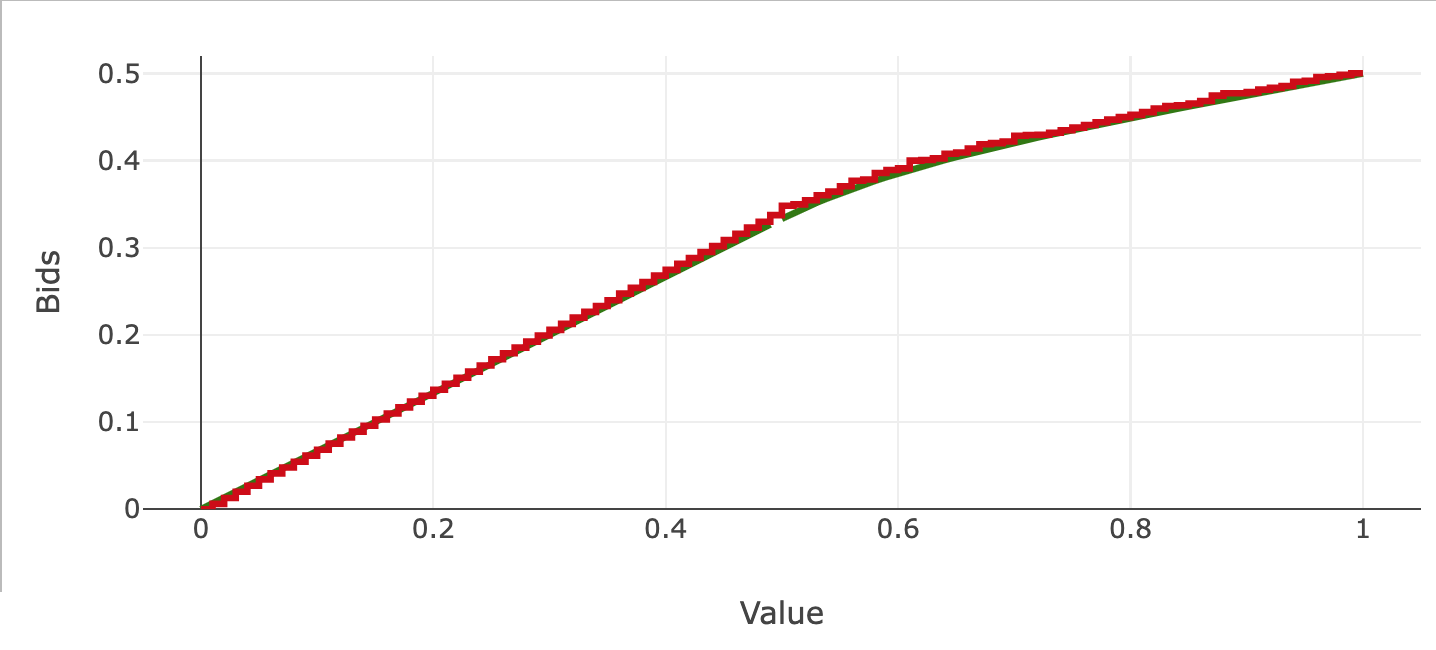}
    \caption{Bidding strategies for first round of the first price sequential sales auction with 3 bidders, 2 goods and reserve prices $r_1=0,r_2=0.5$. Green denotes the theoretical prediction and red our found strategies.}
    \label{fig:res_3_2_fp_1_re}
\end{figure}
\begin{figure}[H]
    \centering
    \includegraphics[trim=1mm 1mm 1mm 1mm, clip,width=\columnwidth]{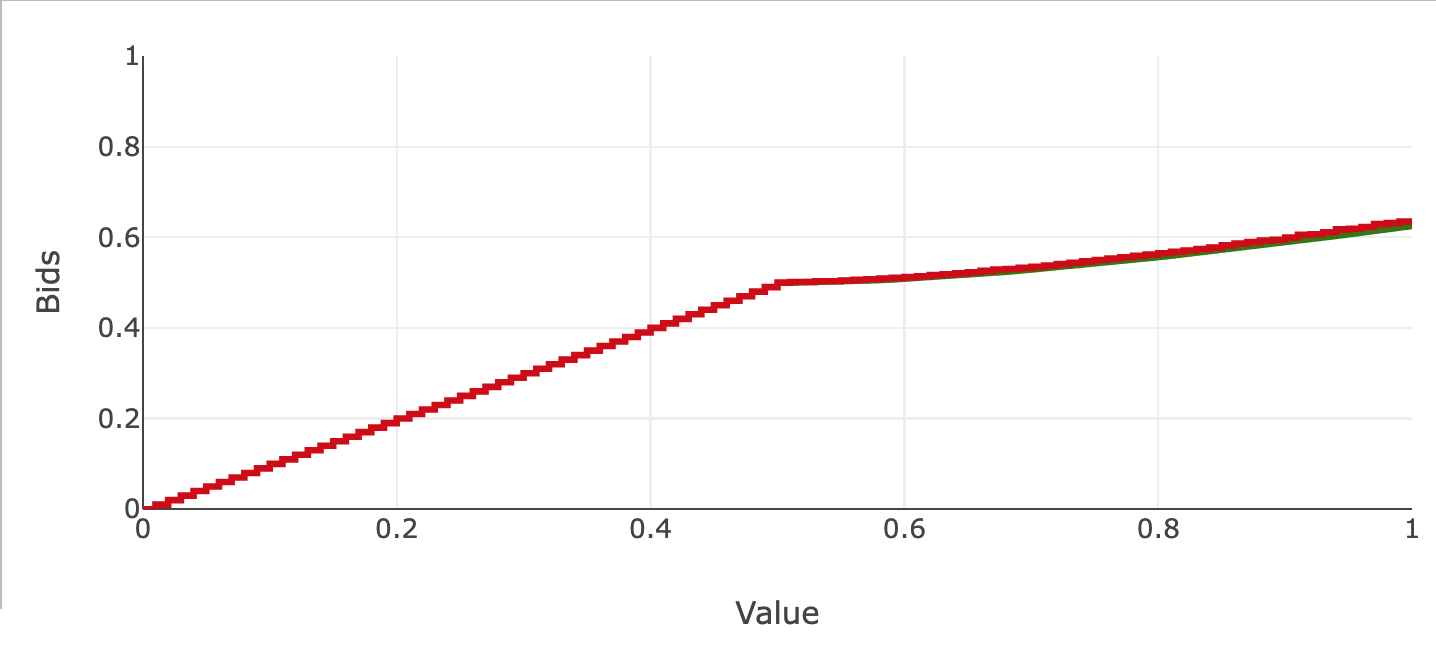}
    \caption{Bidding strategies for second round of the first price sequential sales auction with 3 bidders, 2 goods and reserve prices $r_1=0,r_2=0.5$, when the first item was sold in the first round. Green denotes the theoretical prediction and red our found strategies.}
    \label{fig:res_3_2_fp_2}
\end{figure}
\begin{figure}[H]
    \centering
    \includegraphics[trim=1mm 1mm 1mm 1mm, clip,width=\columnwidth]{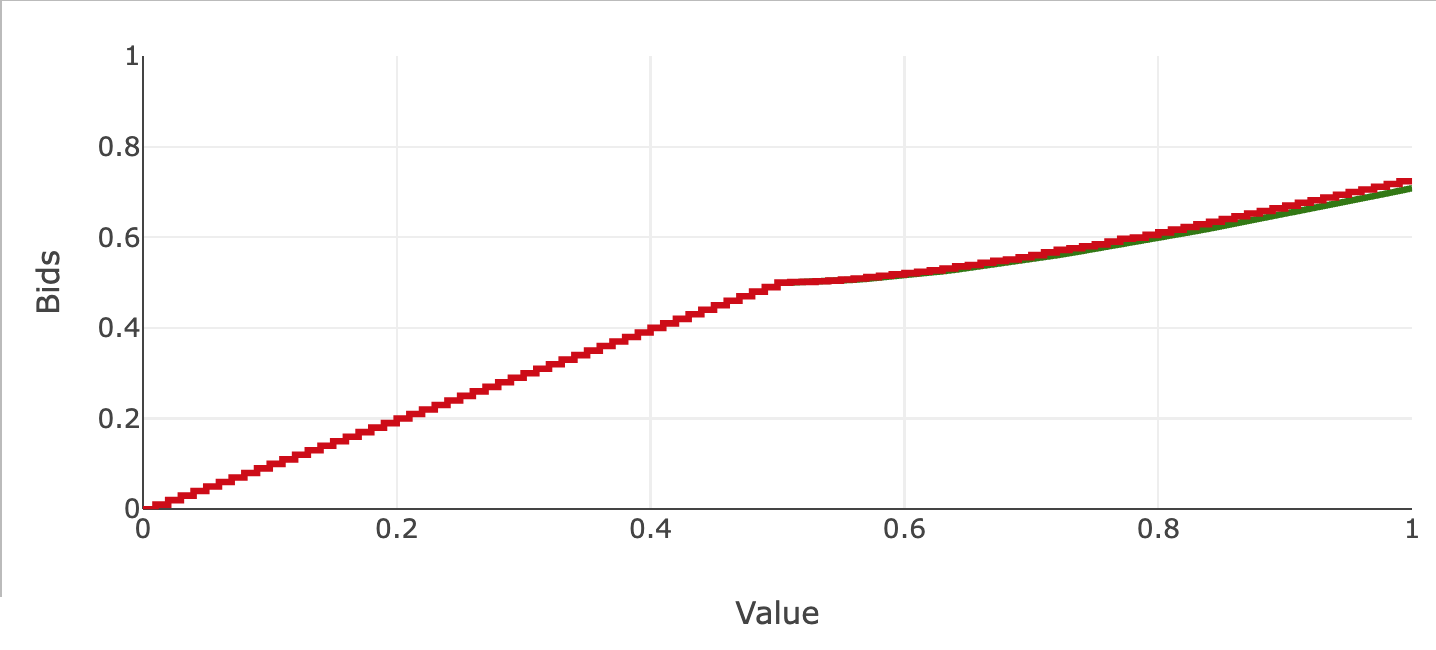}
    \caption{Bidding strategies for second round of the first price sequential sales auction with 3 bidders, 2 goods and reserve prices $r_1=0,r_2=0.5$, when the first item was not sold in the first round. Green denotes the theoretical prediction and red our found strategies.}
    \label{fig:res_3_2_fp_3}
\end{figure}

\begin{figure}[H]
    \centering
    \includegraphics[trim=1mm 1mm 1mm 1mm, clip,width=\columnwidth]{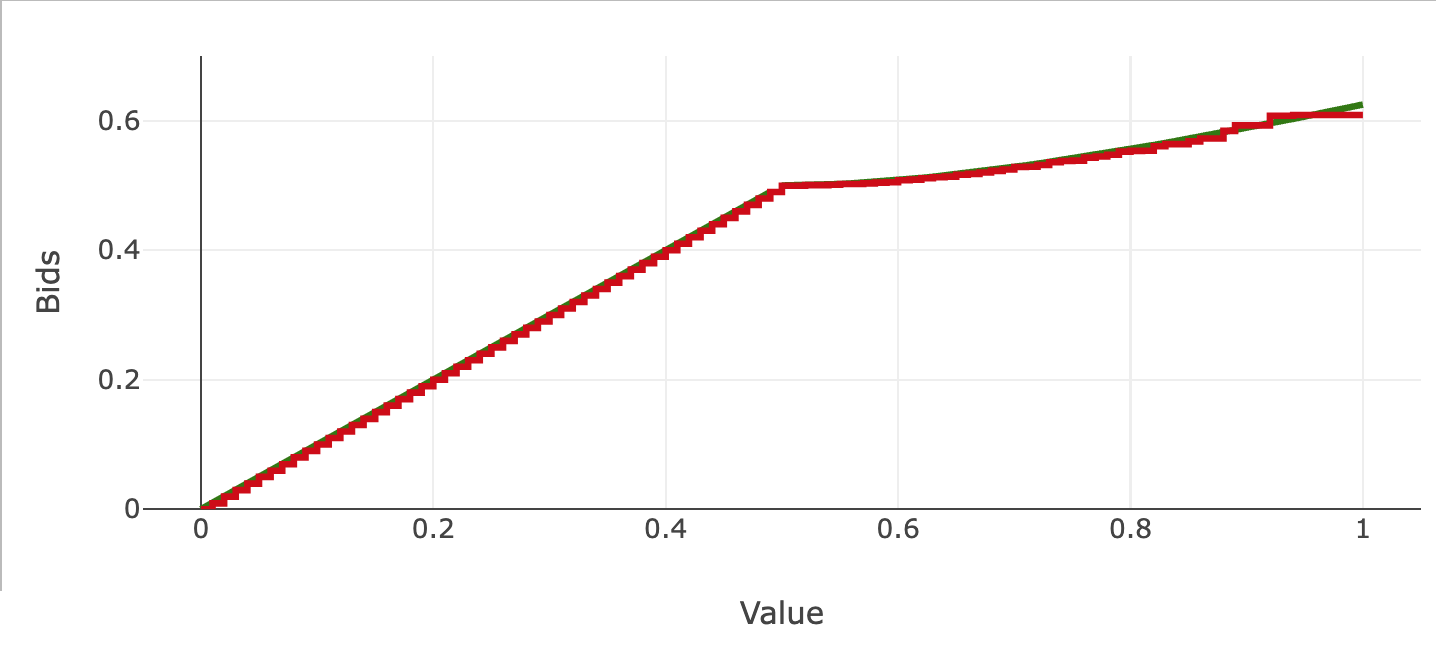}
    \caption{Bidding strategies for first round of the second price sequential sales auction with 3 bidders, 2 goods and reserve prices $r_1=0,r_2=0.5$. Green denotes the theoretical prediction and red our found strategies.}
    \label{fig:res_3_2_sp_1_re}
\end{figure}
\begin{figure}[H]
    \centering
    \includegraphics[trim=1mm 1mm 1mm 1mm, clip,width=\columnwidth]{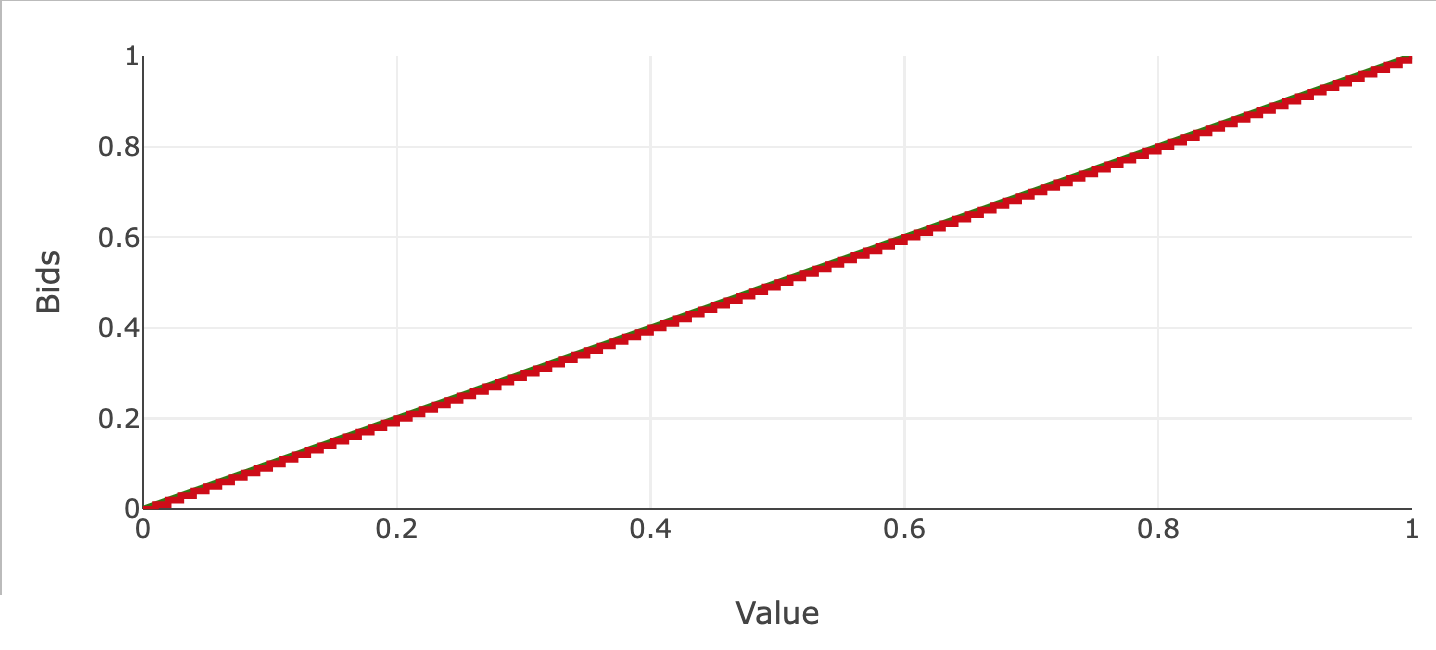}
    \caption{Bidding strategies for second round of the second price sequential sales auction with 3 bidders, 2 goods and reserve prices $r_1=0,r_2=0.5$, when the first item was sold in the first round. Green denotes the theoretical prediction and red our found strategies.}
    \label{fig:res_3_2_sp_2}
\end{figure}
\begin{figure}[H]
    \centering
    \includegraphics[trim=1mm 1mm 1mm 1mm, clip,width=\columnwidth]{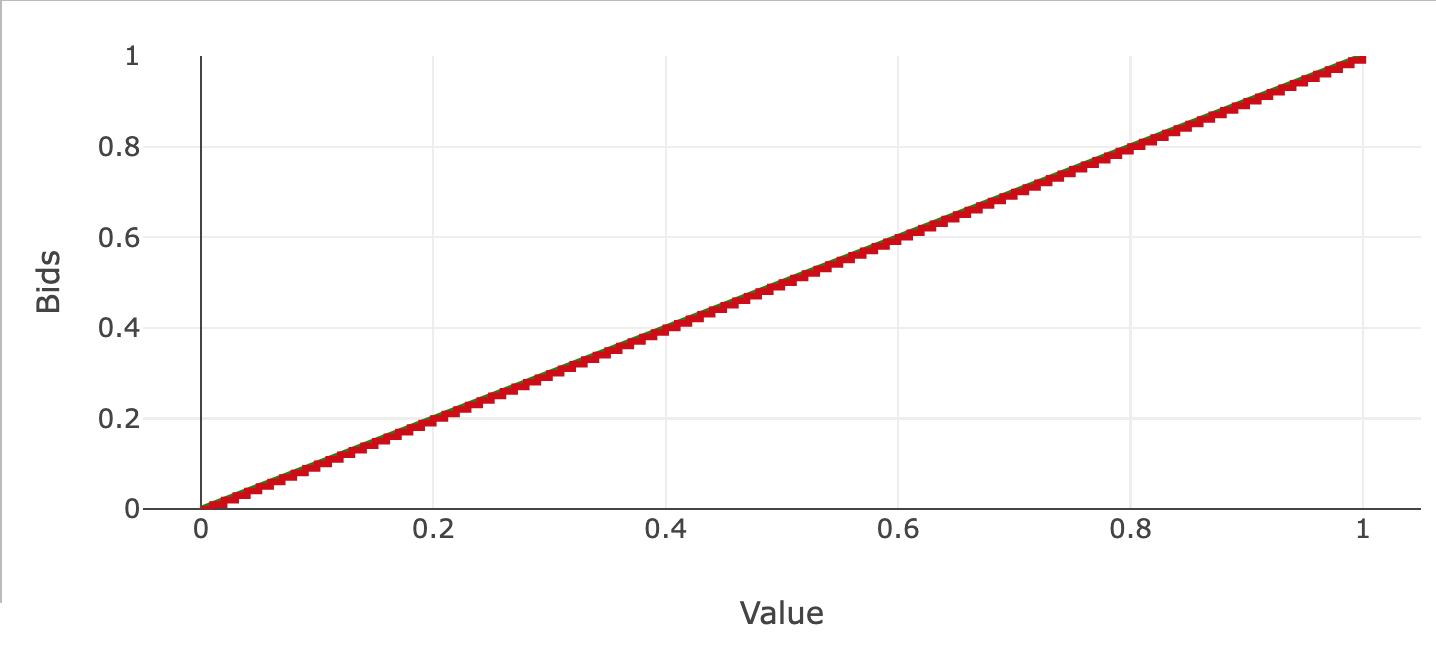}
    \caption{Bidding strategies for second round of the second price sequential sales auction with 3 bidders, 2 goods and reserve prices $r_1=0,r_2=0.5$, when the first item was not sold in the first round. Green denotes the theoretical prediction and red our found strategies.}
    \label{fig:res_3_2_sp_3}
\end{figure}

\begin{figure}[H]
    \centering
    \includegraphics[trim=1mm 1mm 1mm 1mm, clip,width=\columnwidth]{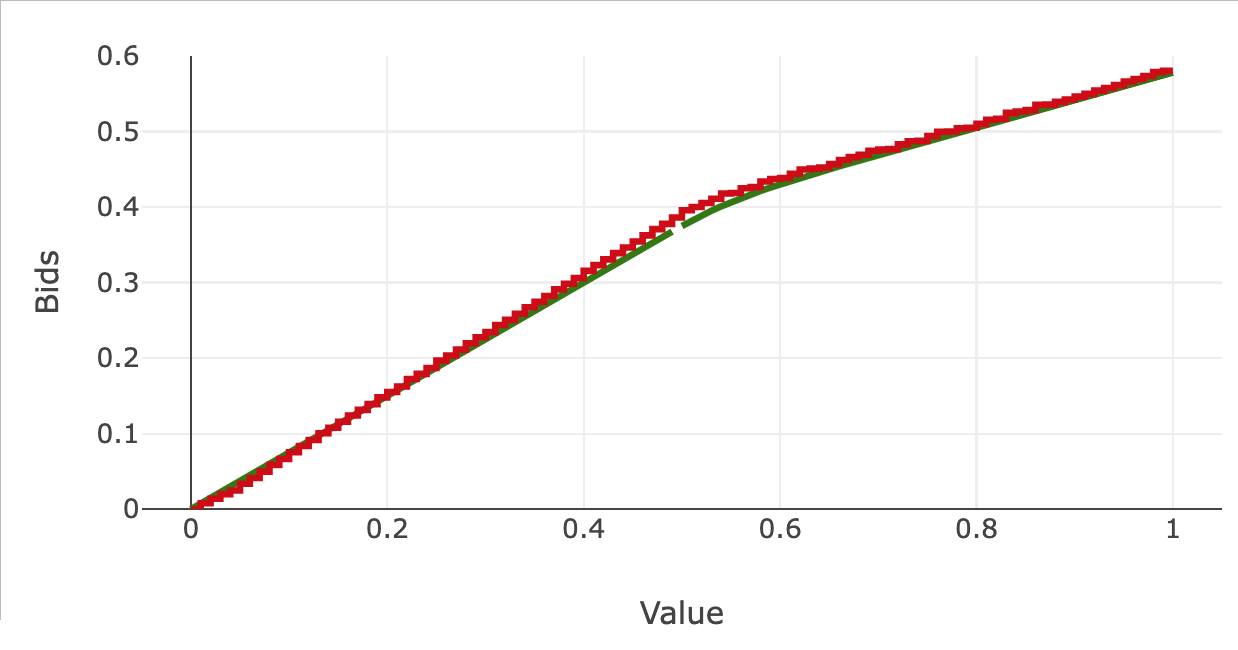}
    \caption{Bidding strategies for first round of the first price sequential sales auction with 4 bidders, 2 goods and reserve prices $r_1=0,r_2=0.5$. Green denotes the theoretical prediction and red our found strategies.}
    \label{fig:res_4_2_fp_1}
\end{figure}
\begin{figure}[H]
    \centering
    \includegraphics[trim=1mm 1mm 1mm 1mm, clip,width=\columnwidth]{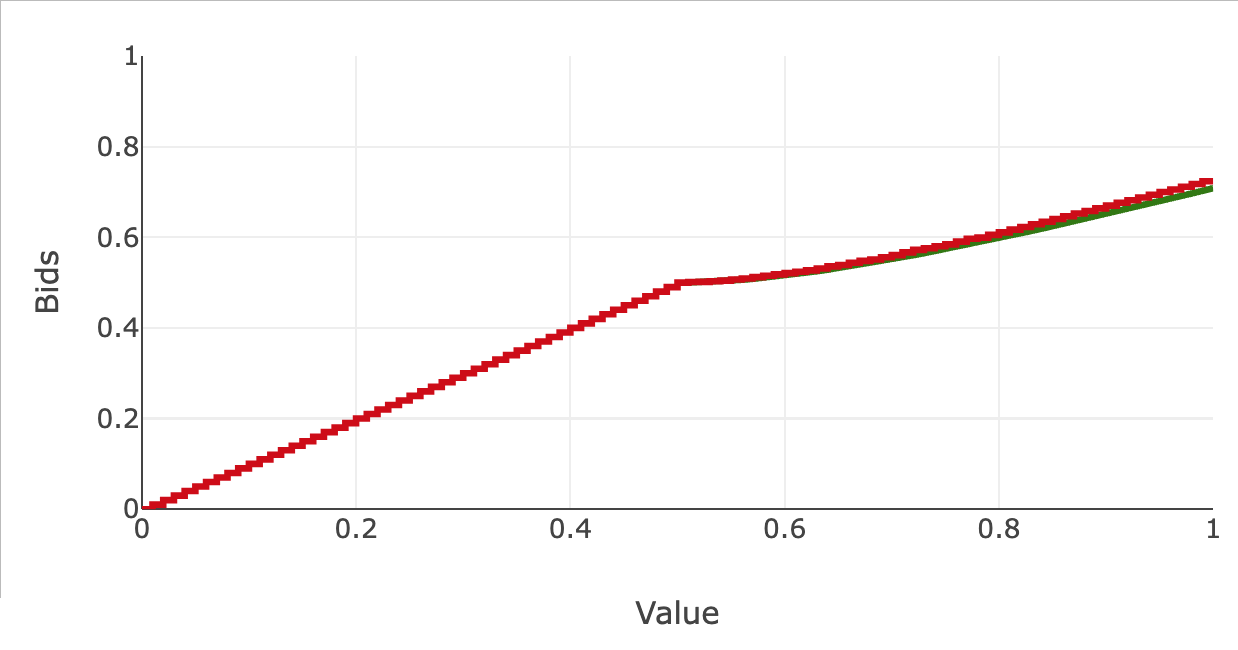}
    \caption{Bidding strategies for second round of the first price sequential sales auction with 3 bidders, 2 goods and reserve prices $r_1=0,r_2=0.5$, when the first item was sold in the first round. Green denotes the theoretical prediction and red our found strategies.}
    \label{fig:res_4_2_fp_2}
\end{figure}

\begin{figure}[H]
    \centering
    \includegraphics[trim=1mm 1mm 1mm 1mm, clip,width=\columnwidth]{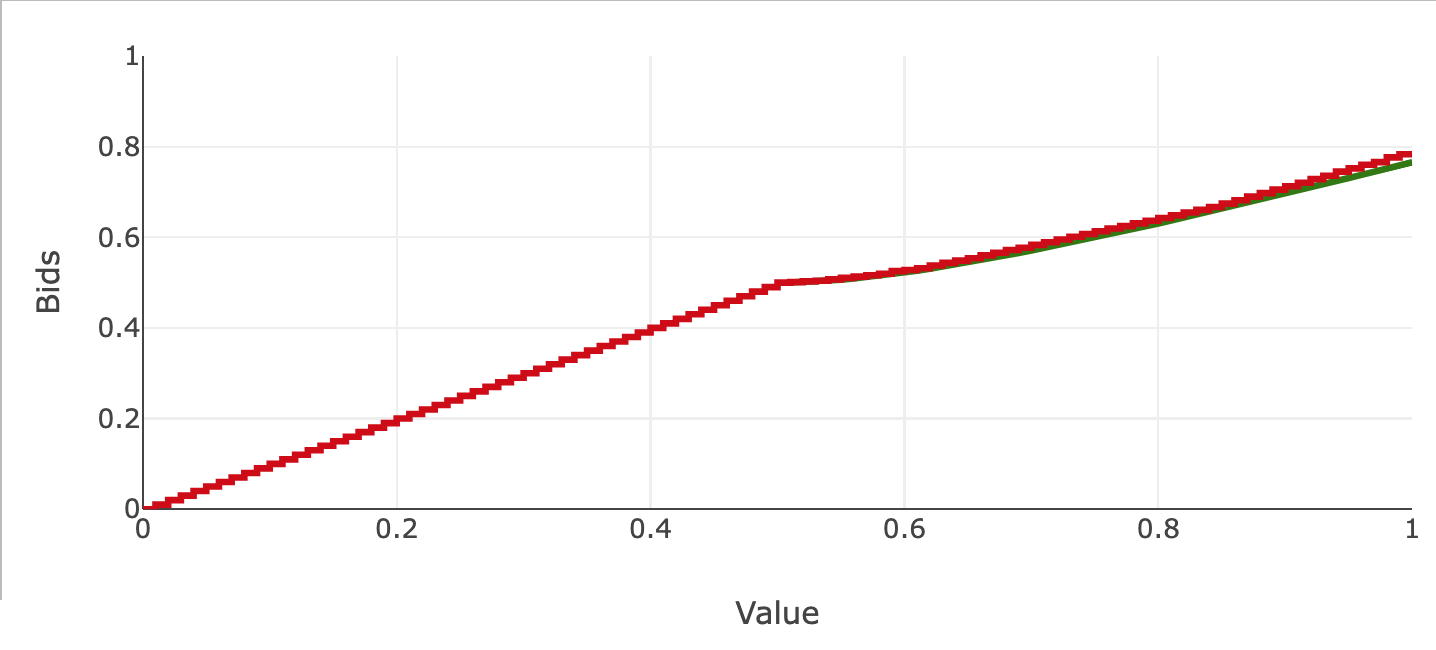}
    \caption{Bidding strategies for second round of the first price sequential sales auction with 3 bidders, 2 goods and reserve prices $r_1=0,r_2=0.5$, when the first item was not sold in the first round. Green denotes the theoretical prediction and red our found strategies.}
    \label{fig:res_4_2_fp_3}
\end{figure}

\begin{figure}[H]
    \centering
    \includegraphics[trim=1mm 1mm 1mm 1mm, clip,width=\columnwidth]{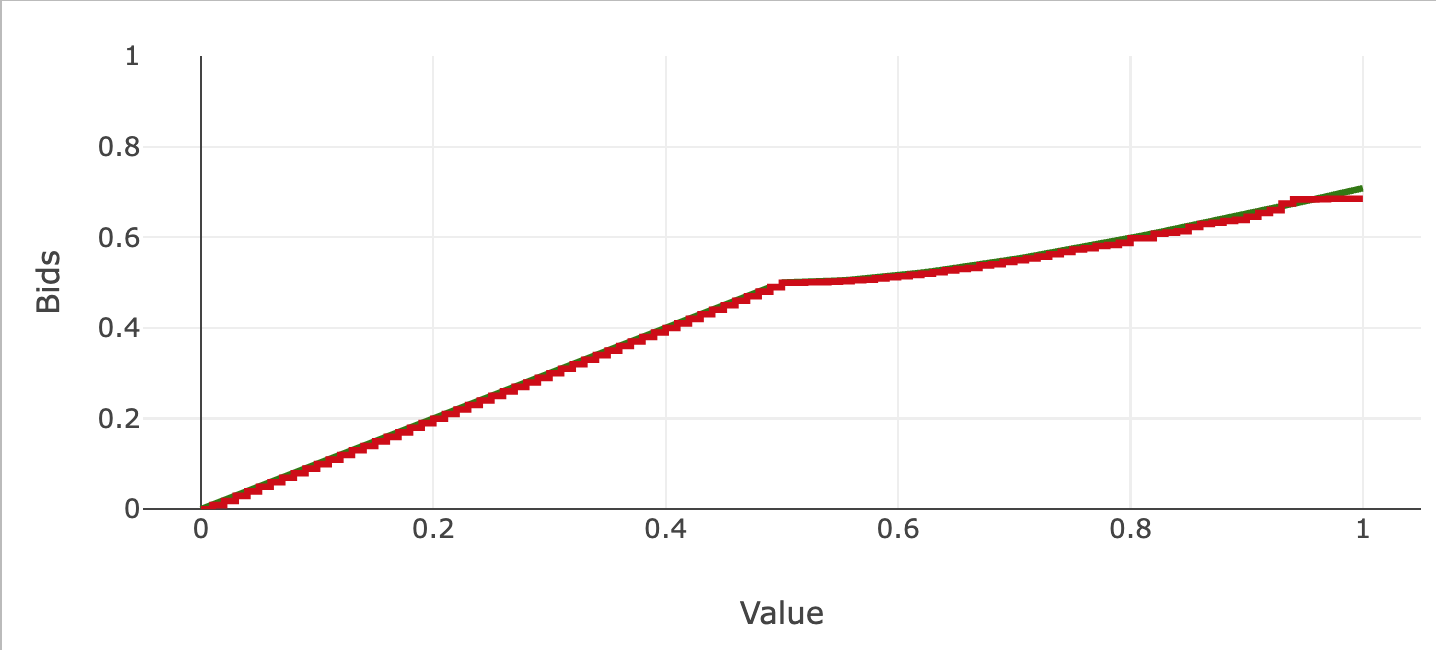}
    \caption{Bidding strategies for first round of the second price sequential sales auction with 4 bidders, 2 goods and reserve prices $r_1=0,r_2=0.5$. Green denotes the theoretical prediction and red our found strategies.}
    \label{fig:res_4_2_sp_1}
\end{figure}
\begin{figure}[H]
    \centering
    \includegraphics[trim=1mm 1mm 1mm 1mm, clip,width=\columnwidth]{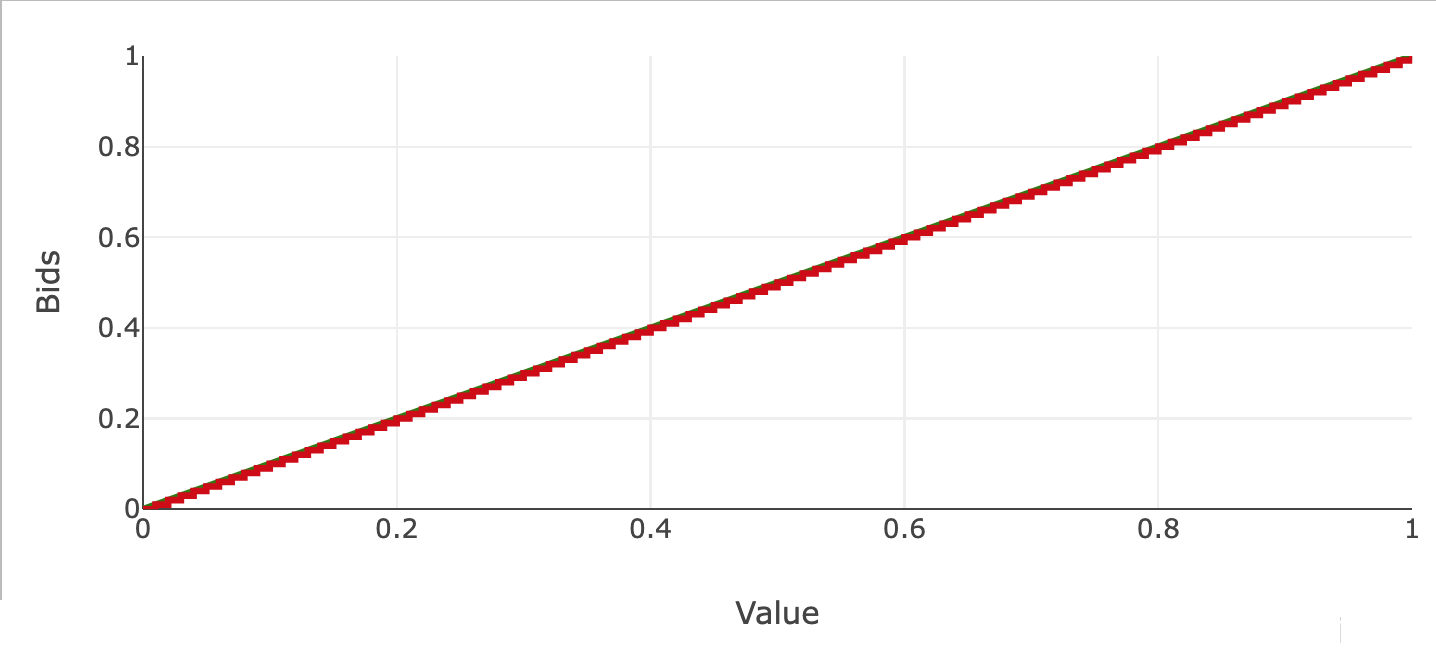}
    \caption{Bidding strategies for second round of the second price sequential sales auction with 3 bidders, 2 goods and reserve prices $r_1=0,r_2=0.5$, when the first item was sold in the first round. Green denotes the theoretical prediction and red our found strategies.}
    \label{fig:res_4_2_sp_2}
\end{figure}

\begin{figure}[H]
    \centering
    \includegraphics[trim=1mm 1mm 1mm 1mm, clip,width=\columnwidth]{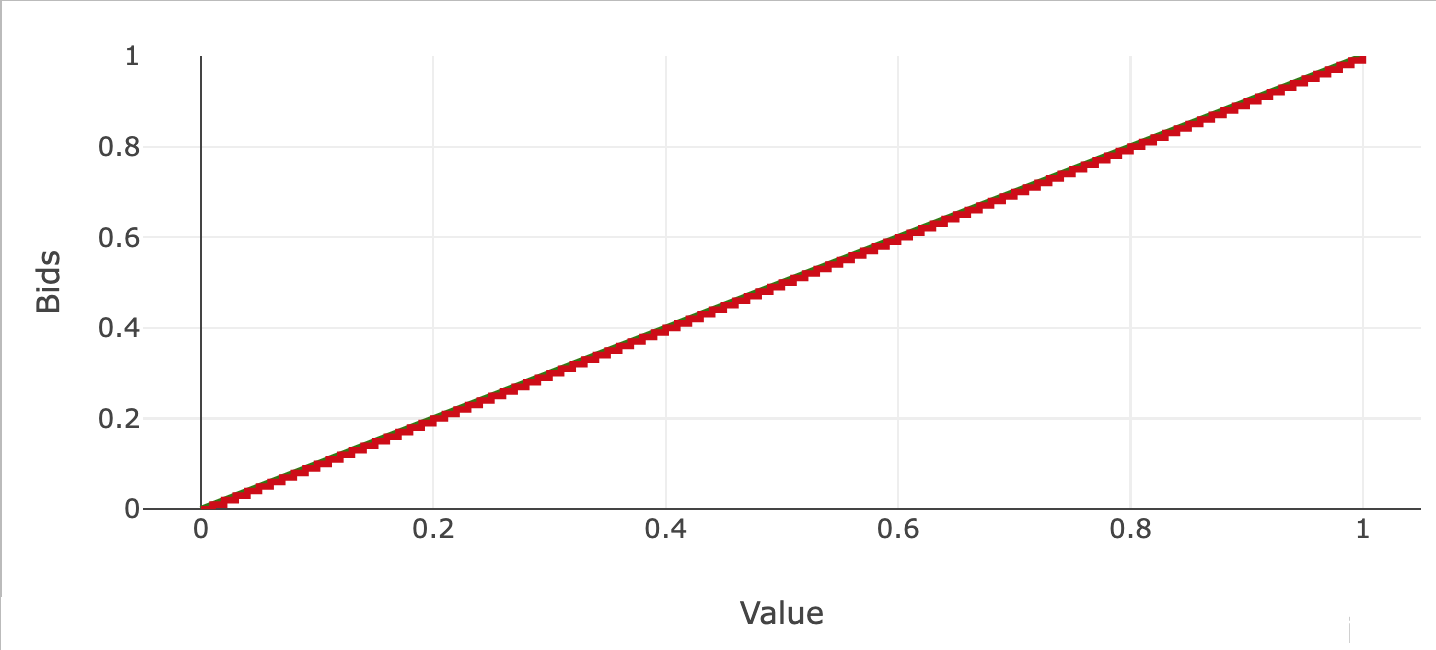}
    \caption{Bidding strategies for second round of the second price sequential sales auction with 3 bidders, 2 goods and reserve prices $r_1=0,r_2=0.5$, when the first item was not sold in the first round. Green denotes the theoretical prediction and red our found strategies.}
    \label{fig:res_4_2_sp_3}
\end{figure}

\begin{figure}[H]
    \centering
    \includegraphics[trim=1mm 1mm 1mm 1mm, clip,width=\columnwidth]{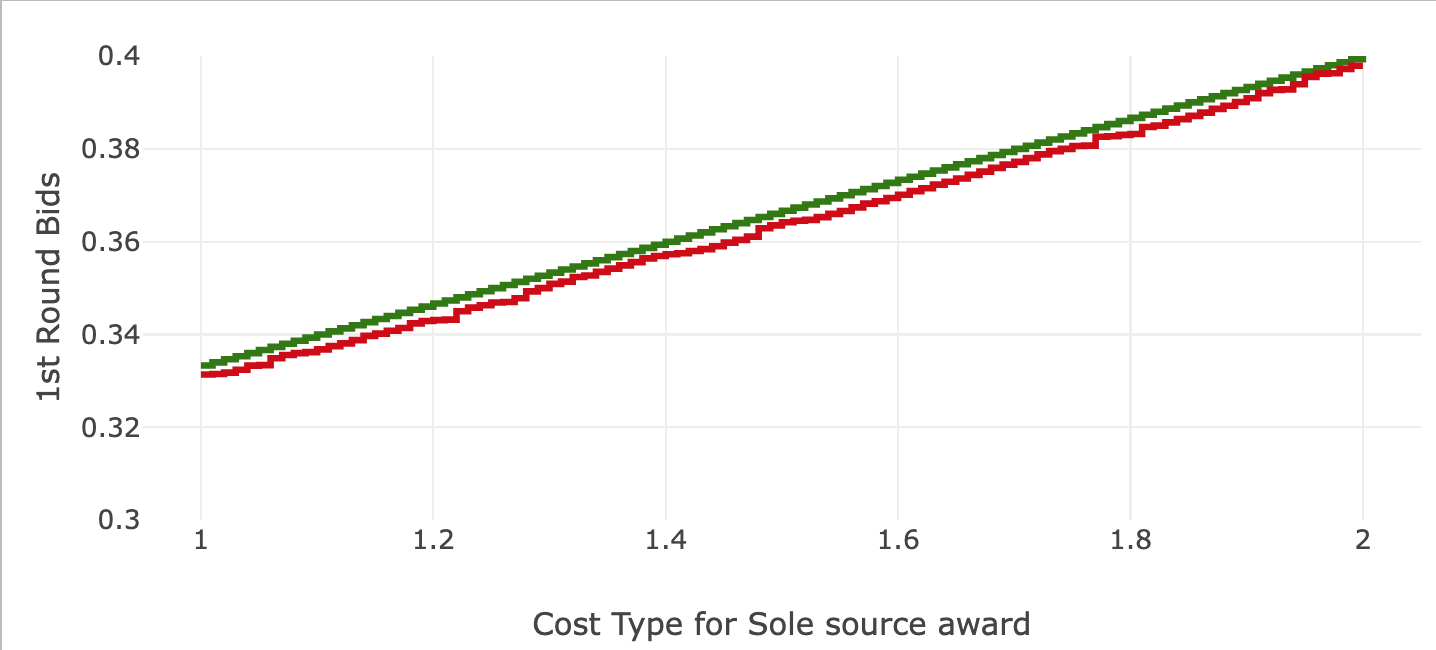}
    \caption{Bidding strategies in the first round of the split award auction with 3 bidders. Green denotes the theoretical prediction and red our found strategies.}
    \label{fig:kokott3_2_1}
\end{figure}

\begin{figure}[H]
    \centering
    \includegraphics[trim=1mm 1mm 1mm 1mm, clip,width=\columnwidth]{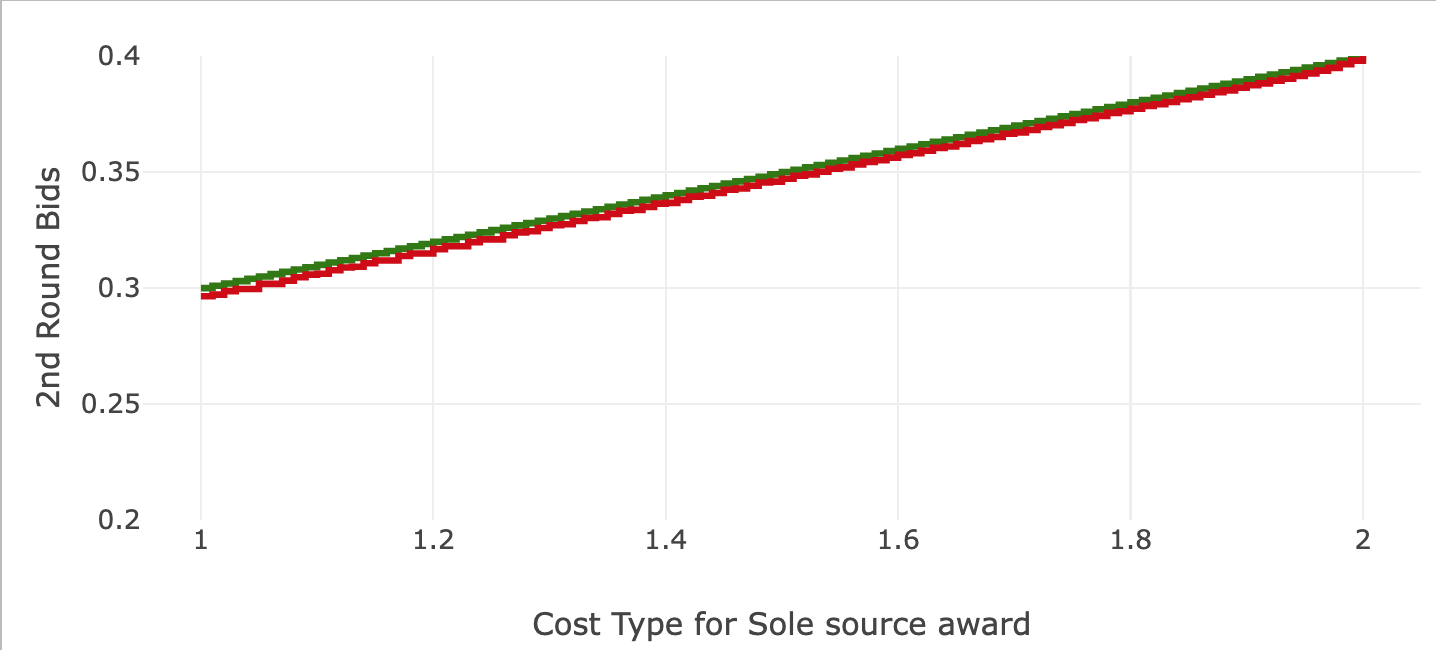}
    \caption{Bidding strategies in the second round of the split award auction with 3 bidders. Green denotes the theoretical prediction and red our found strategies.}
    \label{fig:kokott3_2_2}
\end{figure}

\begin{figure}[H]
    \centering
    \includegraphics[trim=1mm 1mm 1mm 1mm, clip,width=\columnwidth]{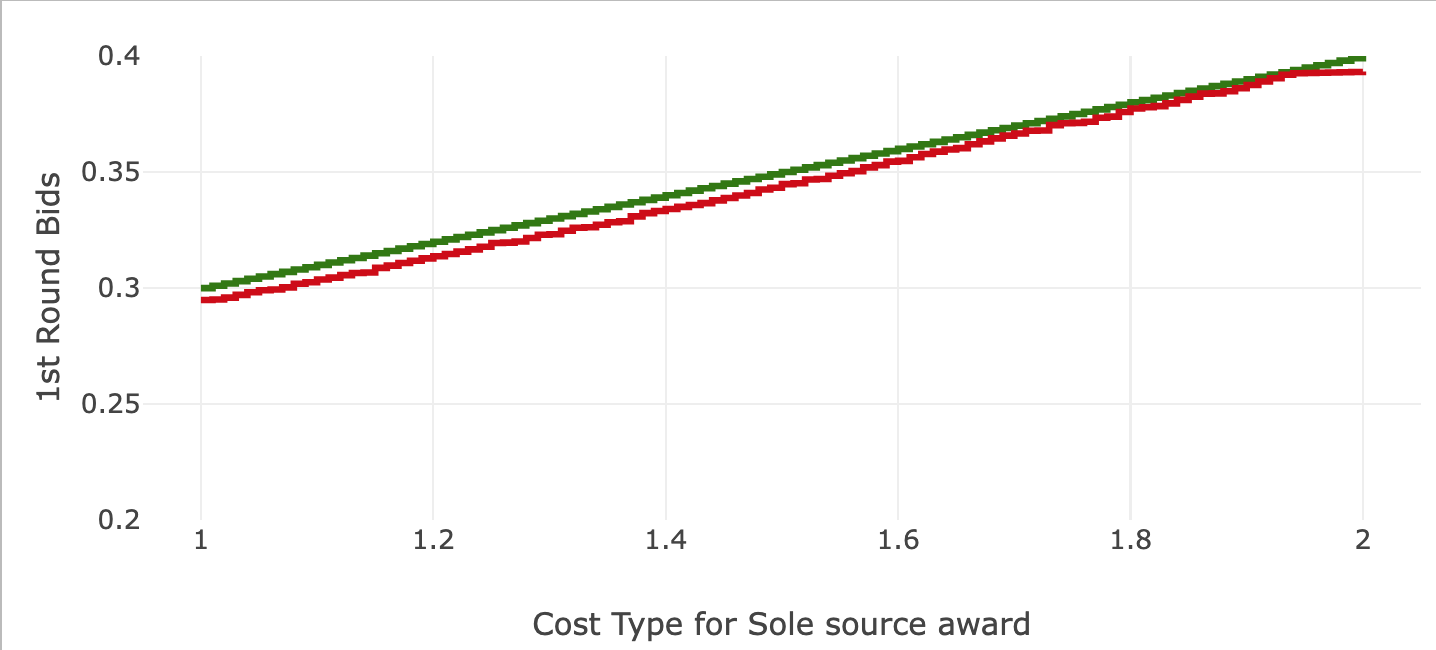}
    \caption{Bidding strategies in the first round of the split award auction with 4 bidders. Green denotes the theoretical prediction and red our found strategies.}
    \label{fig:kokott_4_2_1}
\end{figure}

\begin{figure}[H]
    \centering
    \includegraphics[trim=1mm 1mm 1mm 1mm, clip,width=\columnwidth]{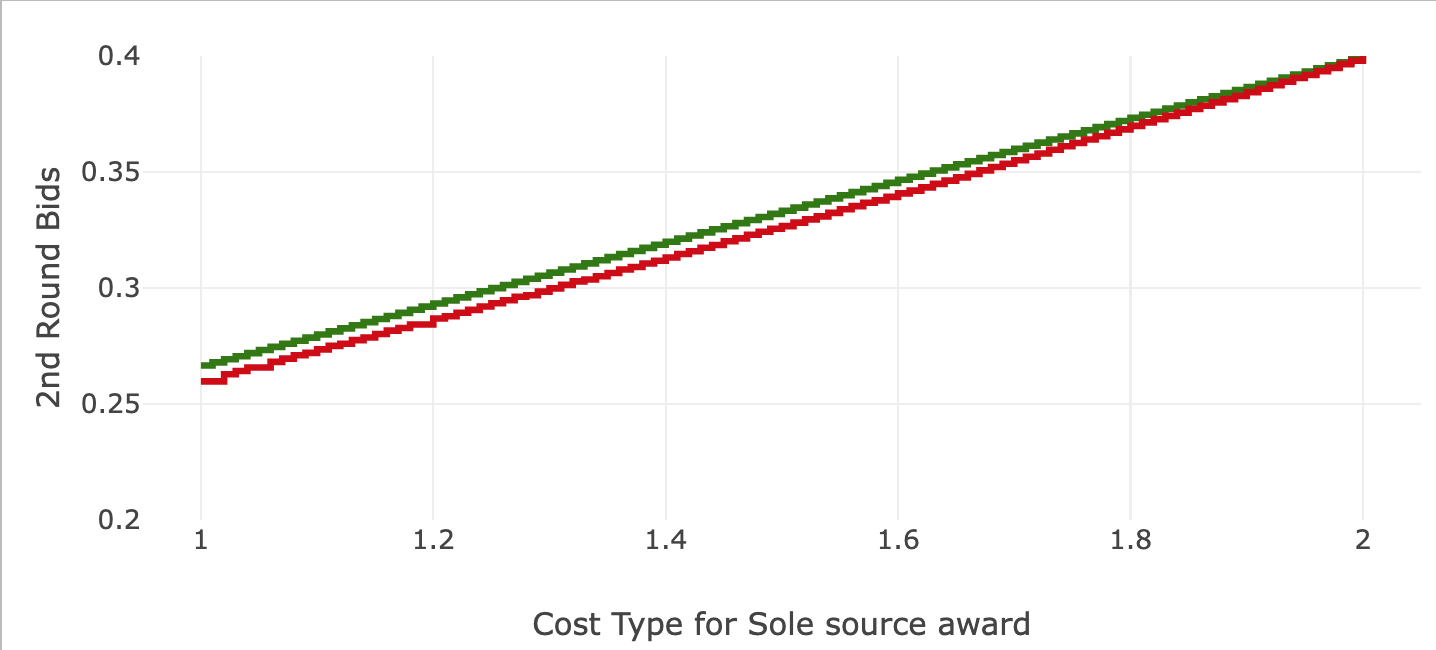}
    \caption{Bidding strategies in the first round of the split award auction with 4 bidders. Green denotes the theoretical prediction and red our found strategies.}
    \label{fig:kokott_4_2_2}
\end{figure}

\begin{figure}[H]
    \centering
    \includegraphics[trim=1mm 1mm 1mm 1mm, clip,width=\columnwidth]{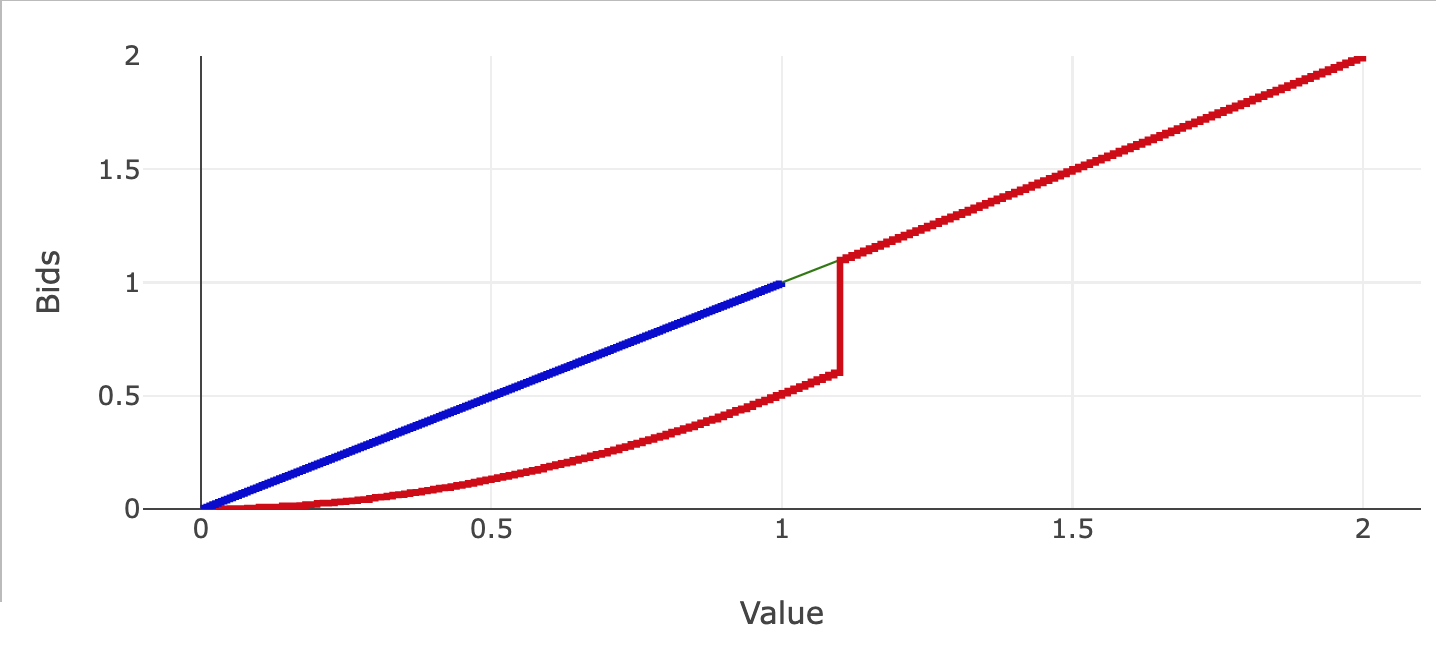}
    \caption{Bidding strategies in the first round of the sequential LLG auction. Green denotes truthful bidding, blue the strategy of the local bidder and red the strategy of the global bidder.}
    \label{fig:llg_1_app}
\end{figure}

\begin{figure}[H]
    \centering
    \includegraphics[trim=1mm 1mm 1mm 1mm, clip,width=\columnwidth]{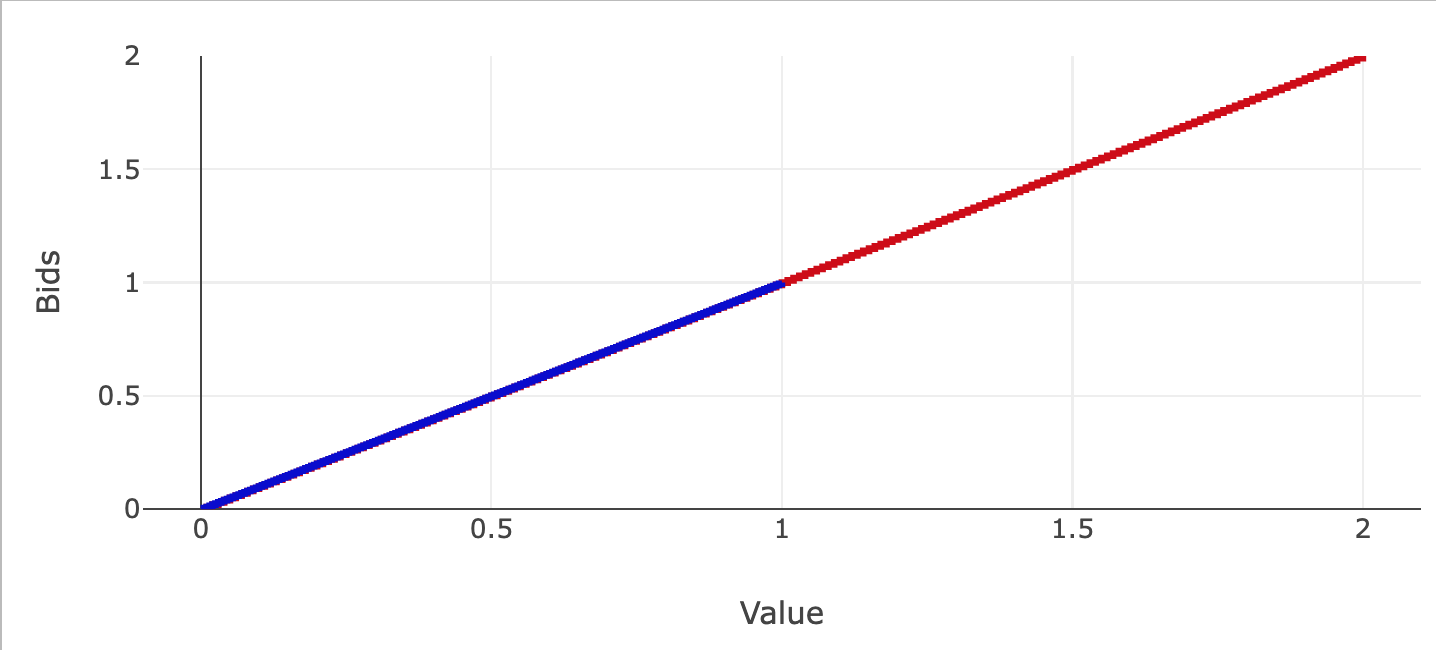}
    \caption{Bidding strategies in the second round of the sequential LLG auction when bidder 3 has won $A$. Green denotes truthful bidding, blue the strategy of the local bidder and red the strategy of the global bidder.}
    \label{fig:llg_2}
\end{figure}

\end{document}